\documentclass[%
mathleft,%
final,%
]{an}
\usepackage{graphicx}
\usepackage{times}
\usepackage[rightcaption]{sidecap}
\newcommand{\Halpha}{H$\alpha$}

\newcommand{\kms}{km\,s$^{-1}$}
\newcommand{\ms}{m\,s$^{-1}$}
\newcommand{\cms}{cm\,s$^{-1}$}

\begin{document}

\Pagespan{324}{}% Document's page range.
% If second parameter is left empty, the last page is computed automatically.
\Yearpublication{2015}%
\Yearsubmission{2015}%
\Month{4}%
\Volume{336}%
\Issue{4}%
\DOI{10.1002/asna.201512172}%

\title{PEPSI: The high-resolution \'echelle spectrograph and polarimeter for the Large Binocular Telescope}

\author{K.\,G. Strassmeier\inst{1}\thanks{Corresponding author:
{kstrassmeier@aip.de}}, I. Ilyin\inst{1}, A. J\"arvinen\inst{1}, 
M. Weber\inst{1}, M. Woche\inst{1}, S.\,I. Barnes\inst{1}, S.-M. Bauer\inst{1},
E. Beckert\inst{2}, W. Bittner\inst{1}, R. Bredthauer\inst{3}, 
T.\,A. Carroll\inst{1}, C. Denker\inst{1}, F. Dionies\inst{1},
I. DiVarano\inst{1},\linebreak D. D\"oscher\inst{1}, T. Fechner\inst{1},
D. Feuerstein\inst{1}, T. Granzer\inst{1}, T. Hahn\inst{1},
G. Harnisch\inst{2}, A. Hofmann\inst{1},\linebreak M. Lesser\inst{4},
J. Paschke\inst{1}, S. Pankratow\inst{1}, V. Plank\inst{1}, 
D. Pl\"uschke\inst{1}, E. Popow\inst{1}, D. Sablowski\inst{1},
 \and\linebreak J. Storm\inst{1}}

\titlerunning{The Potsdam Echelle Polarimetric and Spectroscopic Instrument (PEPSI) for the LBT}
\authorrunning{K.\,G. Strassmeier et al.}

\institute{Leibniz-Institute for Astrophysics Potsdam (AIP), An der Sternwarte 16, 14482 Potsdam,
Germany
\and
Fraunhofer-Institute for Applied Optics and Precision Engineering (IOF), Albert-Einstein-Str. 7, 07745 Jena, Germany
\and
Semiconductor Technology Associates (STA), 27122 Paseo Espada Ste 1004, San Juan Capistrano, CA\,92675, U.S.A.
\and
University of Arizona, Imaging Technology Laboratory (ITL), 325 S. Euclid Ave, Suite 117, Tucson, AZ\,85721, U.S.A.}

\received{2015 Feb 27}
\accepted{2015 Apr 21}
\publonline{2015 Jun 10}

%% bis hierhin, 29.4.15


\abstract{%
PEPSI is the bench-mounted, two-arm, fibre-fed and stabilized Potsdam Echelle Polarimetric and Spectroscopic Instrument for the 2$\times$8.4\,m Large Binocular Telescope (LBT). Three spectral resolutions of either 43\,000, 120\,000 or 270\,000 can cover the entire optical/red wavelength range from 383 to 907\,nm in three exposures. Two 10.3k$\times$10.3k CCDs with 9-$\mu$m pixels and peak quantum efficiencies of 94--96\,\%\ record a total of 92 \'echelle orders. We introduce a new variant of a wave-guide image slicer with 3, 5, and 7 slices and peak efficiencies between 92--96\,\%. A total of six cross dispersers cover the six wavelength settings of the spectrograph, two of them always simultaneously. These are made of a VPH-grating sandwiched by two prisms. The peak efficiency of the system, including the telescope, is 15\,\%\ at 650\,nm, and still 11\,\%\ and 10\,\%\ at 390\,nm and 900\,nm, respectively. In combination with the 110\,m$^2$ light-collecting capability of the LBT, we expect a limiting magnitude of $\approx$\,20th mag in $V$ in the low-resolution mode.  The $R$ = 120\,000 mode can also be used with two, dual-beam Stokes $IQUV$ polarimeters. The 270\,000-mode is made possible with the 7-slice image slicer and a 100-$\mu$m fibre through a projected sky aperture of 0.74\arcsec, comparable to the median seeing of the LBT site. The 43\,000-mode with 12-pixel sampling per resolution element is our bad seeing or faint-object mode. Any of the three resolution modes can either be used with sky fibers for simultaneous sky exposures or with light from a stabilized Fabry-P\'erot \'etalon for ultra-precise radial velocities. CCD-image processing is performed with the dedicated data-reduction and analysis package PEPSI-S4S. Its full error propagation through all image-processing steps allows an adaptive selection of parameters by using statistical inferences and robust estimators. A solar feed makes use of PEPSI during day time and a 500-m feed from the 1.8\,m VATT can be used when the LBT is busy otherwise. In this paper, we present the basic instrument design, its realization, and its characteristics. Some pre-commissioning first-light spectra shall demonstrate the basic functionality. \vspace{-2mm}}

\maketitle

%%%%%%%%%%%%%%%%%%%%%%%%%%%%%%%%%%%%%%%%%%%%%%%%%%%%%%%%%%%%%%%%%%%%%%%%

\section{Introduction}

PEPSI is the fibre-fed high-resolution \'echelle spectrograph for the 11.8\,m Large Binocular Telescope (LBT; e.g. Hill et al. \cite{hill}) in Arizona. The spectrograph is designed to utilize the two 8.4\,m apertures of the LBT in a very advantageous way. Each telescope simultaneously provides light from its optical axis as well as from an off-axis section to the spectrograph. On the CCDs, every single \'echelle order consists therefore of four independent spectra, two from the two optical axes of the telescopes and two from the respective off-axis positions. While the on-axis light is always reserved for a ``target'', the off-axis fibres simultaneously allow either sky light, wavelength-calibration light, or one of the two polarimetric beams. In this way the \'echelle order distribution on the two CCDs remains identical for all of above modi. Because the two LBT eyes can be used independently when in their binocular mode, one may use PEPSI in combination with, e.g., one of the two LUCI's (Seifert et al. \cite{seif:luci}) and then obtain a $R$ = 43\,000 optical spectrum and a $R$ = 20\,000 NIR spectrum simultaneously, covering the wavelength range 0.38--2.45\,$\mu$m. Or, if two targets are within $\approx$\,80\arcsec , PEPSI could expose both targets independently and simultaneously.

%------------------------------ F1  The PEPSI LBT System & Observing modes
\begin{SCfigure*}
%\center
\includegraphics[angle=0,width=127mm,clip]{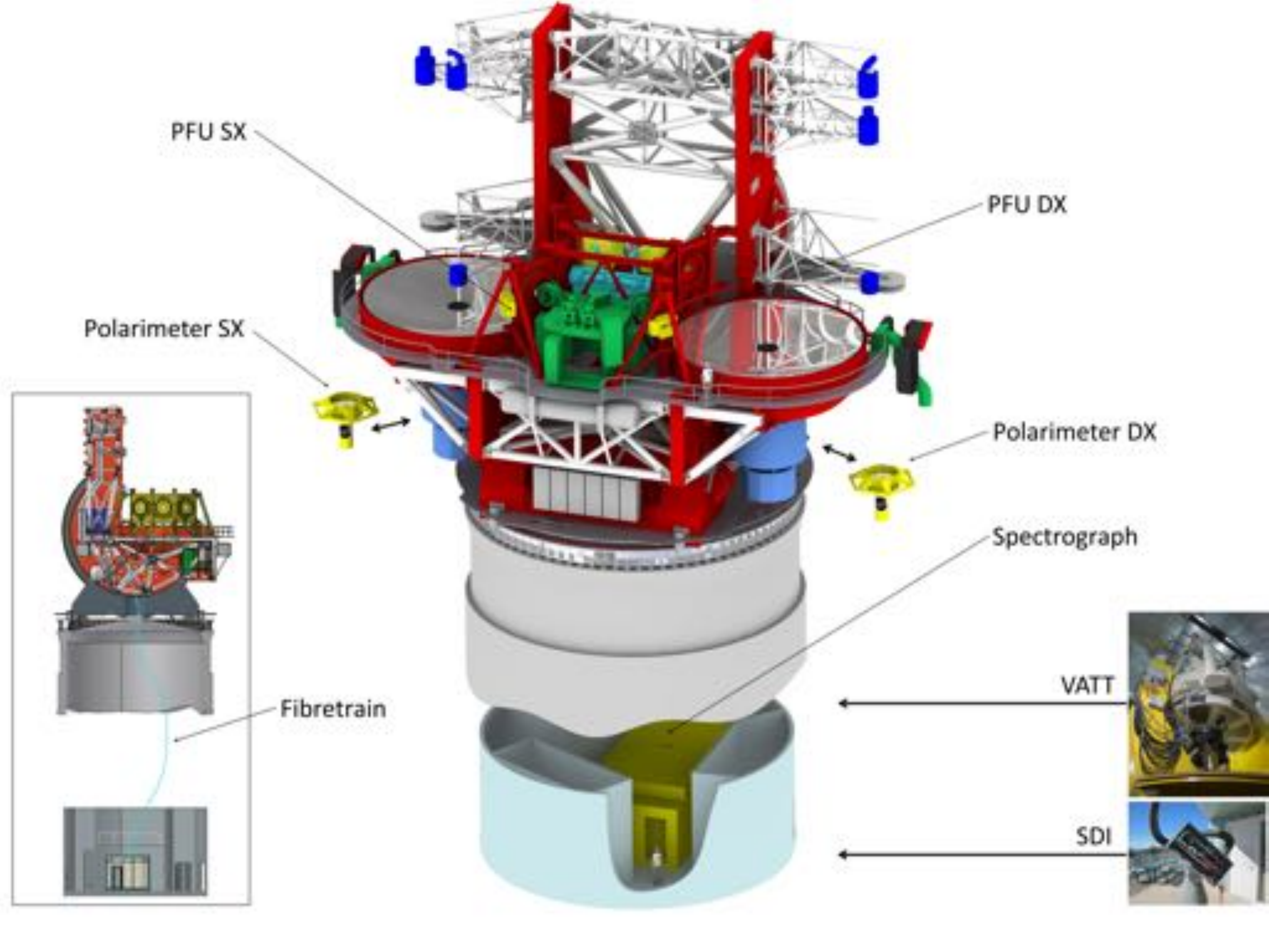}
\caption{System components of PEPSI at the LBT. The fibre-fed bench-mounted spectrograph is located in the pier of the telescope at zero level. It can be used during nighttime by the LBT or the VATT, and during daytime by a small solar telescope dubbed SDI. Integral light is fed to the spectrograph via two Permanent Fibre Units (PFU) while Stokes $IQUV$ light is provided by two polarimeters from the straight-through Gregorian foci of the LBT. }
\label{F-sys}
\end{SCfigure*}

Non-polarized (``integral'') light is fed to the spectrograph via two permanently mounted focal stations in one of the four pairs of bent f/15 Gregorian foci, thereby providing a standby spectrograph with only a 15-min turn-around time in order to get on sky. With two identical but independent polarimeters in each of the direct (rotationally symmetric) f/15 foci, it allows the simultaneous observation of circularly and linearly polarized light with unprecedented high spectral and temporal resolution.

The central scientific objective of PEPSI is to better understand the structure and dynamics of surface magnetic fields of solar-type stars and their impact on stellar formation and evolution. The magnetic field of the Sun is the source of its atmospheric activity which also impacts onto Earth and the other planets. Observations of stars of different mass, rotational period, and evolutionary stage will help to place tight constraints on solar-based theories of stellar dynamos. Many different scientific questions are related to plasma dynamics in general, yet most numerical models can not include the proper MHD effects simply due to a lack of observational material. Ultra-high resolution spectroscopy is required for properly tackling many scientific questions in stellar and interstellar medium physics and will also be crucial on a large telescope to address several problems of extragalactic astronomy.  For details, we refer to the PEPSI science-case paper (Strassmeier et al. \cite{sci:case}).

In this paper, we present the basic technical components of PEPSI and discuss their performance as long as it is defined at this early stage.  The paper is structured as follows.
\begin{enumerate}
\item This section
\item System design and user modes
\item Permanent Focal Units (PFUs)
\item Fibre connections
\item The spectrograph
\item Spectrograph environment
\item Calibration units
\item CCD detector system
\item Polarimeters
\item PEPSI control- and data-reduction systems
\item VATT fibre feed
\item Solar-Disk-Integration (SDI) telescope
\item Pre-commissioning spectra
\item Summary.
\end{enumerate}

% ------------------------------------------------------------------------------------------------------------------------------
\section{System design and user modes}\label{S-System}

The PEPSI system design was driven by the binocular structure of the telescope and the wish to operate a polarimeter together with a permanently available fibre feed for integral light. Because the prime and the straight-through foci were already occupied, i.e. the two rotationally symmetric foci of the LBT, we split the entire PEPSI feed into an integral-light part and a polarimetric-light part. The latter is in a symmetric focus and the former got its own new and permanent focus. Therefore, the main operations burden for spectropolarimetry with PEPSI is to dismount the two MODS instruments (Multi Object Double Spectrograph; Pogge et al. \cite{mods}) from the two straight-through foci whenever the polarimeters are used. The integral-light feed, on the other hand, is available on very short notice within approximately 15 min. The integral-light mode of PEPSI is also available in remote control from Potsdam.
%The VATT-mode will be available in remote control from Tucson.

% ------------------------------ Table "System specs"
\begin{table}
\caption{Overall spectrograph characteristics. }\label{T1}
\begin{tabular}{lllllll}
\hline  \noalign{\smallskip}
Fibre    & Sky        &  & \multicolumn{2}{c}{Slicer}  & ~~$R=$ & ~~~$R$       \\
Diam.   & Diam.     & $n$ & Width & Width & ~$\lambda/\Delta\lambda$    & Sampl. \\
($\mu$m) & ~(\arcsec) & & ($\mu$m) &  ~~(\arcsec) & & (pix)\\
% \noalign{\smallskip}
 \hline  \noalign{\smallskip}
 100 & 0.74 & 7 & ~~70  & 0.11 & 270\,000 & ~~2.0  \\
 200 & 1.5  & 5 & 175 & 0.30 & 120\,000 & ~~4.2 \\
 300 & 2.3  & 3 & 500 & 0.77 & ~~43\,000  & 12.1 \\
% \noalign{\smallskip} 
\hline
\end{tabular}

\vspace{1.5mm}Notes: $n$ = number of slices.
\end{table}

\begin{figure}
{\emph{a)} Wavelength coverage:}

\includegraphics[angle=0,width=82mm,clip]{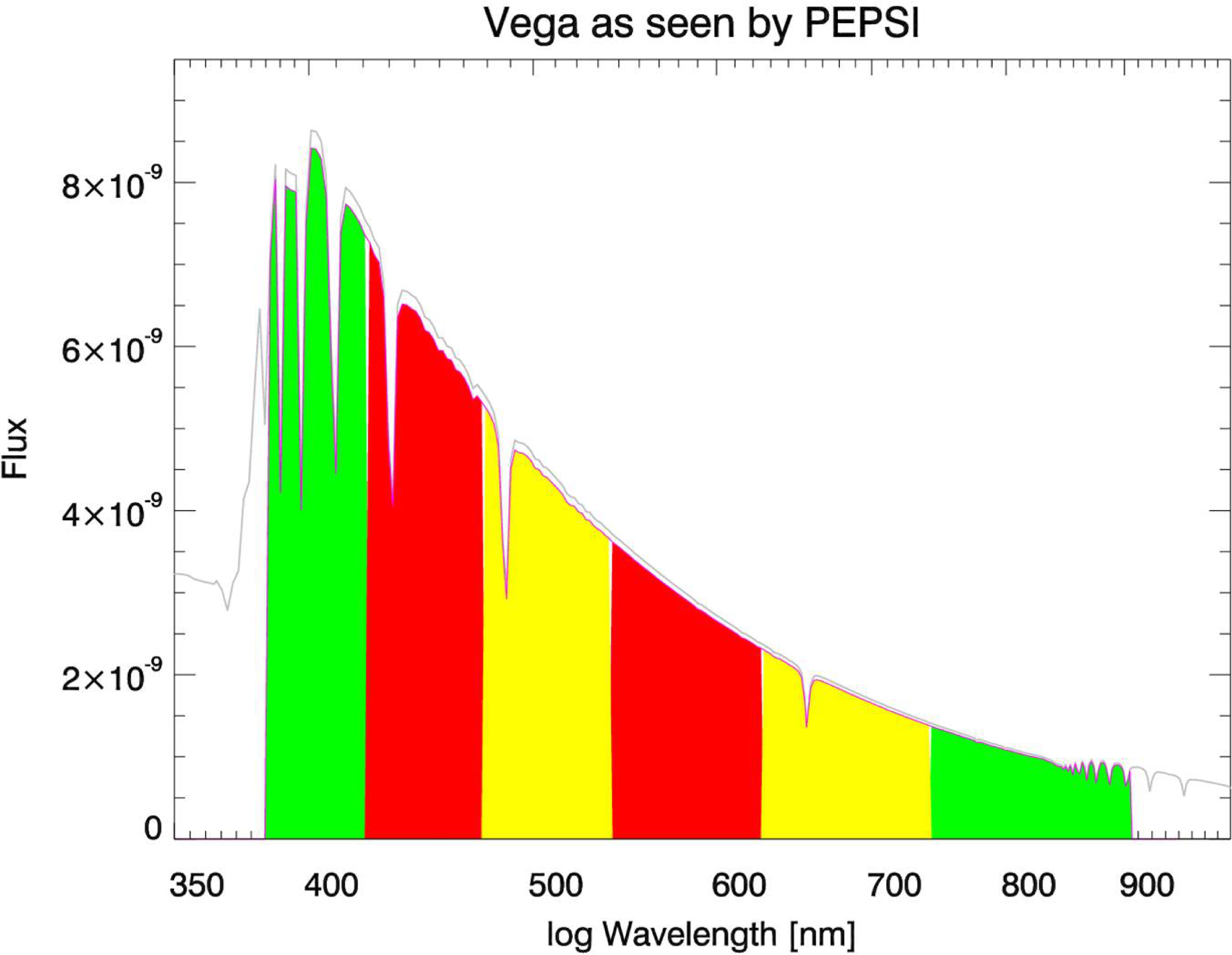}\\
{\emph{b)} Resolving power:}

\includegraphics[angle=0,width=82mm,clip]{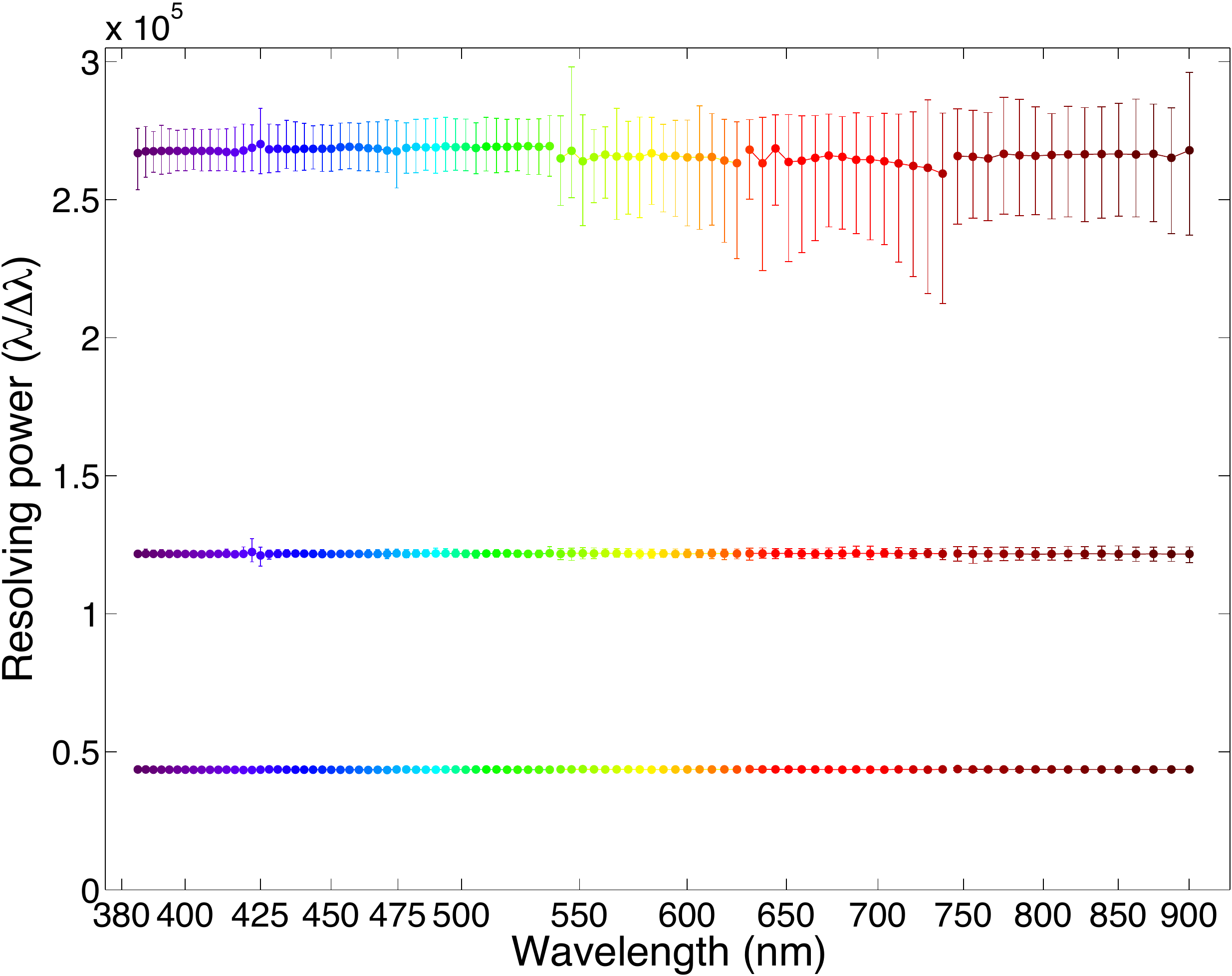}
\caption{\emph{a}) Shown is the wavelength coverage (colored area) of each cross disperser for a spectrum  example of Vega. \emph{b}) Resolving power as a function of wavelength. Each dot is the mean resolving power within an \'echelle order. The bars denote the range of resolutions for the respective free spectral ranges. }\label{F-syseff}
\end{figure}

Figure~\ref{F-sys} shows the system components of PEPSI and their locations on the telescope. Table~\ref{T1} summarizes the spectrograph characteristics. The basic specifications and operations modes are:
\begin{itemize}
\item Integral-light spectroscopy in the wavelength range 383--907\,nm with spectral resolving powers $R=\lambda/\Delta\lambda$, of either 43\,000 (2.3\arcsec\ aperture on the sky), $R$ = 120\,000 (1.5\arcsec ), or $R$ = 270\,000 (0.74\arcsec ). Figure~\ref{F-syseff} shows the wavelength coverage and the spectral resolving power as a function of wavelength (for definitions see Sect.~\ref{grating}).
\item Polarized-light $IQUV$ spectroscopy at a fixed spectral resolution of $R$ = 120\,000.
\item Full wavelength coverage in three exposures for all of above modes. Note that the two cross dispersers in the ``blue'' and the ``red'' arm of PEPSI that are adjacent in wavelength can not be used simultaneously because of the dichroic beam splitting.
\item Simultaneous  two target spectra from both LBT eyes, the integral-light mode provides the user the choice between two additional background-sky spectra per \'echelle order or two wavelength-calibration spectra from a Fabry-P\'erot \'etalon.
\item Each polarimeter provides simultaneous spectra of the ordinary and the extra-ordinary beam. Two polarimetric science modes are currently defaulted to circular Stokes $V$ and linear Stokes $QU$. It takes two spectra for Stokes $V$ and four spectra for Stokes $Q$ and $U$.
\item Daily solar disk-integrated (SDI) spectra with a resolution of $R$ = 270\,000 are provided from an   automated 1-cm binocular telescope.
\item A light feed from the 1.8\,m Vatican Advanced Technology Telescope (VATT) for a fixed spectral resolution of $R$ = 120\,000. This is mostly recommended for the red arm of PEPSI due to the strong blue attenuation in the 500m-long fibre path.
\end{itemize}

The overall efficiency of \'echelle spectrographs is typically $\approx$10\,\%\ including the telescope (and typically at the laser wavelength of 532\,nm). Dual-pass camera-collimator designs are more efficient than white-pupil designs but have other issues (see, e.g., the discussion by Bernstein et al. \cite{mike} and Vogt et al. \cite{apf}). The PEPSI optical train contains up to 66 optical surfaces from the telescope's primary mirror down to the CCDs (64 in the blue train, and 66 in the red train).  However, most of these surfaces are immersed, e.g., the PFUs have 22 surfaces for the 200 and the 300-$\mu$m fibres of which only one surface is a reflection in air and two more are transmissions with air. The 100-$\mu$m fibres have no reflection in air at all. Similar is the case for the spectrograph with 34/32 surfaces (red/blue) but only 5 reflections with air. Its efficiency alone (with CCD) is 32\,\%\ and includes aperture losses, Fresnel scatter, coating transmission, vignetting, and internal transmissions. A detailed listing of efficiencies for all optical elements is given later in Table~\ref{T-effall}.  The overall PEPSI peak efficiency,  including the telescope but excluding the atmosphere, is 15\,\%\ at 650\,nm and still 11\,\%\ at 390\,nm and  10\,\%\ at 900\,nm.

%------------------------------ F2  lam coverage and resolving power   - ARTO w/ Vega

%------------------------------ F3  PFU system design
\begin{SCfigure*}[][!tbh]
%\center
\includegraphics[angle=0,width=126mm,clip]{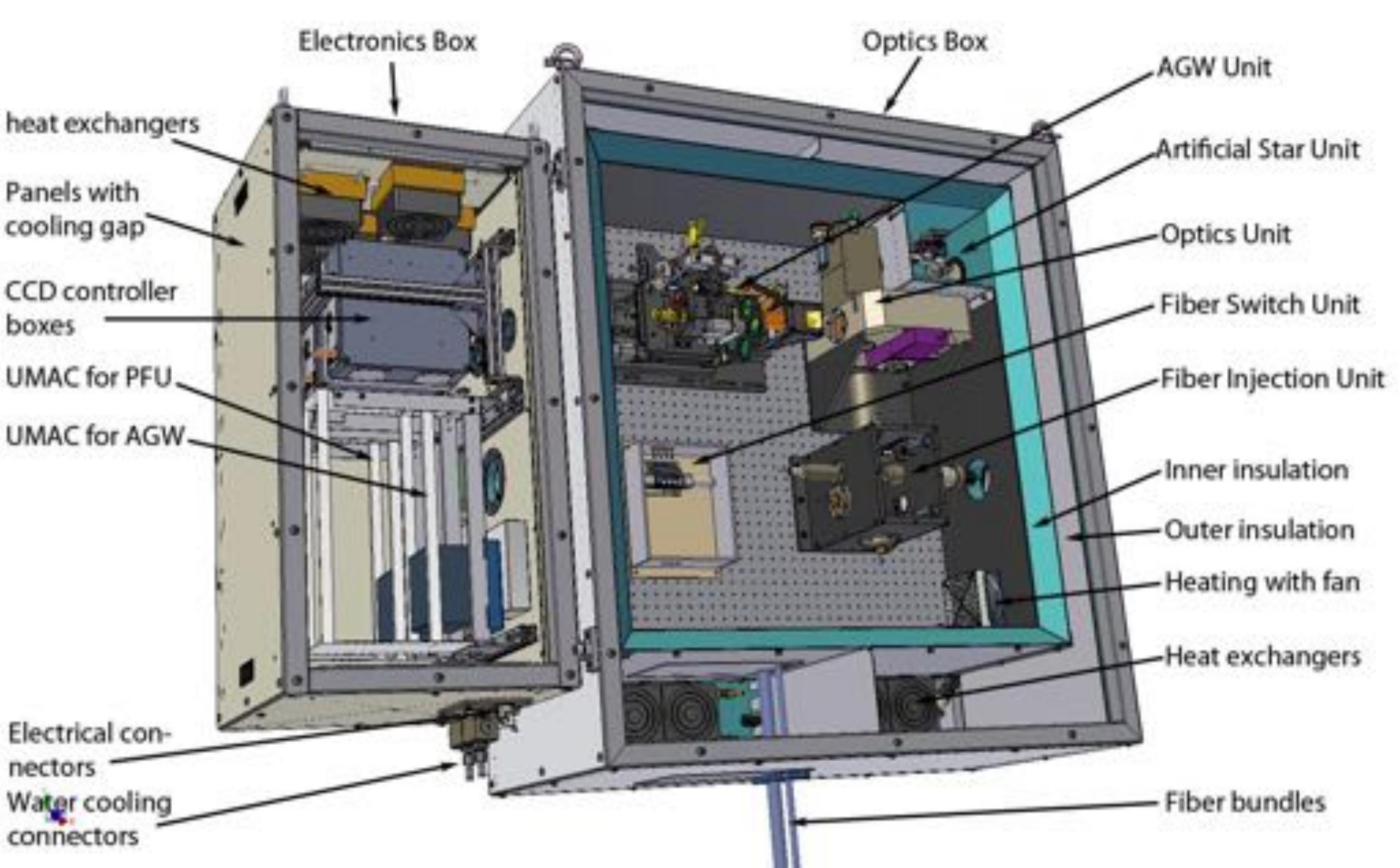}
\caption{System layout of the Permanent Fiber Unit (PFU). Shown is the unit for the DX telescope. The SX unit appears mirrored to the DX units. Two Jakodur-insulated boxes (with an inner insulation and an outer insulation) are wrapped around each other with an air gap in between. Two heat exchangers circulate the air within the gap. The electronics box is attached to the optics box and has its own water cooling system. }\label{F-pfu1}
\end{SCfigure*}

%------------------------------------------------------------------------------------------------------------------
\section{Permanent Focal Units (PFUs)}\label{S-PFU}

PEPSI has two permanently mounted telescope focal stations. These focal stations occupy a dedicated 4th bent Gregorian focus of the LBT located on the instrument platform consecutive to the three regular foci. These PFUs carry their own Acquisition-, Guiding-, and Wave-front sensing (AGW) units for on-axis, low-frequency wavefront sensing as well as target acquisition and guiding. By rotating the telescopes' M3 mirrors to an angle of 47.05$^\circ$ from the optical axis of the central station, the PFUs will see the sky. A look-up table makes sure that the focus is reached and collimated within $\approx$\,10 min. No image de-rotation is required because the guiding system uses only light from the on-axis target. A summary of recent developments for the LBT guiding control subsystem was given by Golota et al. (\cite{golota}).

\subsection{System layout}

\begin{figure*}
{\it a) \hspace{95mm} b)}

\includegraphics[angle=0,width=90mm,clip]{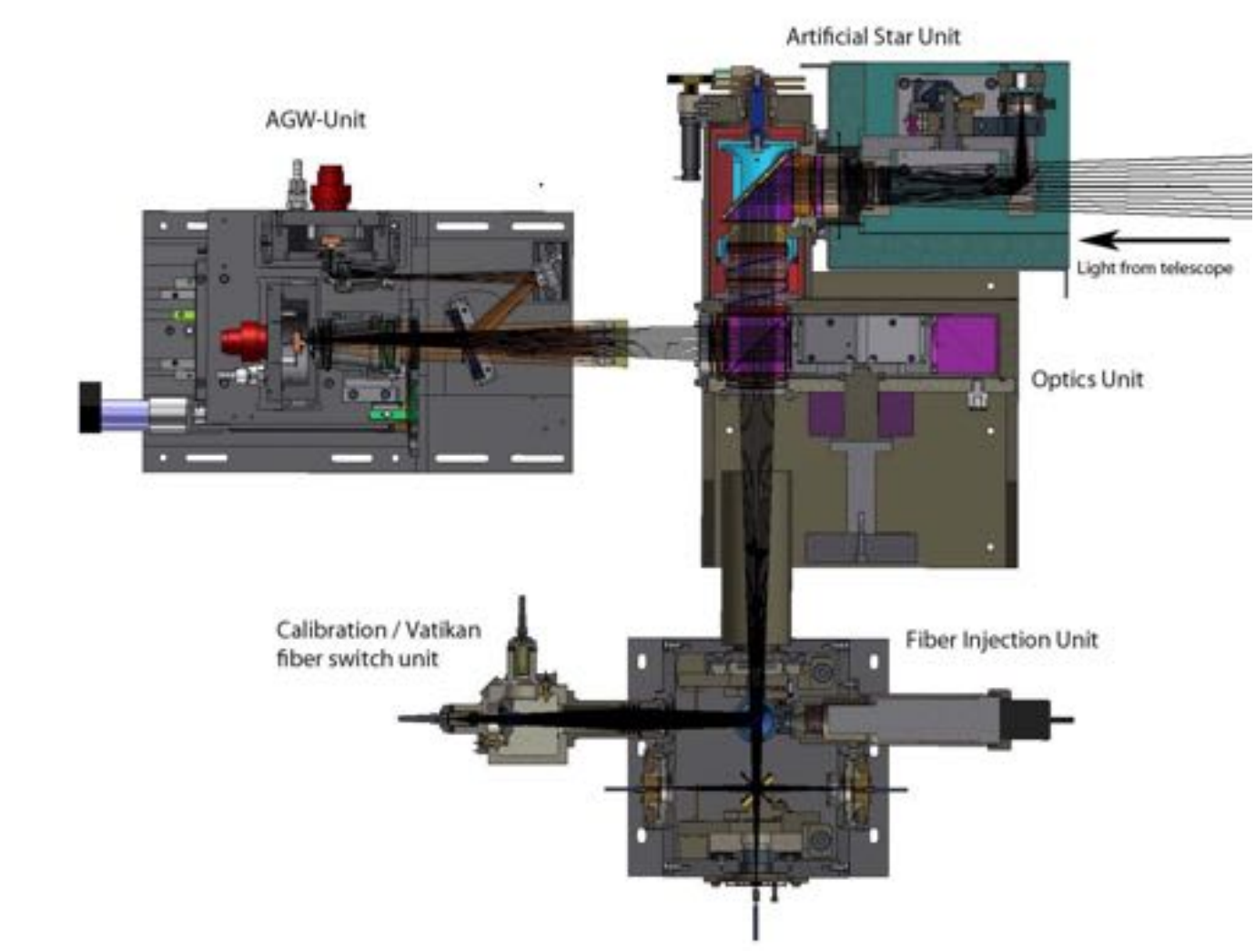}\hspace{10mm}
\includegraphics[angle=0,width=50mm,clip]{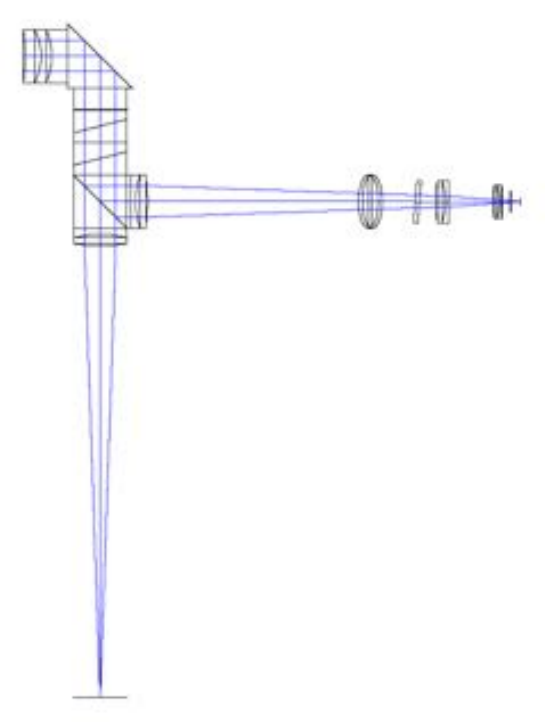}
\caption{Close-up of the PFUs and its optical design. \emph{a}) Zoom on the various optical-mechanical subunits. Top left is the AGW unit, top right the artificial star unit, in the middle the optics unit with its dichroic beamsplitter wheel, and below is the fibre injection unit. Just below the AGW unit is the calibration-light and VATT fibre injection. Light from the LBT enters from the right. \emph{b}) The horizontal beam is the AGW beam, the vertical beam is the fiber feed. Shown are the three-lens collimator, the folding prism, the 4-element ADC, and the beamsplitters with their re-imaging systems (one for the AGW CCDs and one for the fiber injection to the spectrograph). Note that the AGW beam is split into two beams (not shown); one for guiding and one for wavefront sensing. Enlargements of the AGW and the fibre-injection units are shown in Figs.~\ref{F-agw1} and \ref{F-pfu4}, respectively.}\label{F-pfu3}
\end{figure*}

Figure~\ref{F-pfu1} shows the layout of one of the two PFUs\footnote{Note that the two primary LBT mirrors are identified as SX (sinistra) and DX (diritto), and accordingly named are the PFUs.}. Fig.~\ref{F-pfu3}a is a close-up of the optical part of the PFU.  Dividing the light between the science focus and the guiding system takes place in an optical module located 400\,mm ahead of the telescope focal plane. This module consists of an afocal lens creating a parallel beam in which an Atmospheric Dispersion Corrector (ADC) and various (dichroic) beamsplitters are located. Behind this wheel, a camera doublet recreates the f/15 telescope beam. All optics in the parallel beam are immersed with an index-matching liquid reducing reflection losses by a factor of about 4 to less than 3\,\%\ for the entire PFU optics. Glass-to-air and air-to-glass surfaces are broad-band anti-reflection (AR) coated. Note that the on-axis guiding and wavefront sensing systems are identical to the off-axis guiding systems for the polarimeters in the straight-through foci of the telescope, described later in Sect.~\ref{S-Pol}. Acquisition and guiding frames are recorded by the AGW with separate CCD cameras. The guiding camera views the telescope aperture through yet another dichroic beamsplitter simultaneously with the fibre entrance aperture. The latter is redirected through a back reflection of the fibre focal plane via an aluminized diaphragm. Note that the sky fibre is off axis by 6\arcsec\ on the sky. It rotates around the on-axis target with the parallactic angle during integration and one should make sure not to have a second target in its direct path. The wavefront sensing also takes into account that the Shack-Hartmann sensor and the image of the main mirror (pupil image) do not rotate because the sensor is fixed to the main mirror, i.e. there is no need for a rotation of the Zernike polynomials. Once the user has chosen the spectral setting of the spectrograph, i.e. has chosen one of the three cross dispersers in each arm, the system moves automatically to the dichroic beamsplitter that fits the wavelength range \emph{and} the target brightness (see Sect.~\ref{S-bs}). In this layout the guiding wavelength depends on the spectral setting of the spectrograph. Alternatively, a grey beam splitter may be chosen.

%------------------------------ F4  PFU optical design & close-up

\subsection{Optical design}

Fig.~\ref{F-pfu3}b shows the Zemax design. The first optical surface from the telescope's point of view is the three-lens collimator. It collimates the f/15 beam from the telescope via two Ohara FPL51 and one NSL36 lens with free diameters of 50\,mm. Its entrance lens L1 is AR coated and MgF$_2$ protected for the entire wavelength range of the spectrograph with a $<$\,1\,\%\ reflectivity. A prism is optically contacted to the exit surface of lens L3 and operated in total reflection by maintaining an air gap on its backside. It folds the beam by 90$^\circ$  downwards to the ADC. A 19.6\,mm plane-parallel spacer of Ohara NSL36 is contacted to the prism and bridges the distance to the first ADC optical surface. The ADC itself is a Risley-prism mechanism with two optical components, each consisting of two prisms made of Schott N-Pk51 and Ohara PBL6Y glass, respectively, that are rotated with respect to each other in the opposite direction. Below the ADC is a turret wheel with six beam splitters that send  part of the light to the AGW unit for acquisition, guiding and wave front sensing. The last optical elements are two camera doublets tilted by 90$^\circ$ with respect to each other, each made of an Ohara FPL51 and a NSL36 lens. The first doublet re-images the telescope aperture onto the fibres while the second doublet re-images it to the AGW-unit. Their respective exit lenses are again AR coated and MgF$_2$ protected.

To minimize light losses by reflection on the optical surfaces, the entire optics units of the PFUs are filled with immersion oil. Its refraction index is matched to the main glass component and a predefined operation temperature of 12\,$^\circ$C.  Because changes of the focus position on the AGW influences the wave-front sensing, which controls the telescope main mirror, the temperature of the optics unit is kept constant to within $\pm$0.5\,$^\circ$C for a particular day-night cycle for an absolute range 2--3\,\degr C above ambient. The absolute temperature inside the PFUs optics unit is allowed to range between 5--30\,$^\circ$C for the extremes, i.e. a cold winter night and a hot summer day.

\subsection{Dichroic beam-splitter specifications}\label{S-bs}

Each beamsplitter (BS) is designed as a 50\,mm cube cut in halve by 45$^\circ$. The total of 12 beamsplitters are thus made of 24 right-angle prisms of glass N-BK7, with quality of $\lambda/8$ at 633\,nm and 80--50 Scratch \& Dig, and an angular precision better than $\pm$30\arcsec . The diagonal (hypotenuse) surface is used for the dichroic, or grey, coating and is optically glued together with its counterpart prism. Beam diameter at the dichroic layer is 40\,mm. There are six different beam splitters with coatings designed according to the wavelength range of the spectrograph cross dispersers.  A wedge plate with an angle of 1\degr\ is glued to the ``back'' of each cube to redirect the reflected light from the fibre surrounding to the acquisition CCD.

\noindent
\emph{Beamsplitter \#1} is a grey beamsplitter that diverges 1\,\%\ of the light to the AGW and 99\,\% to the spectrograph. This BS is intended to be the default workhorse. At a practical bright limit of 0th magnitude with a standard filter, guider saturation sets in at an exposure time of roughly 10\,ms. By increasing the exposure time to 1\,s and accepting an integrated exposure level of 2$\times10^3$ photoelectrons (centroid accurate to about 0.05\arcsec ), we can observe a 13\fm5 star with this beamsplitter. By using no filter in the guider filter wheel, one reaches  15th magnitude with this BS. The wave-front sensor then gives useful measurements at this limiting magnitude with 5\,min integrations.

\noindent
\emph{Beamsplitters \#2 and \#3}. These are designed to divert 10\,\%\ of the light in passbands not recorded by the spectrograph while affecting the actual cross-disperser passbands by less than 1\,\%. In this mode, the normal AGW dichroic (the one for splitting the acquisition and the wave-front sensing beams) is replaced by a more or less grey 50/50 beamsplitter. This takes us down to about 16th magnitude with 1\,s and 5\,min exposure times on the guider and wave-front sensor, respectively. The respective wavelength ranges of the two beamsplitters fall in the red arm of the spectrograph and are then close to the peak QE of the guiding e2v CCD while the blue-arm wavelengths remain unaffected.  BS\#2 redirects all light redder than 742\,nm (for use with CD\#4 and 5) while BS\#3 redirects light between 544--741\,nm (for use with CD\#6).

\noindent
\emph{Beamsplitters \#4 and \#5}. These are designed for the same wavelength ranges as \#2 and \#3 but divert more than 50\,\%\ of the light in above passbands while affecting the actual cross-disperser passbands as little as possible. This takes us to 18th magnitude with still 1\,s/5\,min exposures. Reaching 20th magnitude by increasing exposure time in the guider is possible, but wave-front sensing would not work properly anymore (tbd).

\noindent
\emph{Beamsplitter \#6}. This is an engineering slot for alignment purposes. It is basically a mirror and redirects all light to the AGW and no light to the spectrograph.

% ------------------------------ Table CDs
\begin{table}
\caption{Default combinations of cross dispersers (CD) and guider beam splitters (BS) as function of target brightness and dichroic field lens (FL).} \label{T-CD}
\begin{tabular}{rccc}
 \hline \noalign{\smallskip}
 $V$ range   &CD I \& VI & CD II \& IV & CD III \& V \\
%  \noalign{\smallskip} 
  \hline \noalign{\smallskip}
0\fm--14\fm   &BS\#1 &BS\#1 &BS\#1\\
15\fm--17\fm &BS\#2 &BS\#3 &BS\#3\\
18\fm--20\fm &BS\#4 &BS\#5 &BS\#5\\
%  \noalign{\smallskip} 
  \hline \noalign{\smallskip}
FL & Blue & Blue & Red \\
%  \noalign{\smallskip}
  \hline
\end{tabular}
\end{table}

\subsection{AGW units}

\begin{figure}[tbh]
%\center
\includegraphics[angle=0,width=83mm,clip]{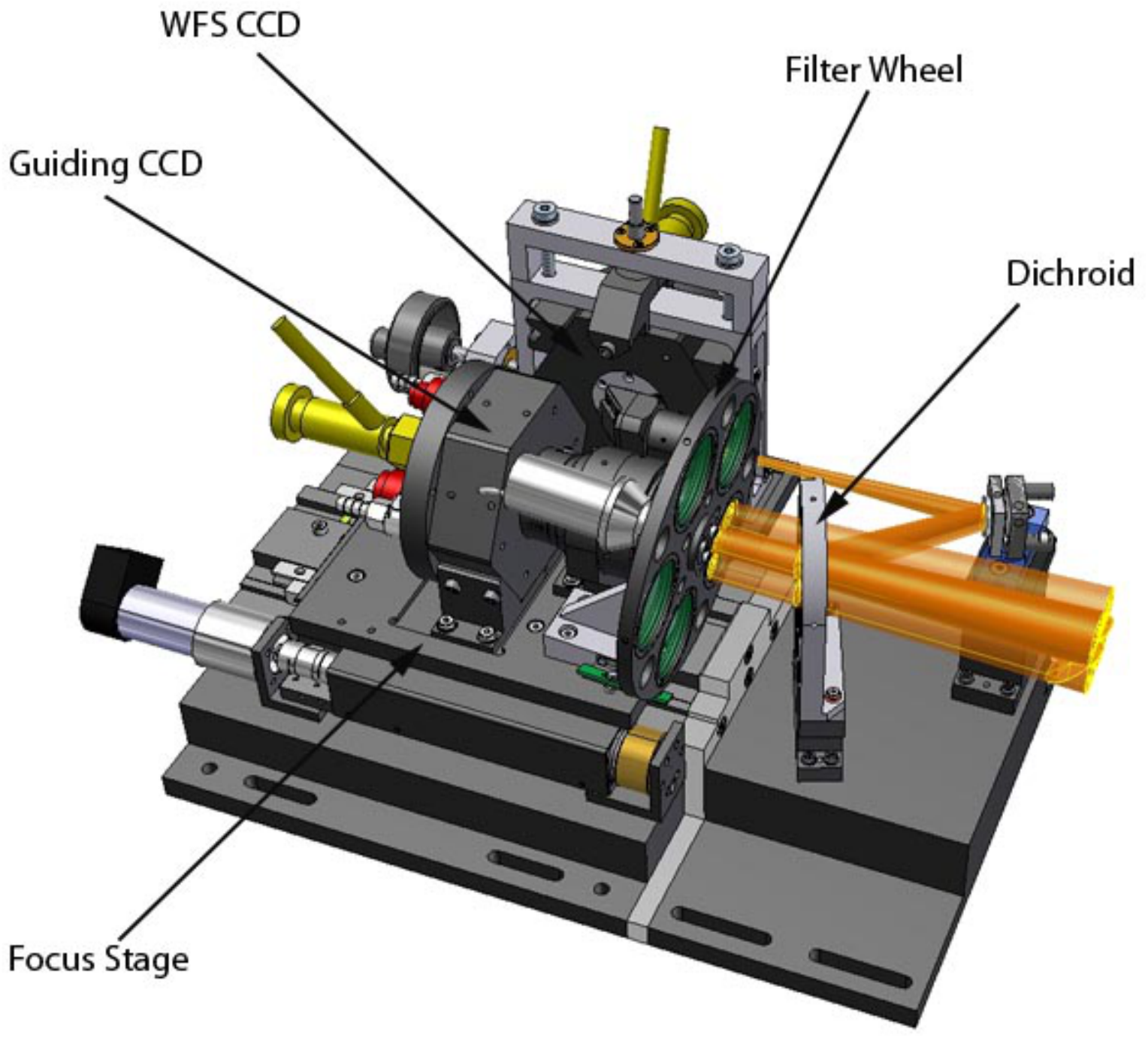}
\caption{Detailed view of one of the acquisition, guiding and wavefront sensing (AGW) units within the PFUs. The AGWs are used for guiding the telescope and controlling the active telescope optics. }
\label{F-agw1}
\end{figure}

The Acquisition, Guiding and Wavefront-sensing (AGW) units for the PFUs are hard copies of the off-axis part of the regular telescope-control units (Storm et al. \cite{agw12}) as for LUCI, but used on axis. Note that the AGWs are not mirrored for the PFU SX and PFU DX. Figure~\ref{F-agw1} shows a drawing of the mechanical layout. The PFU AGWs do not need a guide-star selection mechanism and have only a focus stage as its sole moving part. The original AGW dichroic for LUCI, which is used to split light between the guider and the wave-front sensor, can not be used in the PEPSI PFU. The wave-front sensor has a broad response in wavelength, and we replaced the dichroic beamsplitter with a grey beamsplitter.

The two CCD cameras, one for guiding and one for wave-front sensing, are identical units, are thermo-electrically cooled and equipped with one e2v CCD\,57-10 frame-transfer device each. Two Magellan controllers under control of ITL's AzCam software (see also Sect.~\ref{AzC}) running on a Windows-XP PC are used always for a pair of CCDs. The wave-front sensing cameras have a Shack-Hartmann lenslet array glued onto the CCD surface.

%------------------------------ F5  AGW unit

The real guiding hot-spot position is defined so that the back-reflected image is located on the pinhole of the entrance fibre of the spectrograph. It is slightly different for the SX and the DX sides depending on the optical alignment of the PFUs and the choice of dichroic and fibre being used. A look-up table of hot-spot positions for each of the six dichroics for three entrance fibres, and for two PFU sides, is provided by the PEPSI control system.

PEPSI analyzes the back-reflected pinhole image by using the Gaussian weighted center of gravity which also calculates the second-order momentum as a proxy for the seeing image and focus estimation. If the seeing image size is different for the two focal stations or too different with respect to the direct reflected image, PEPSI may perform re-focusing of the PFU.

%------------------------------ F6  PFU fiber injection
\begin{SCfigure*}
%\center
\includegraphics[angle=0,width=133mm,clip]{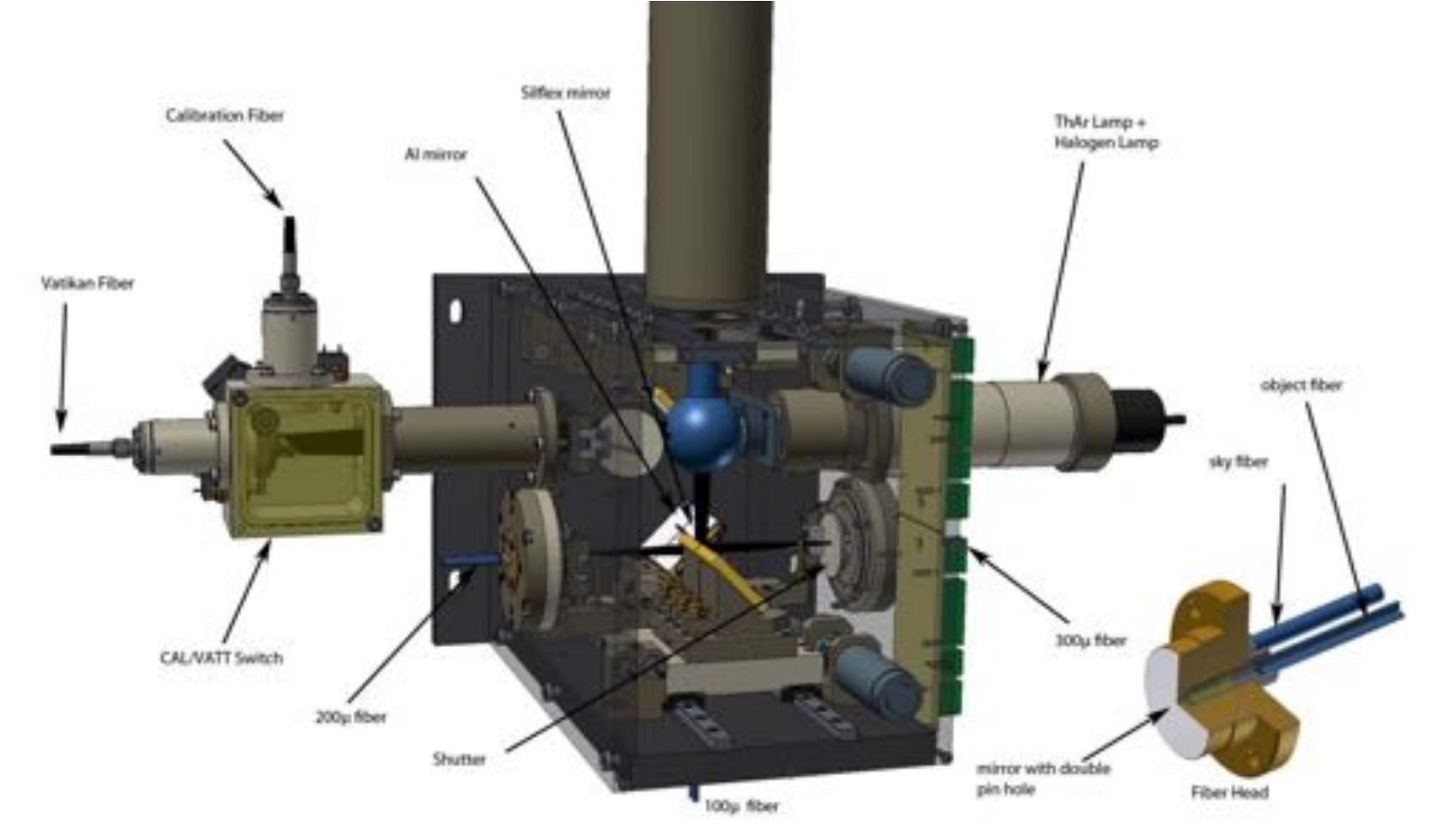}
\caption{View of the PFU fiber injection unit. LBT-light enters the unit from above. Auxiliary light from the calibration unit or the VATT is injected via fibres from the left side of the unit. The right side holds an integration sphere that can be moved into the regular beam. It is fed either directly with an attached Th-Ar hollow cathode or with halogen light via an extra fibre from the calibration unit on the bridge. The insert (small image \emph{bottom right}) shows one of the three regular fiber heads, see text.}\label{F-pfu4}
\end{SCfigure*}

\subsection{Fibre injection unit}\label{S-FIU}

This unit couples the beam into the fibres (Fig.~\ref{F-pfu4}). Three pairs of fibres are available; 100$\mu$m, 200$\mu$m, and 300$\mu$m (see Table~\ref{T-fibres}). 
%%%(see Table~\ref{T-microlens}).
A fibre pair always consists of an off-axis sky fibre and an on-axis target fibre with one (hot) spare mounted behind a reflective plate with two holes.  The insert in Fig.~\ref{F-pfu4} shows one of the three regular fiber heads produced by FISBA. Once the spectral resolution was chosen by the user, a linear stage moves a flat mirror in the position to select either the 200-$\mu$m or the 300-$\mu$m fibre head. At this point, the user may also select between two flat mirrors with either an aluminum coating or a Silflex UV-enhanced coating from Optics Balzers, the latter being optimized for blue light and represents the default. If the UHR-mode (100$\mu$m fibre) is selected, no mirror is needed because the light continues without an extra reflection straight to the fibre head.

% ------------------------------ Table Microlens data for fibre injection
\begin{table}
\caption{Microlens data for the fibre-injection design in Fig.~\ref{F-microlens}. All dimensions in mm. $d1$ Stop diameter, $d2$ lens diameter, $d3$ pupil diameter, $f$ focal length, $l1$ length of L1, $l2$ length of L2, $r=r1=r2$ radius of curvature of L1 and L2.} \label{T-microlens}
\begin{tabular}{lccccccc}
 \hline \noalign{\smallskip}
Fibre  &	$d1$	&$d2$	&$d3$	&$f$	& $l1$	&$l2$	&$r$	 \\
%  \noalign{\smallskip} 
  \hline \noalign{\smallskip}
100 $\mu$m	&0.45&	1.00	&0.09	&1.26	&0.45	&1.74	&1.01  \\
200 $\mu$m      &0.90&	1.60	&0.18	&2.52	&0.88	&3.50	&2.03 \\
300 $\mu$m	&1.35&	2.40	&0.28	&3.93	&1.30	&5.52	&3.16  \\
%  \noalign{\smallskip}
  \hline
\end{tabular}
\end{table}

\begin{figure}
\center
\includegraphics[angle=0,width=82mm,clip]{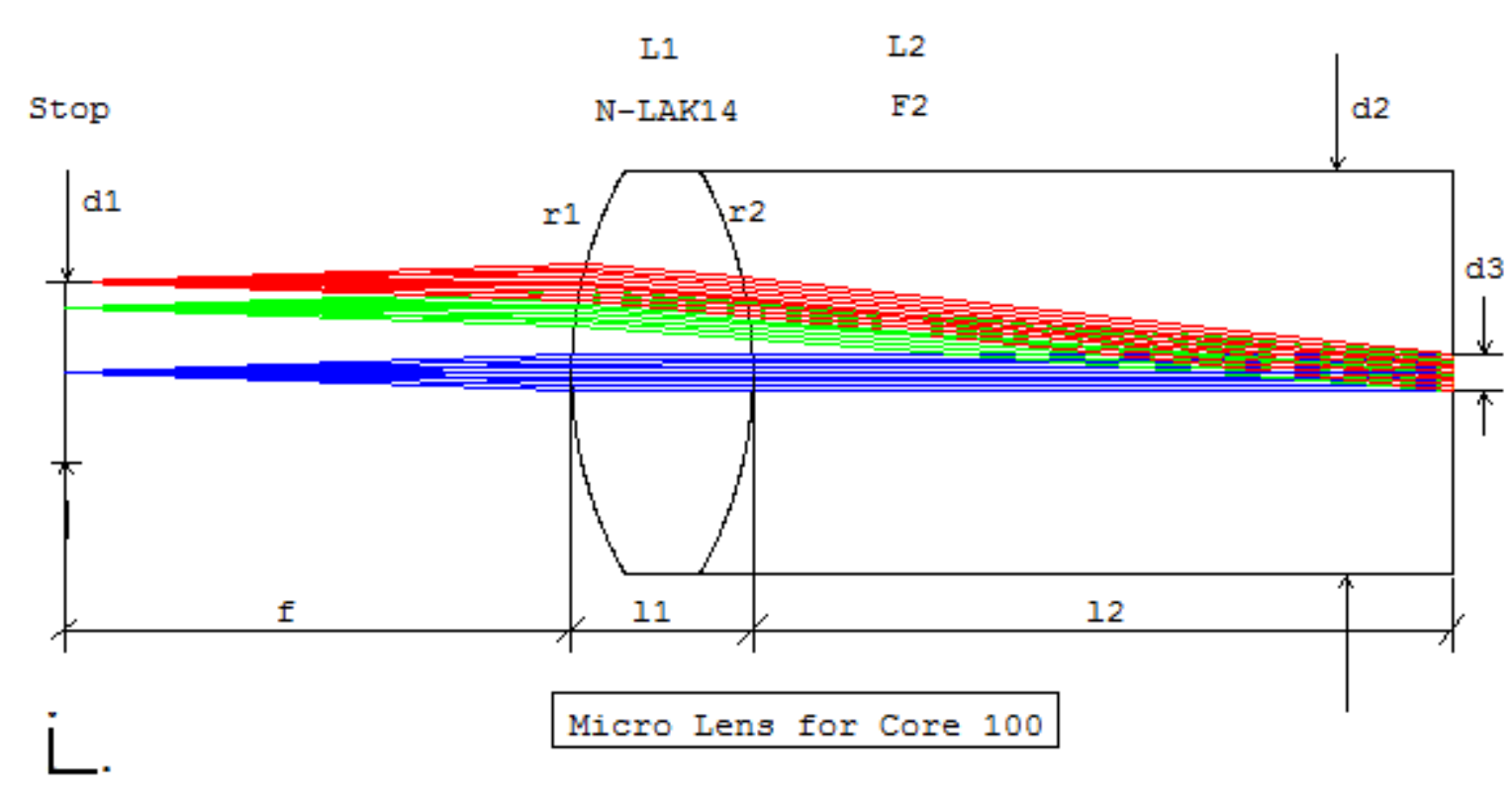}
\caption{Micro-lens design for the octagonal fibre injection. Shown is the design for the 100-$\mu$m core fibre based on a F2 rod lens and a convex N-LAK14 lens. The respective dimensions for all three fibre cores are given in Table~\ref{T-microlens}. }\label{F-microlens}
\end{figure}

The injection problem itself is solved by the method of pupil imaging with rod lenses on the fibre core (e.g. Hopkins~\cite{hop}). In this case the telescope pupil is located on the fibre entrance. The change of aperture from fibre entrance to fibre exit causes most of the known problems of connecting telescopes with spectrographs. The PEPSI fibres are 44\,m in length. The solution is thus to operate the fibre as near as possible to its specified numerical aperture. A further reason to operate the fibre at a high numerical aperture is the large image scale of large telescopes. A large image scale requires a fast f-ratio in the rod lens in order to accept an area on the sky that is as large as possible. Therefore, fibres with a numerical aperture (NA) of 0.22 are used. The optical design for the fibre injection is based on this NA and the rod-lens data in Table~\ref{T-microlens}. The rod-lens radius $R_{\rm rl}$ is given in paraxial approximation by
\begin{equation}
R_{\rm rl} = f_{\rm eff} \ d_{\rm core} \ (n_{\rm rl} - 1) \ ,
\end{equation}
where $f_{\rm eff}$ is the f-ratio in front of the rod lens and in our case identical with the telescope f-ratio, $d_{\rm core}$ is the diameter of the imaged pupil on the fibre core, and $n_{\rm rl}$ is the rod-lens material refractive index. In order to simplify the mechanical design, we set $f_{\rm eff}$ = 15 for all observing modes and change the rod-lens radius according to the fibre-core size. Fig.~\ref{F-microlens} shows the pupil imaging with a rod lens connected with the fibre. The f-ratio at the entrance of the fibre (f/4.37 in silica) is such that the fibre exit f/ratio is always a little less than f/3 (in air) corresponding to roughly 70\,\%\ of the nominal numerical fibre aperture of NA = 0.22. We are thus always underfilling the fibre core by between 13--19\,\%\ depending on core diameter. Table~\ref{T-microlens} lists the detailed lens parameters.

%------------------------------ F7  PFU fiber injection 2: microlens

% ------------------------------ Table   Fibre dimensions
\begin{table}
\caption{Fiber dimensions (diameter core, cladding, buffer), transmission $T$, and focal-ratio degradation FRD.} \label{T-fibres}
\tabcolsep=12pt
\begin{tabular}{lcccc}
 \hline \noalign{\smallskip}
Core   &Cladding & Buffer &$T^{\rm\, a}$ & FRD \\
($\mu$m)   & ($\mu$m)  & ($\mu$m) &  (\% ) & \\
%  \noalign{\smallskip} 
  \hline \noalign{\smallskip}
 100   & 115 & 140 & 76.1\rlap{$^{\rm b}$} & 91 \\
 200   & 230 & 250 & 86.1 & 92 \\
 300   & 345 & 380 & 90.5 & 96\\
% \noalign{\smallskip}
 \hline
\end{tabular}
\vspace{1.5mm}

$^{\rm a}$ Measured transmission at 635\,nm.\\
$^{\rm b}$ This value is likely influenced by the injection coupling in our test set-up and might be significantly higher.

\end{table}

The unit also serves to redirect calibration light to the science fibres. This constitutes the regular calibration set up for PEPSI and allows two options. Option~1 brings calibration light through a dedicated 300-$\mu$m fibre from the main calibration unit. A triplet lens transforms its exit into an f/15 beam analog to the telescope. A linear stage simultaneously moves a flat mirror into the beam that redirects the calibration light downwards to where again the appropriate fibre head (diameter) can be selected. The calibration-light fibre can carry any type of calibration light, even sunlight from the solar-disk integration telescope (see Sect.~\ref{S-SDI}).  Option~2 can move the same stage so that it instead brings light directly from an integration sphere into the beam. It can be used with either another fibre dedicated to a halogen lamp on the bridge or a local Th-Ar lamp. The purpose of this second option with its integration sphere is to have a validation source with no modal-noise (or any other) pattern from the support fibre. To prevent stray light within the PFU volume the fibre-injection unit has five light-tight mechanical shutters that will be closed an opened automatically depending on the option chosen.

\newpage

\subsection{Fiber switch unit}

Each sky fibre is interrupted by a mechanical precision switch so that either of these fibres can be fed with wavelength calibration light if the spectrograph is to be used in high-precision radial-velocity mode. It is indicated in Fig.~\ref{F-pfu1} as a separate box within the PFU's ``Optical Box''. The fibre alignment is achieved by a precision linear stage, specially modified for our purpose, that brings the respective fibre ends nearly into contact. Precision stainless-steel SMA plugs are used for the terminations. The interception takes place within the space of the PFU after 1.2\,m of fibres behind the rod lens assembly of the fibre-injection unit. The calibration light itself is brought up to the PFU by a dedicated 300-$\mu$m fibre via the calibration unit. The default is the  stable interference pattern from the Fabry-P\'erot \'etalon on the spectrograph table, although Th-Ar could be used instead as well. To keep the geometric light loss caused by the fibre interruption as small as possible the positioning accuracy of matching the two fibre ends is $\approx$\,10\,$\mu$m peak-to-valley, mostly in direction of the separation. The errors in the other two dimensions are so small that they were not easily measurable. The separation of between 10--20 $\mu$m leads to light losses for the calibration light in the range 5--10\,\%\ inversely depending on core diameter.

\subsection{Artificial star unit}

A light-emitting diode (LED) and its re-imaging optics can be inserted into the telescope optical axis right in front of the collimator, as seen from M3, and act as an artificial star for alignment and other engineering purposes. The insertion mechanism consists of a flat mirror tilted by 45$^\circ$ and a f/15 re-imaging lens group on a wedge-like rotating structure. In its ``in'' position it also acts as a shutter for the entrance opening of the optics box towards M3. This is its default position. The facility is used in day time for fine alignment of the entire PFU optics but can also be used for quick alignment verification prior to an observing run without the need for precious night time. The alignment procedure particularly concerns the beam-splitters that direct light to the AGW unit.

%------------------------------ F  fibre train
\begin{figure}
\center
\includegraphics[angle=0,width=83mm,clip]{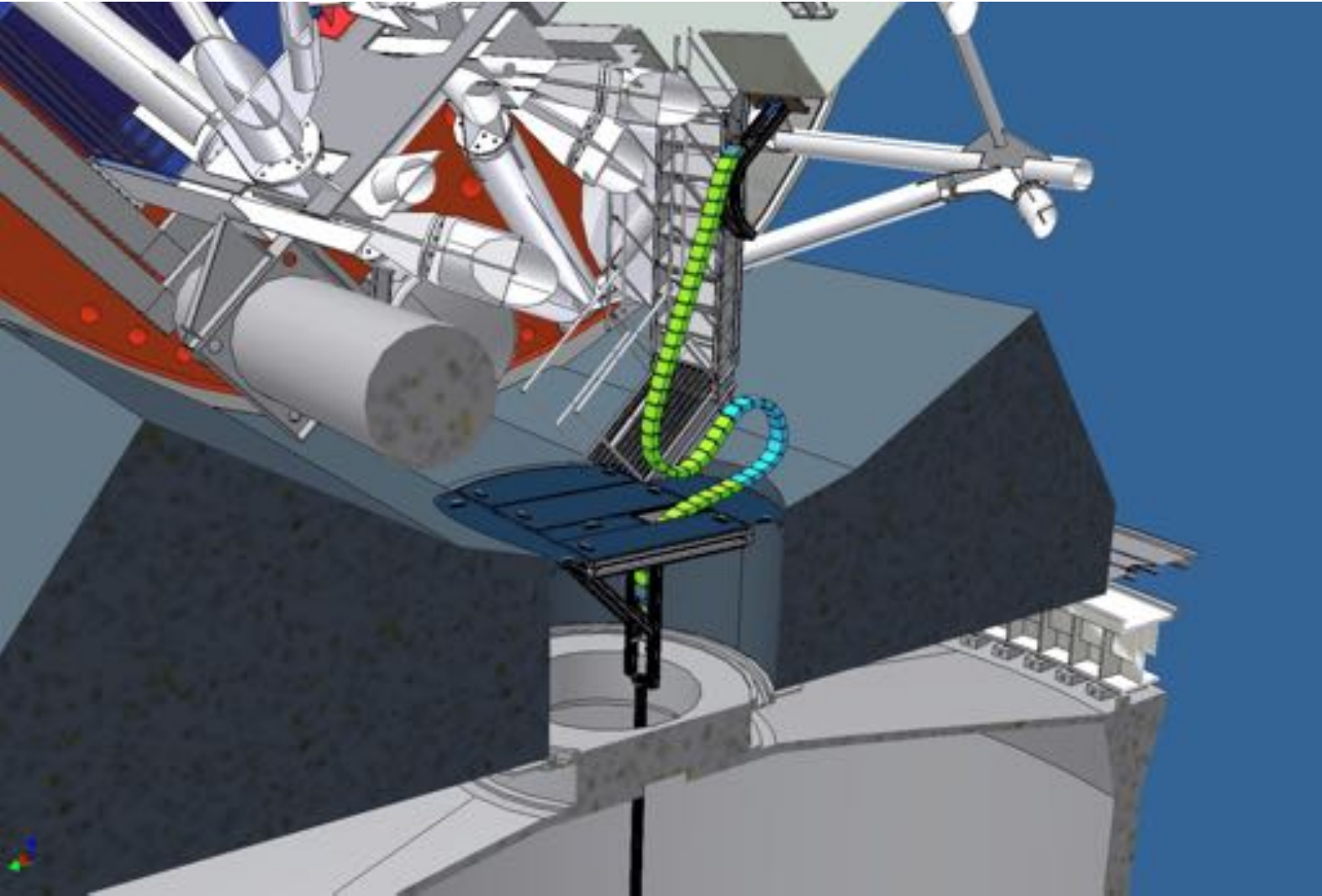}
\caption{The fibre train at its most critical location. Shown is the telescope's azimuth-hole structure (mid to bottom side) that guides the fibre-chain (the S-like structure) during telescope elevation motion. Note the dimension of the azimuth hole is $\approx$\,3\,m in diameter.}\label{F-fibretrain}
% The telescope pointing shown is for an elevation angle of 45$^\circ$. Fibres continue from the dual-axis S-chain % into the round tri-axial chain at the top end of the azimuth hole that then leads to the PEPSI support bridge
% just on top of the spectrograph chamber.
\end{figure}

%------------------------------ F  near- and far-field for PEPSI fibres
\begin{figure}
\center
\includegraphics[angle=0,width=83mm,clip]{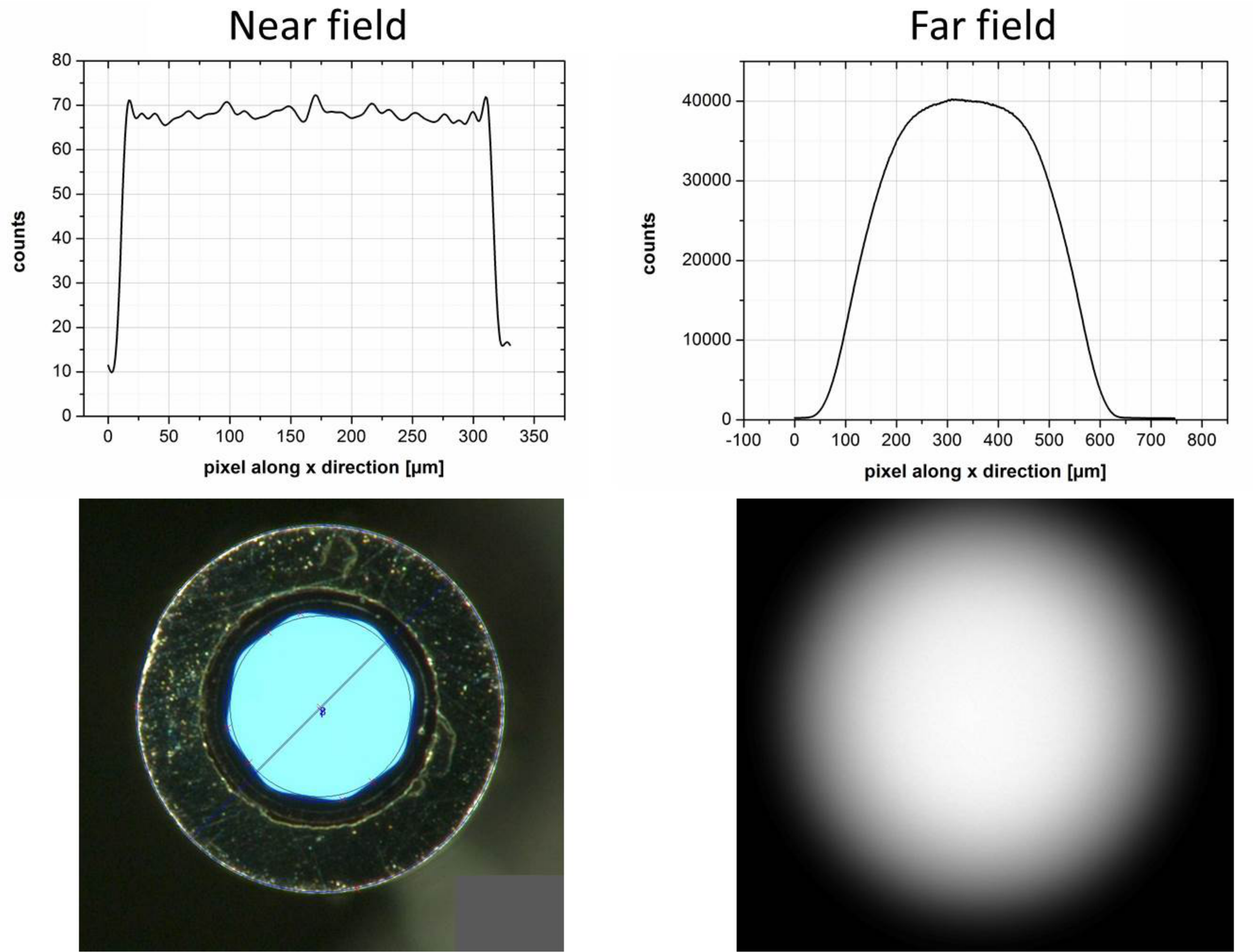}
\caption{Near and far field of one of the 300-$\mu$m PEPSI science fibres. Fibre length is 44\,m and the output beam f-ratio is f/3. }\label{F-ffnf}
\end{figure}

%------------------------------------------------------------------------------------------------------------------------------
\section{Fibre connections}\label{S-Fibers}

All science fibres are of octagonal shape and of PolyMicro silica FBP type with a numerical aperture of 0.22. They have a silica core and a doped silica cladding with a polyimid buffer. Table~\ref{T-fibres} summarizes its dimensions and measured characteristics.

%------------------------------ F  fibre logistics
\begin{figure*}
\center
\includegraphics[angle=0,width=165mm,clip]{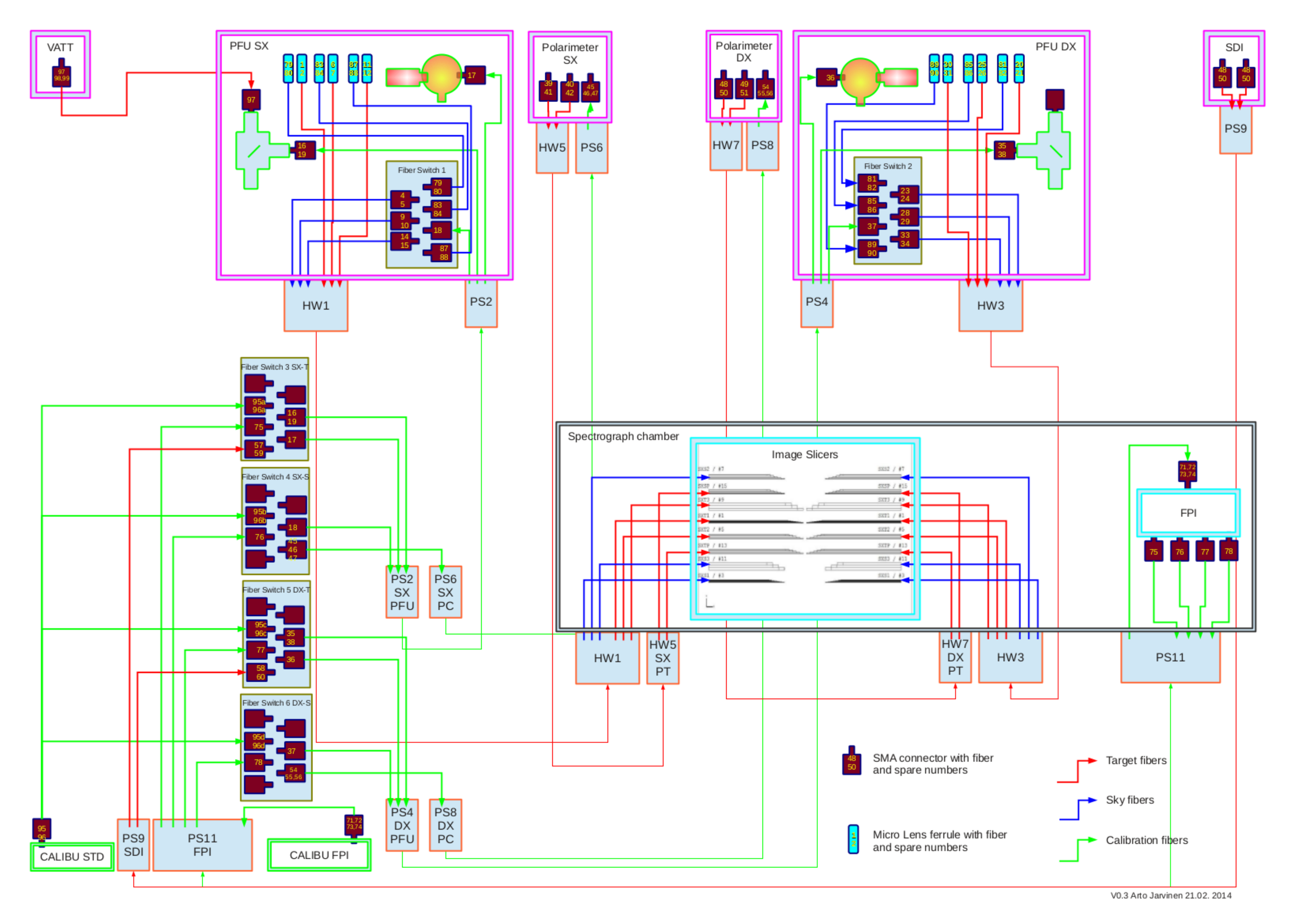}
\caption{The fibre distribution for PEPSI. There are 9 subunits with fibre connection: the image-slicer unit in the spectrograph, the Fabry-P\'erot unit on the spectrograph table, the calibration unit on the bridge, the two PFUs (SX and DX), the two polarimeters (SX and DX), and the SDI and VATT  telescopes. Four ``Helawrap'' (HW) hoses carry a total of 26 octagonal FBP science fibers, five Panzerhoses (PS) carry a total of 20 circular FBP calibration-light fibres. One PS with 8 fibres (4 spares) connects the calibration unit with the Fabry-P\'erot unit. Four fibre switches in the electronics rack (in the figure left) and two switches within the PFUs (on the top) gate calibration-, sky- and sunlight to wherever it is needed. }\label{F-fibrelogistics}
\end{figure*}

\subsection{Fibre train}

The way fibres are fed through a moving telescope structure decides on their longevity and thus to a certain degree also on their performance. Because the telescope is populated by other instruments and many subsystems that need to be accessed by human beings, special attention was given to the housing and location of the fibres, which we call the fibre train. Each fibre rests in a PVC jacket of 3\,mm diameter with a Kevlar strength-member insert. The jackets are organized into larger protection hoses of type ``Helawrap'' with 16\,mm diameter.  Standard conduits on the telescope structure additionally protect the fibres. A bit of a special case are the polarimeter fibres because they must be connected, disconnected, and locally stored when the polarimeters are on and off the telescope. A clean storage box on the C-rings of the telescope structure close to the straight-through foci hosts the fibre-heads when the polarimeters are not in use.

Figure~\ref{F-fibretrain} shows part of the fibre train with all relevant PEPSI fibres. A ``dual-flex energy chain'' was implemented on the telescope to connect the elevation structure of the telescope with its azimuth structure. The chain's lower end is held by an aluminum structure within the 3m-diameter azimuth hole of the telescope mount that rotates with the telescope. With this design the telescope elevation motors have to compensate only the off-axis momentum of the weight of the filled energy chain plus the chain-internal friction. The chain's upper end is bolt to the telescope staircase-support structure where the first of two (fibre-)weight compensators are mounted. The fibre path from the 3\,m azimuth hole to the spectrograph 20\,m below is bridged by a ``tri-flex energy chain''. The tri-flex chain compensates the azimuthal rotation of the telescope but also a residual pendulum migration because the fibre bundle could not enter the azimuth hole in its exact center (which is occupied for boogie alignment\footnote{The boogie wheels carry and rotate the entire LBT building.}) but off axis by 50\,cm. Before the fibres enter the tri-flex chain the second weight-compensator is mounted. These just consist of an interception of the Helawrap hoses with a fixture of each individual fibre jacket within a soft-foam clam. The lower end of the tri-flex chain feeds into a fibre-logistics box on the support bridge from where fibres are directed to the spectrograph or the calibration unit. Note that the VATT fibre is fed into one of the PFUs and from there the light continues into the regular science fibres. The SDI fibre on the other hand enters the calibration unit on the bridge and from there the light goes up to both PFUs and down again through the octagonal science fibres.

%------------------------------ F  FRD
\begin{figure}
\includegraphics[angle=0,width=83mm,clip]{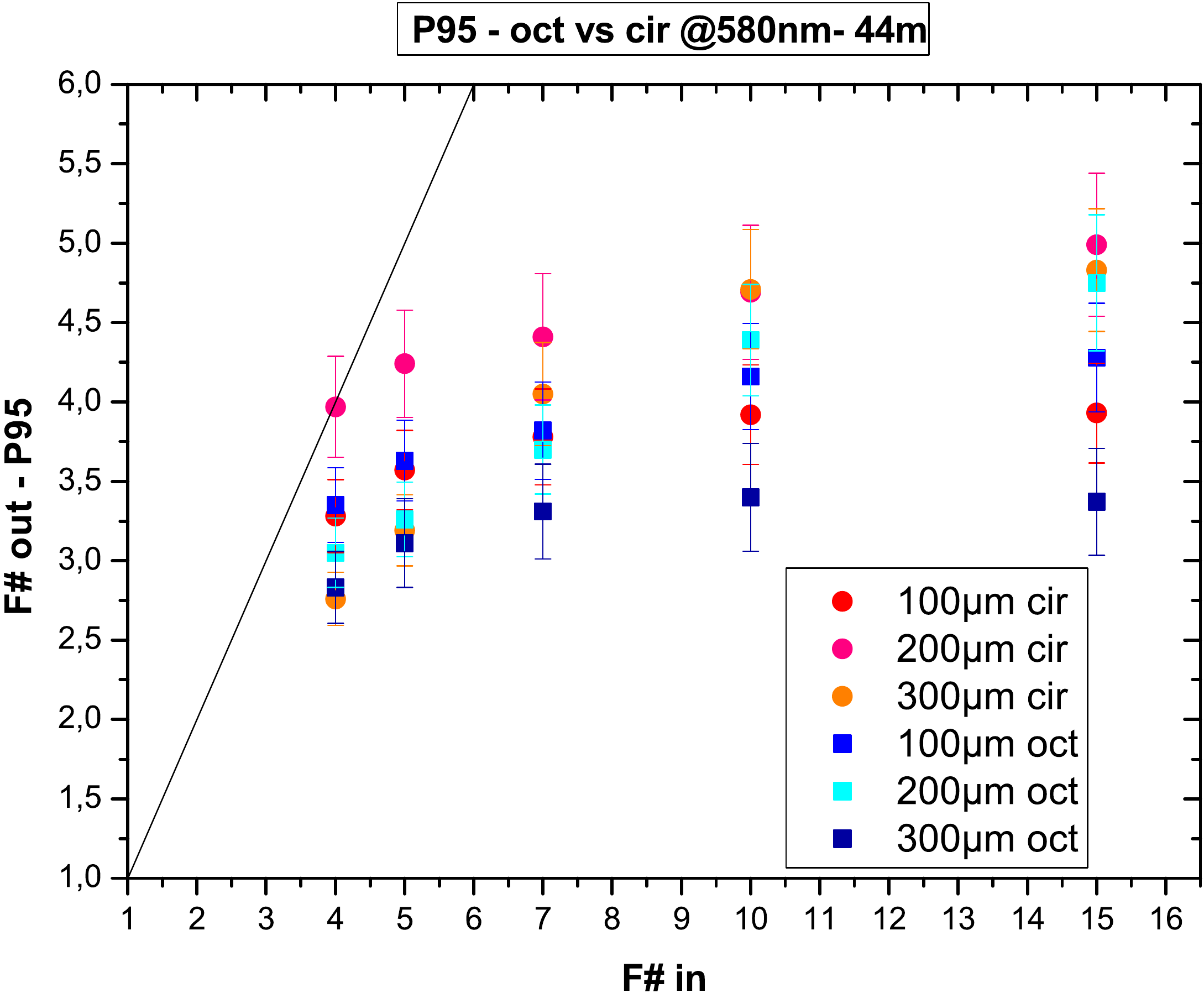}
\caption{Focal-ratio degradation for the PEPSI fibres. The three octagonal fibres are also compared with analog circular fibres. Measurements were made for the central wavelengths of the six cross dispersers and for a large range of input f-ratios. In PEPSI, the input beam f-ratio into the fibre is f/3.3. }\label{F-FRD}
\end{figure}

\subsection{Fiber logistics}

Figure~\ref{F-fibrelogistics} shows the overall fibre logistics. A total of 26 science fibres connect the LBT to the spectrograph and its subunits. These are arranged in four hoses, one hose per PFU and one hose per polarimeter. Fiber length is 44\,m from the PFUs as well as from the polarimeters. One additional hose has a slightly different length (40\,m) and is used to connect the SDI telescope (Sect.~\ref{S-SDI}) to the calibration unit. The VATT hose is 453\,m in length and described in detail in Sect.~\ref{S-VATT}. Fibers that feed just calibration light to a particular unit are all circular FBP types of 300 $\mu$m core diameter.

%------------------------------ F modal noise (SNR)
\begin{figure}
{\it a)}

\includegraphics[angle=0,width=83mm,clip]{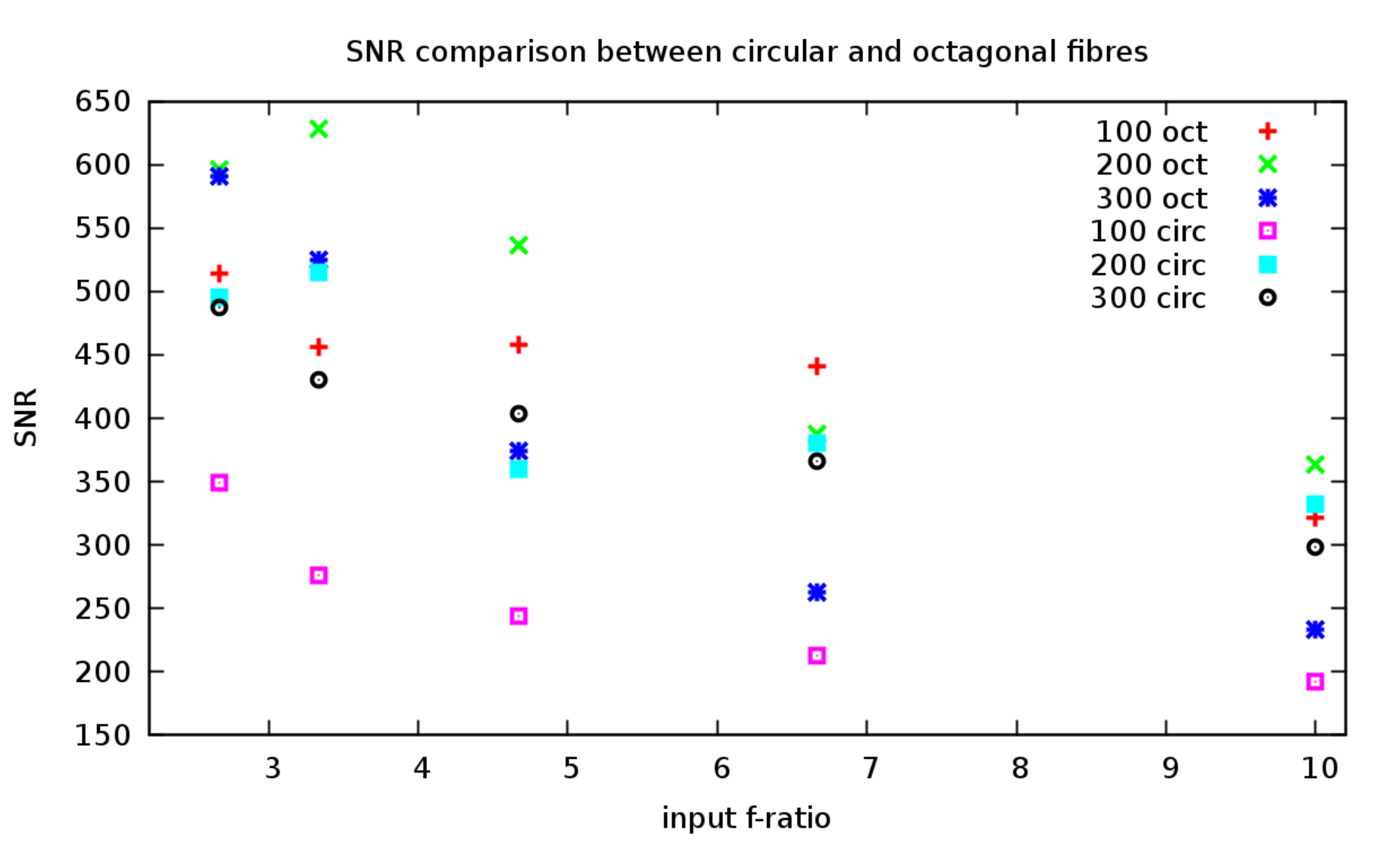}\\
{\it b)}

\includegraphics[angle=0,width=83mm,clip]{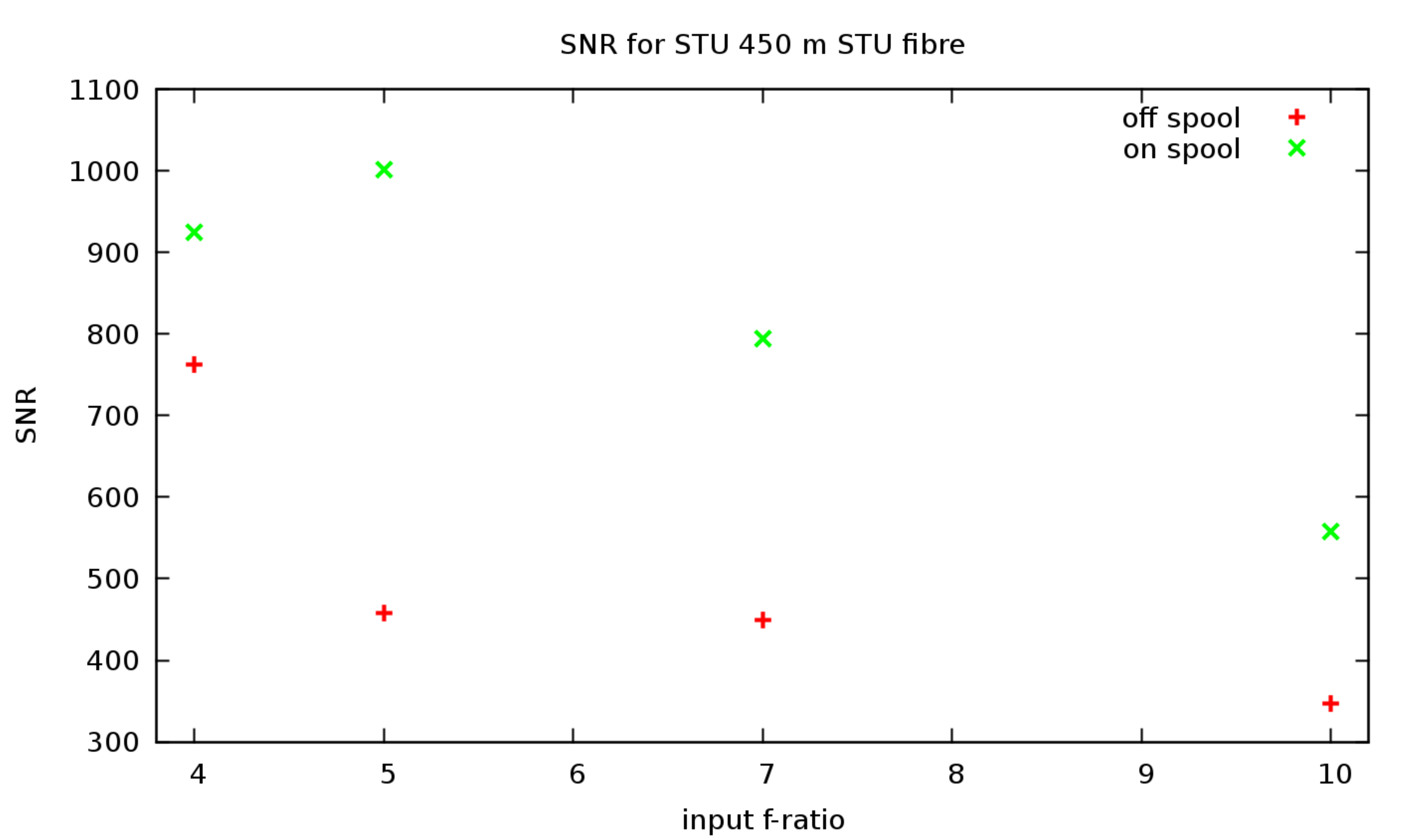}
\caption{The modal-noise limited signal-to-noise Ratio (SNR) as a function of fibre input f-ratio. \emph{a}) Modal noise for 44-m long fibres of circular and octagonal shape for the three PEPSI core diameters (100, 200, 300 $\mu$m). Note, in PEPSI the input beam f-ratio into the fibre is f/4.37 in silica ($\approx$\,f/3 in air). \emph{b}) The same for a  \hbox{450-m} circular STU fibres with 200\,$\mu$m core (spooled and unspooled). This fibre is similar to the FBP fibre in use for the VATT mode.}\label{F-SNR}
\end{figure}

\subsection{FRD and transmission}

The focal-ratio degradation (FRD) of fibres must be treated together with their transmission efficiency. Bending a fibre decreases the angle between the ray and the normal of the interface between core and
cladding. This results in an increase of the angle at the output. Also the microscopic structure of the interface leads to FRD. It is also well known that FRD is introduced by mechanical stresses within the bifurcation tube and by the fibre termination (Avila et al. \cite{avila2010}). The transmission is strongly dependent on the fibre entrance f-ratio. FRD for our fibres was measured including the FISBA fibre heads with an f/15 entrance diaphragm and an exit diaphragm that is optimized for the output f-ratio of f/3. Measurements were made with and without the diaphragm and the intensity ratio defined as the FRD shown in Fig.~\ref{F-FRD} (for more details, see Sablowski et al.~\cite{sab}). The FRD is smaller (better) for larger core sizes and also better for octagonal fibres compared to the corresponding circular FBP fibres. Small fibres suffer from stronger scatter of the FRD in wavelength for the same input aperture. A scatter in wavelength is explainable by roughness of the interface between core and cladding which causes wavelength dependent scattering. This could be due to the production process, which is more critical for small core fibres. Representative measurements are also listed in Table~\ref{T-fibres}.

\subsection{Modal noise and scrambling properties}

A homogeneous light beam entering a fibre is transformed to a discrete intensity distribution at the output due to interference. This interference pattern can also appear in the image plane of the instrument, which images the fibre output. Its structure, or better its intensity fluctuations caused by the interference, is called modal noise (e.g. Baudrand \& Walker \cite{bau:wal}). Stabilized spectrographs now achieve a radial-velocity precision of $\approx$1\,\ms\ using fibers, but precisions of near 10\,\cms\ are required to detect terrestrial planets orbiting solar-like stars. One limiting factor in multi-mode fiber coupling is the modal noise of the fibre.
%It appears to limit the radial-velocity precision.

Recent advances with non-circular fibres were described by Avila (\cite{avila2012}). Figure~\ref{F-ffnf} shows the near and the far field of one of our octagonal 300-$\mu$m science fibres. Its far field is flat and reasonable uniform. We notice that none of our fibres shows the central brightening reported by Avila (\cite{avila2012}) and all have a flat near-field with a transmission rms of below three counts across the core. All of PEPSI's science fibres exit first into the collimator f-ratio-transformation optics before the light enters the respective wave-guide image slicers. While the transfer optics act as a mild optical scrambler, the waveguide slicer acts as a mild descrambler due to its slit geometry. The filling factor of the fibre-core entrance is between 81--87\,\%\ in our design (see Sect.~\ref{S-FIU}). This works in favor of the near-field scrambling (Avila~\cite{avila2012}).

Figure~\ref{F-SNR}a shows the signal-to-noise ratios (S/N) measured from the 44\,m PEPSI octagonal fibres for various input f-ratios in comparison with analog circular fibres. The 300-$\mu$m octagonal fibre is basically photon-noise limited within the precision of our laboratory set-up. Practically, this means that modal noise is not the limiting factor for the 300 nor the 200-$\mu$m fibres, while modal noise in the 100-$\mu$m fibre limits above S/N of $\approx$\,500\,:\,1 per exposure. Sablowski et al.~(\cite{sab}) show a comparison of the modal-noise amplitudes for a larger set of different fibres. Figure~\ref{F-SNR}b shows the S/N for a 450\,m circular Polymicro STU fibre (spooled and unspooled) and various input f-ratios. This fibre behaves very similar, if not identical, to the FBP type. It shows that modal noise is much less of an issue in the longer fibres due to nearly perfect scrambling. A fibre agitator for further increase of the scrambling gain is available on demand but currently not in use.

%------------------------------- Fig   Zemax spectrograph
\begin{figure*}
\center
\includegraphics[angle=0,width=170mm,clip]{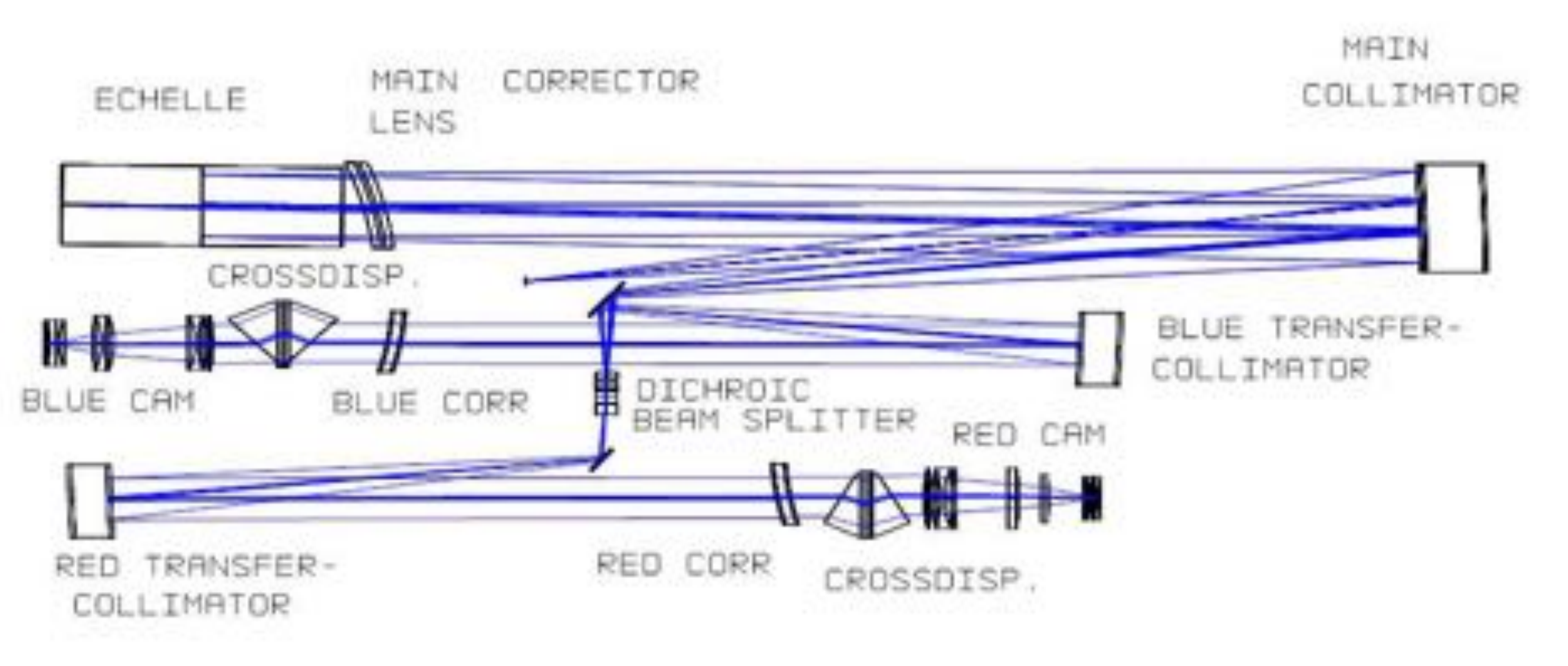}
\caption{Spectrograph optical design. Light from two fibre pairs of either 100, 200, 200(pol), and 300-$\mu$m diameter is injected into the spectrograph through a total of 16 fixed image slicers. A Maksutov-type main collimator is used in double pass. The grating is an 80$\times$20\,cm $R$4 \'echelle in Littrow with a blaze angle of 76\degr . The field lens in the intermediate focus carries a dichroic beam splitter for a blue arm  (383--544\,nm) and a red arm (544--907\,nm). Each arm consists of an optimized Maksutov transfer collimator, a set of three cross dispersers, each a combination of a volume phase holographic grating and two prisms, an optical f/3 camera with up to 11 lenses, and a monolithic 10.3k$\times$10.3k CCD as detector.}\label{F-specdesign}
\end{figure*}
%------------------------------ Fig  Mechanical design
\begin{figure*}
%\center
\includegraphics[angle=0,width=165mm,clip]{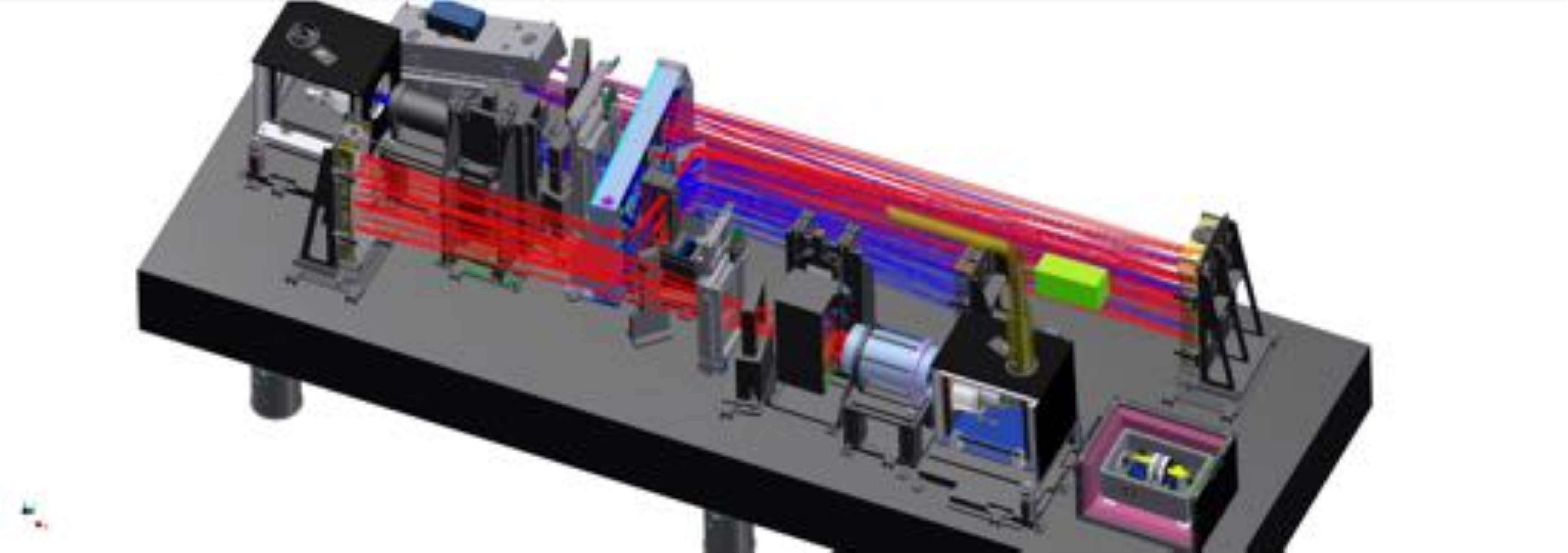}
\caption{Spectrograph mechanical design. Also shown is the Fabry-P\'erot unit on the optical table (the box to the very right).}\label{F-mechdesign}
\end{figure*}

%------------------------------------------------------------------------------------------------------------------------------
\section{The spectrograph}\label{S-Spec}

PEPSI is an asymmetric white-pupil \'echelle spectrograph in Littrow configuration inspired by earlier concepts put forward by Baranne et al. (\cite{elodie}), Gratton et al. (\cite{sarg}) a.o.. The PEPSI \'echelle grating has a blaze angle, $\theta$, of 75.27\degr\ (i.e. $R$3.80, further referred to as $R$4) and a collimated beam diameter, $d$, of 200\,mm which convert to the standard resolution times slit product of $R\alpha$ = $2 d \tan\theta / D$ = 34\,600, where $D$ is the telescope diameter of a single LBT eye (\hbox{$D$ = 8.4\,m}).

\subsection{Optical design}

The optical Zemax design (Fig.~\ref{F-specdesign}) is based on a dual-arm white-pupil configuration where the main collimator and the two transfer collimators are off-axis Maksutov systems, reaching a near diffraction limited optical quality over a large field of view. This enables relatively long slits as required for recording four sliced spectra simultaneously (two per telescope). At the field lens, located at the intermediate focal plane between the main- and the transfer-collimators, the light is divided between the blue and the red arm by either one of two  selectable dichroic lenses.

The spectrograph is physically located on a 6$\times$2\,m optical bench (Fig.~\ref{F-mechdesign}) in the observatory basement inside the telescope pier within a pressure- and temperature-stabilized chamber. It receives light from the polarimeters or the permanent focal stations via 44\,m long fibres coupled with image slicers. A 200\,mm main beam diameter is employed based on the currently largest available monolithic \'echelle grating. The off-axis Maksutov systems includes a beam-diameter reduction from 200\,mm to 125\,mm in the blue-transfer and the red-transfer collimators. A set of three VPH grisms for cross dispersion are available per arm. The two spectrograph cameras are each equipped with a monolithic 10.3k$\times$10.3k Semiconductor Technology Associates (STA) CCD with 9\,$\mu$m pixels. For each spectral order, four spectra are recorded, i.e. either the two polarization states in polarimetric mode or object/sky and object/Fabry-P\'erot light in integral mode for each of the two telescope. The entire spectral range from 383\,nm to 907\,nm can be covered in three exposures, for all three resolution modes.

\begin{figure*}[!bth]
\includegraphics[angle=0,width=\textwidth,clip]{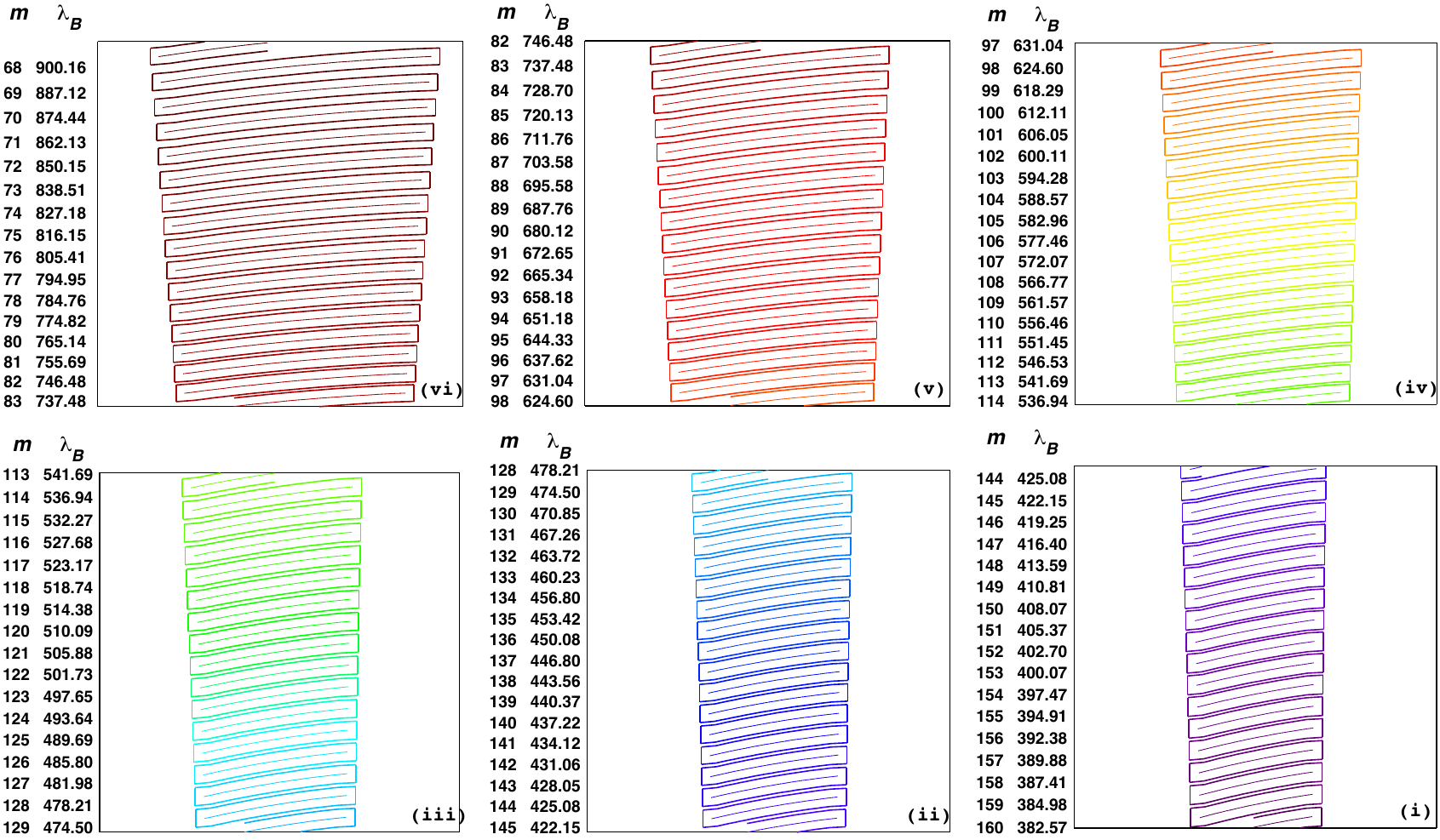}
\caption{Spectral formats on the CCD. Each plot shows the format for one cross disperser from the red-most wavelength (\emph{top left}; CD\,VI) to the blue (\emph{bottom right}; CD\,I). Only the free spectral range and only one slice are shown. Indicated on the left are the order number $m$ and the blaze wavelength $\lambda_{\rm B}$ in nm. The squared box outlines the CCD and the number in parenthesis indicates the cross disperser number (I to VI). }\label{F-formats}
\end{figure*}

The spectral formats of PEPSI are shown in Fig.\,\ref{F-formats}. These have been ray-traced using the Zemax model. In each order, a single free spectral range is shown assuming a source placed at the center of a pseudo entrance slit. The order number, $m$, and the blaze wavelength, $\lambda_B$, are indicated on the left-hand side of each panel. The bold-line box around each order shows the extent of the PEPSI slit.

\subsection{Collimators}

As shown in Fig.~\ref{F-specdesign}, the $R$4 \'echelle is used in Littrow configuration with a small, 0.65\degr , off-plane angle, fed by the 200\,mm beam diameter f/13.9 Maksutov main collimator ($f$~=~2881\,mm) in double pass. A transfer Maksutov collimator couples to the first one through a field lens that casts a white pupil onto the camera entrance pupil, where a grism acts as the cross-disperser. The Maksutov primary and transfer collimators have unequal focal lengths, which makes it possible to reduce the beam diameter of the transfer collimators to 125\,mm, thereby making very significant savings on the optical spectrograph cameras.

%------------------------------- Fig   spectral formats

Due to the dispersion of the \'echelle grating, the overall dimensions of the three spherical collimator mirrors are quite large (blank diameters are M1 = 880\,mm, M2 = M3 = 700\,mm). The (optical)length of the spectrograph is about 5.5\,m, just fitting within the 6\,m optical bench on which it is installed. M1 is of rectangular shape (880$\times$320\,mm) with round ends and is cut from a circular mirror with a diameter of 880\,mm. Its radius of curvature is 4144\,mm. The mirrors are coated with a protected silver coating, giving a reflectivity of 97\,\%\ in the visible/red part of the spectrum. The N-BK7 corrector lens is an off-axis portion of a spherical meniscus lens with a central thickness of 40\,mm. Its size (400$\times$300\,mm) is cut from a circular lens with a diameter of 500\,mm. The two spherical surfaces of the lens segment are inclined with respect to each other, i.e. the lens is a spherical wedge. The edge thickness of the lens therefore had to be controlled carefully during grinding.

The two transfer collimators (blue and red) are very similar to but smaller than the main collimator. The mirror dimensions are 700$\times$200\,mm (red collimator) and 400$\times$200\,mm (blue collimator), and the N-BK7 corrector lenses are 300$\times$250\,mm, with a nominal central thickness of 40\,mm. Focal length is 1773\,mm.

\subsection{\'Echelle grating}\label{grating}

The grating is a $R$4 mosaic of two halves on a joint Zerodur carrier of thickness 125\,mm produced by Richardson Grating Lab. It has a total weight of 55\,kg. The groove density is 31.6\,gr/mm over the width of 214\,mm and for a length of 840\,mm (with a central 15\,mm gap due to the mosaic). The total wave-front error is 0.142$\lambda$ at 632.8\,nm which limits the maximum spectral resolution to $R$ = 1\,250\,000. The blaze angle is measured to 75.3$\pm$0.4\degr\ and the reflectivity on average 67\,\%\ (380--680\,nm). The off-plane angle of incidence on the \'echelle is 0.65\degr\ (confirmed to 0.70$\pm$0.03\degr\ after alignment). The CCDs record a total of 92 \'echelle orders, from the 68th to the 159th.

When quoting the resolving power of a spectrograph, it is standard practice to use the ``full width at half maximum". Three components are folded into this. Firstly, the ``geometric'' resolving power as computed from the Zemax ray tracing simulations. Secondly, the ``diffraction limited'' resolving power, $mN$, with $m$ being the order number and $N$ being the number of illuminated grooves per \'echelle grating segment and, thirdly, the effect of finite pixel binning. Its value is computed by determining the width of a single pixel in wavelength space at the center of the current resolution element. All computations were done with the EchMod package (Barnes \cite{echmod}). The final resolving power, $R$, in Fig.~\ref{F-syseff}b is computed by assuming that each of the above components are effectively cumulative convolutions.

%------------------------------- Fig   CD design
\begin{figure}
{\it a)}

\includegraphics[angle=0,width=82mm,clip]{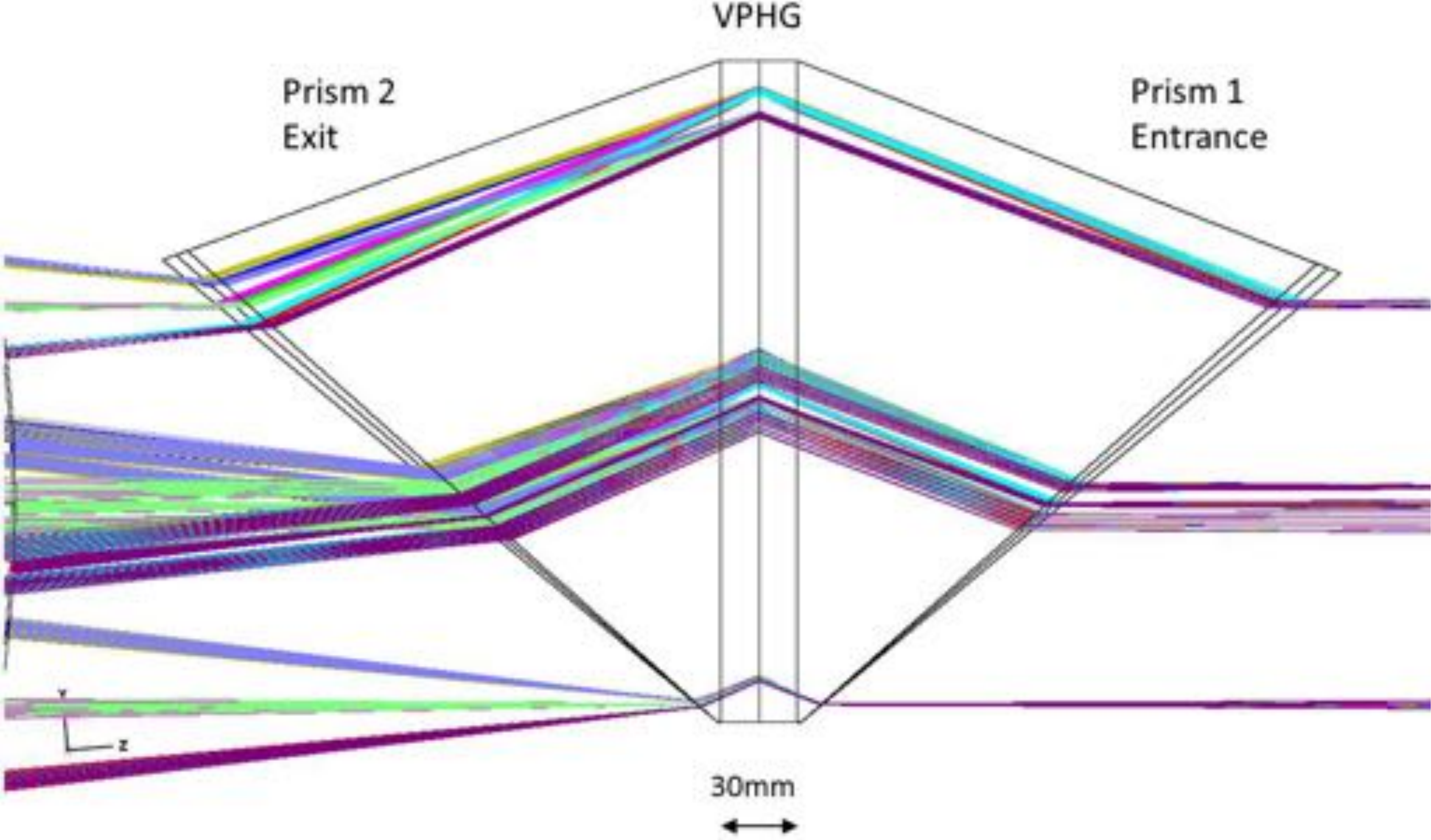}\\
{\it b)}

\includegraphics[angle=0,width=82mm,clip]{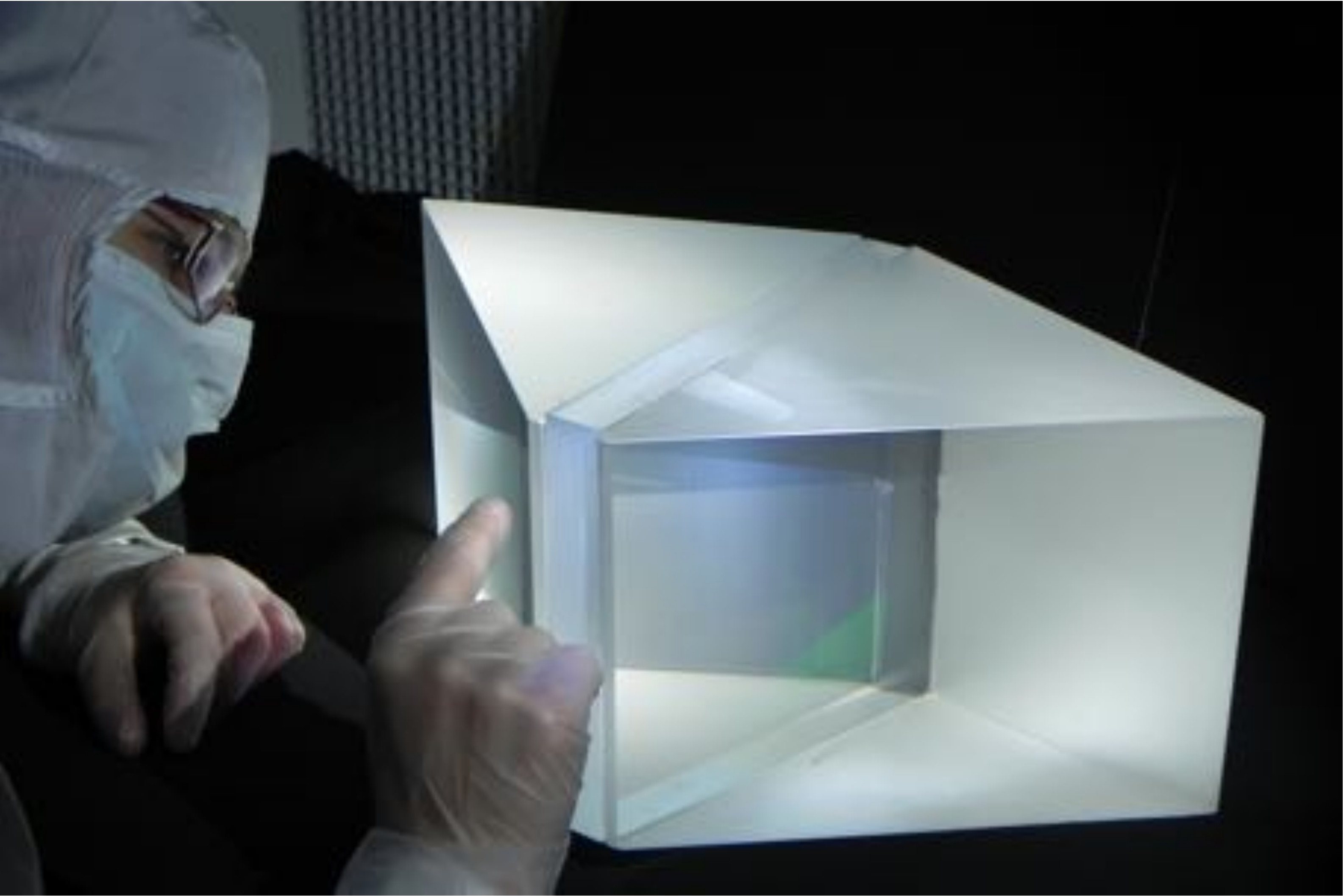}
\caption{Cross disperser design. \emph{a}) Shown is an overlay of the beam for all three cross dispersers of the blue arm. Light enters the CD from the right. \emph{b}) CD-III during final inspection at Thales-SESO.}\label{F-cddesign}
\end{figure}

\subsection{Cross dispersers}

In order to maximize spectrograph throughput, we have selected grisms as cross-dispersers (CD). Grisms have a more uniform inter-order separation, allowing us to enlarge the minimum-order separation. A larger inter-order separation was required so that we can arrange for a larger number of (image) slices on the CCD and then, for a fixed slice width, a larger effective slit width with respect to the sky. Three grisms per spectrograph arm cover the full wavelength range. The blue arm is compromised of CDs~I--III, the red arm of CDs\,IV--VI. Table~\ref{T3} summarizes the CD properties. Figure~\ref{F-cddesign}a shows the optical design.

One cross-disperser grism consists of a sandwiched Volume-Phase Holographic (VPH) grating between two glass prisms of Ohara PBL25Y for CDs II and III, Schott N-SF57 for the red arm, and PBM2Y for CD\,I. Two air-glass surfaces exist per grism which are anti-reflection coated for the respective wavelengths  and entrance and exit angles. Typical transmission efficiencies of 98.2\,\%\ were achieved for angles near 50\degr . Wavefront errors on the individual prism surfaces were always around 34\,nm rms. The VPH gelatine itself is sandwiched within two plane parallel substrate plates. In this way each VPH grating is illuminated with the right incidence angle that maximizes the grating's efficiency. The beam behind cross dispersion proceeds straight through towards the respective optical cameras. The wavelength range from 383 to 907\,nm is covered by six cross dispersers for the total of 92 \'echelle orders.  The number of orders per CD setting is either 15 or 16, and was driven by the desire to completely cover the last of the blue orders (159th) and still being not too close to the edge of the CCD. The average order separation is $\approx$\,40\arcsec . Table~\ref{T4} lists the (fixed) order coverage and wavelengths for all cross dispersers.

% ------------------------------ Table 3
\begin{table}
\caption{Cross disperser VPH grating efficiency measurements. } \label{T3}
\tabcolsep=10pt
\begin{tabular}{lllll}
 \hline \noalign{\smallskip}
CD \#   & $\lambda_{\rm c}$  & VPH & Prism & Eff. \\
           & (nm)  & ($\ell$\,mm$^{-1}$) & Glass & (\% ) \\
%  \noalign{\smallskip} 
  \hline \noalign{\smallskip}
 I     & 404 &  3160 & PBM2Y  & 74 \\
 II    & 450 &  2625 & PBL25Y & 81 \\
 III   & 508 &  2258 & PBL25Y & 85 \\
 IV   & 584 &  1871 & N-SF57 & 93 \\
 V    & 685 &  1512 & N-SF57 & 95 \\
 VI   & 825 &  1110 & N-SF57 & 90 \\
% \noalign{\smallskip}
 \hline
\end{tabular}
\end{table}

% ------------------------------ Table 4
\begin{table}
\caption{Wavelength coverage and order numbering per cross disperser. } \label{T4}
\begin{tabular}{lcccc}
 \hline \noalign{\smallskip}
CD \#  & $\lambda$ range & $\Delta\lambda$ & Order  &  Order \\
           & (nm)                  & (nm)                  &  Range & Sum    \\
%  \noalign{\smallskip} 
  \hline \noalign{\smallskip}
 I     &  383.7--426.5 & 42.8 & 159--145 & 15 \\
 II    &  426.5--480.0 & 53.5 & 144--129 & 16 \\
 III   &  480.0--544.1 & 64.1 & 128--114 & 15 \\
 IV   &  544.1--627.8 & 83.7 & 113--98~~   & 16 \\
 V    &  627.8--741.9 & 114.1~~ & 97--83   & 15 \\
 VI   &  741.9--906.7 & 164.8~~ & 82--68   & 15 \\
% \noalign{\smallskip}
 \hline
\end{tabular}
\end{table}

The individual substrate plates are made of N-BK7 for CD~II--VI and Silica for CD\,I, and are each 15\,mm thick and bonded together. Their surface quality was specified to $\lambda$10th (but never measured).  These substrate plates are cemented to their respective entrance and exit prisms. This procedure was done under optical control at Thales-SESO (Fig.~\ref{F-cddesign}b) and the achieved alignment of the dispersion-axes between the two prisms were 14\arcsec$\pm$10\arcsec\ and between the prisms and the VPH grating $<$1\arcmin . Parallelism was measured to be better than 20\,$\mu$m for all units. All assemblies show a beam displacement between 54\arcmin--79\arcmin\ (peak to valley) with respect to the vertical (gravity) axis due to a systematic difference between the diffracted dispersion axis of the VPH grating and the physical edges of the substrate plates. We compensate this by tilting the entire optical camera by 63\arcmin .

Each CD rests on a wiffle-tree support in order to minimize stress forces on the substrate plates that carry the VPH gelatine. The CDs weight 26\,kg (CD\,I), 24.5\,kg (CD\,II), 24.3\,kg (CD\,III), 23.9\,kg (CD\,IV), 22.3\,kg (CD\,V) and 18.5\,kg (CD\,VI).  A mechanical selector based on a large precision linear-dish positioner allows to select the proper CD and move it into the optical axis of the respective transfer collimator. The individual CDs are arranged vertically and moved up and down as a whole. In principle, this is the only moving part of the spectrograph.

%------------------------------- Fig  BLUE & RED camera
\begin{figure*}
{\it a)} Blue camera

\includegraphics[angle=0,width=85mm,clip]{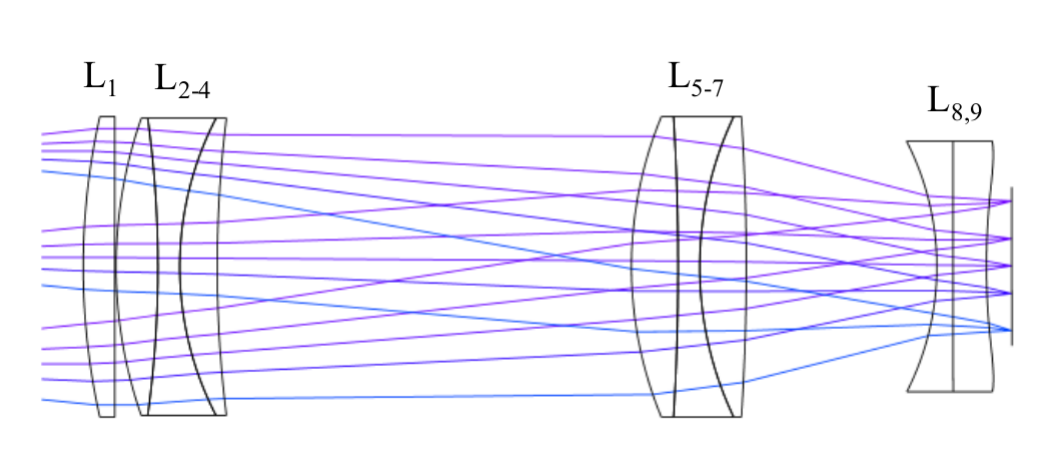}\hspace{3mm}
\includegraphics[angle=0,width=80mm,clip]{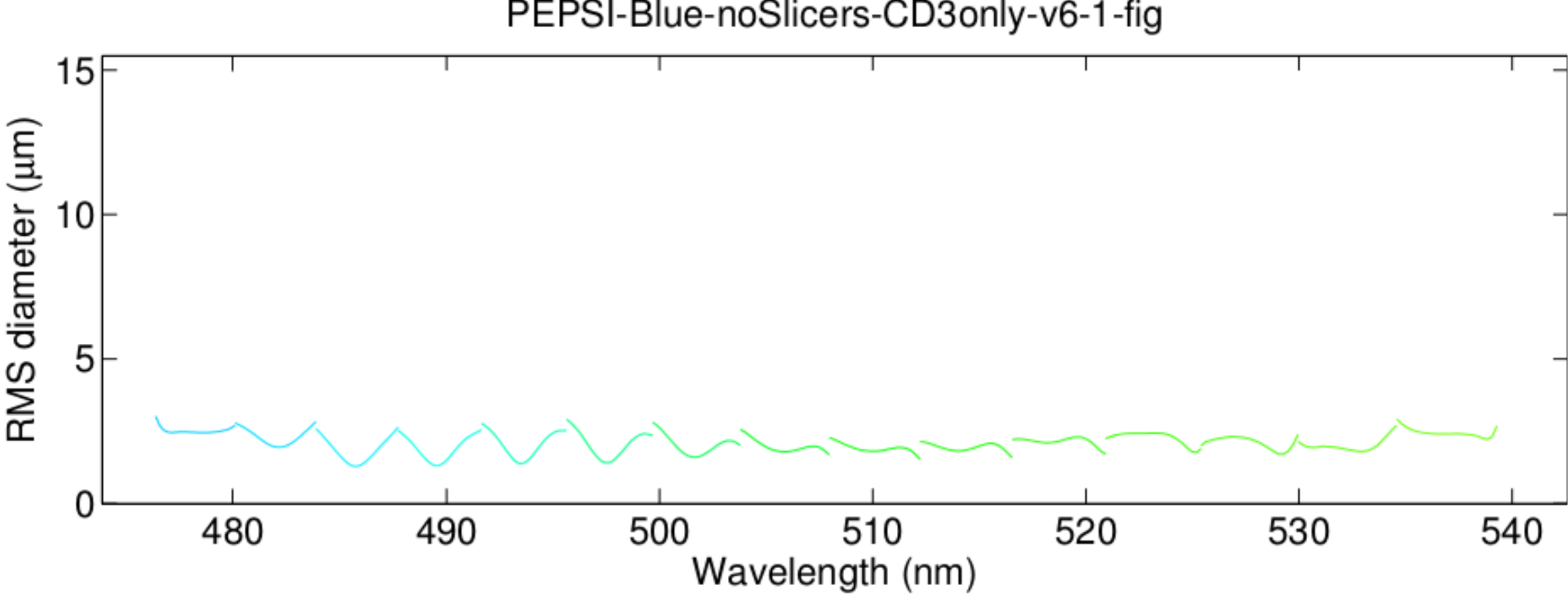}\\
{\it b)} Red camera

\includegraphics[angle=0,width=85mm,clip]{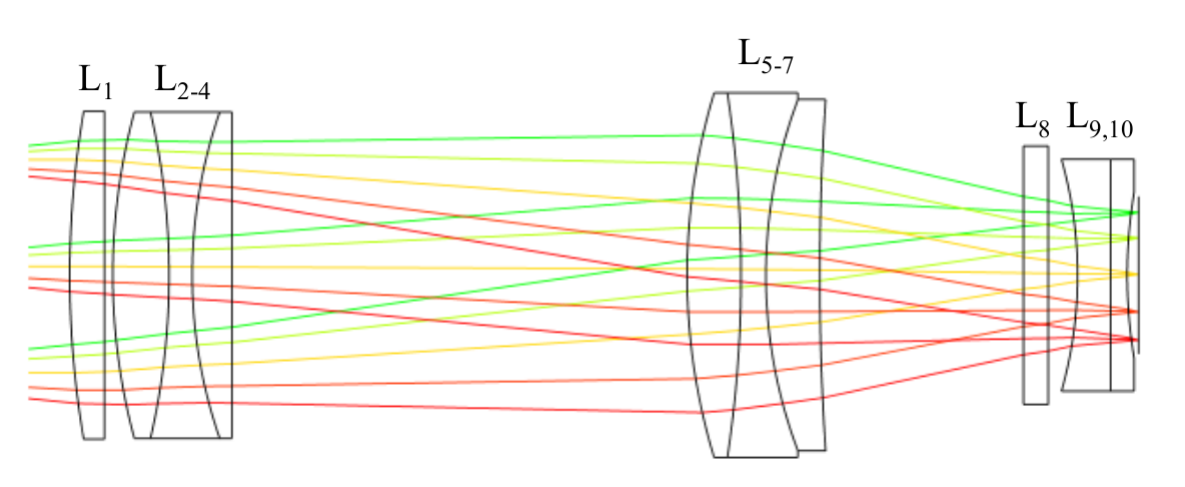}\hspace{3mm}
\includegraphics[angle=0,width=80mm,clip]{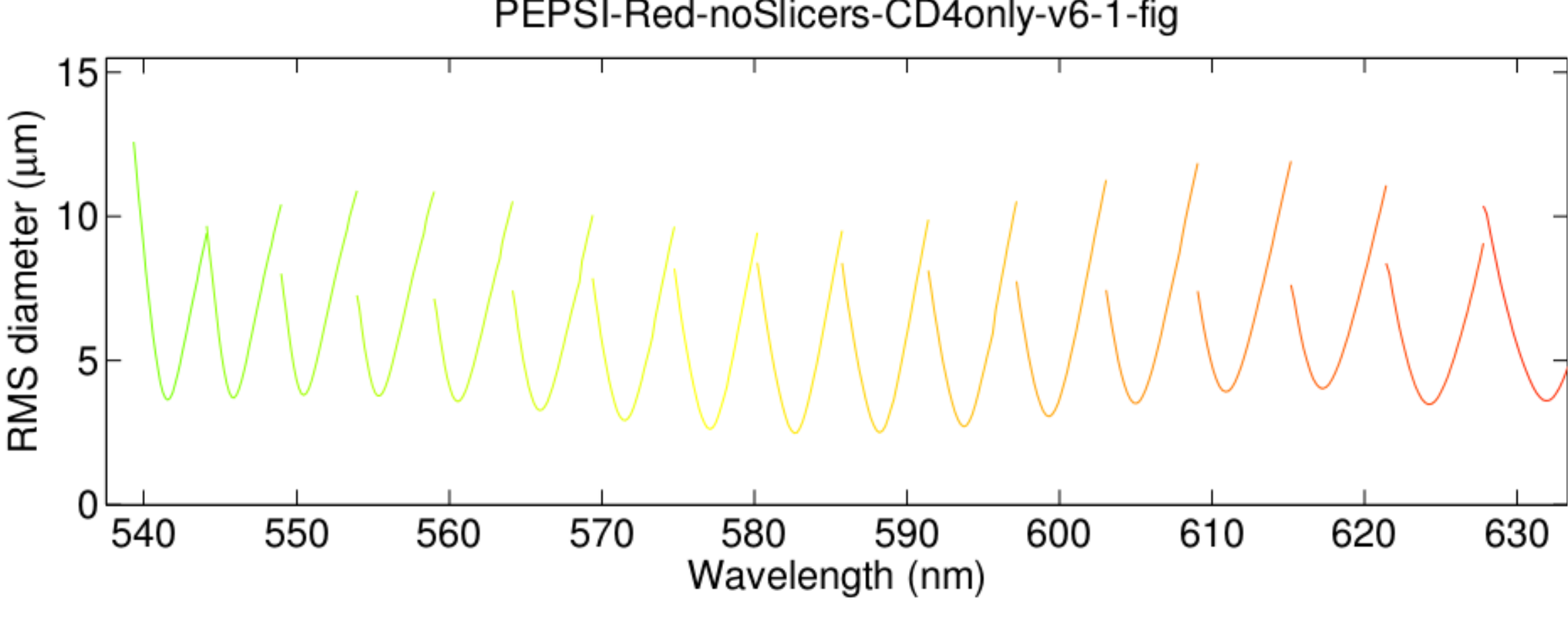}
\caption{Optical camera designs. ßemph{a}) Blue camera. \emph{b}) Red camera. Each row shows the lens arrangement on the left side and the spot rms diameter on the right side (for one representative cross disperser). }\label{F-camera}
\end{figure*}

\subsection{Optical cameras}

Both spectrograph cameras are all-refractive and have an f/3 focal ratio with lenses arranged in four groups. Their individual optical design has been wavelength optimized. Both cameras have an asphere in the field-lens group to further compensate for a two-dimensional field curvature due to the tilt of the pupil introduced by the \'echelle in one direction and introduced by the grisms in the perpendicular direction. Note that the last optical surface of the field-lens group is the last surface in front of the CCD; it is number 64/66 (blue/red camera) in the entire optical train. Therefore, the field-lens group of the optical cameras also act as the respective dewar windows.

\subsubsection{Blue camera}

The blue-optimized camera consists of 9 lenses with a free entrance aperture of 200\,mm (Fig.~\ref{F-camera}a). The 9 lenses are arranged in four groups of one (L1), three (L2--L4), three \hbox{(L5--L7)} and two (L8--L9)  lenses of various glasses and add up to 8 air-to-glass surfaces. The asphere on the last surface of L9 has  a mid-point separation to the CCD surface of 15.58\,mm and an edge-point separation of 9.5\,mm in all four corners. Table~\ref{T-bluecam} summarizes the lens data.

Its final Zemax merit function was 0.002393. The camera is nearly diffraction limited at 435\,nm. Spots range from a diameter of 6 to 10\,$\mu$m for the central wavelength on axis to the shortest wavelength at the edge of the order. The 80\,\%\ encircled energy also ranges from 6 to 10\,$\mu$m diameter. The maximum wavefront deviation is 0.61\,$\lambda$. The beam diameter that still contains 80\,\%\ of the energy (EE80)  is 7.0\,$\mu$m on axis and widens up to 11$\mu$m at the field edges. Longitudinal chromatism causes on-axis focal shifts of $\pm$10\,$\mu$m between the two cut-off wavelengths while it amounts to --100\,$\mu$m at the field edges  averaged for all wavelengths. Lateral chromatism on axis ranges between --10\,$\mu$m and +25\,$\mu$m and up to --120\,$\mu$m for the field edge at the shortest wavelength. The camera transmission peaks at 94\,\%\ at the red cut-off wavelength, 82\,\% at 400\,nm, and 70\,\% at 383\,nm. The measured focal length is 385.4$\pm$2.0\,mm at 546\,nm.

% ------------------------------ Table 7 Linsen Blaue Kamera
\begin{table}
\caption{Blue camera lens data. $D$ diameter, $R_c$ radius of curvature, $d$ mid thickness (AS = asphere).} \label{T-bluecam}
\tabcolsep=8pt
\begin{tabular}{lcccl}
 \hline \noalign{\smallskip}
Lens   & $D$ & $R_{\rm c}$ & $d$    & Glass \\
         & (mm)   & (mm) & (mm) & \\
%  \noalign{\smallskip} 
  \hline \noalign{\smallskip}
L1 &200 &411.21 &19.0 &S-FPL51-P1\\
    &200 & $\infty$& 0.5 & Air\\
L2 &200 &270.03 &25.5 &S-FPL53-P1\\
L3 &200 &--649.38~~ &13.0 &S-NBM51-P1\\
L4 &200 &193.04 &22.0 &S-FSL5-P1\\
    &200 &720.65& 248.8 & Air\\
L5 &200 &235.04 &28.0 &S-FTM16-P1\\
L6 &200 &--1468.38~~~~ &13.0 &PBM18Y-P1\\
L7 &200 &208.91 &28.0 &S-FTM16-P1\\
    &200 &--1304.31~~~~~& 114.4 & Air\\
L8 &150 &--164.4~~~~ &9.75 &PBL6Y-P1\\
L9 &140 &$\infty$ &20.0 &F-Silica\\
    &140 &AS 488.14& 15.2& Vacuum\\
% \noalign{\smallskip}
\hline
\end{tabular}
\end{table}

% ------------------------------ Table 8 Linsen Rote Kamera
\begin{table}
\caption{Red camera lens data. $D$ diameter, $R_c$ radius of curvature, $d$ mid thickness (AS = asphere).} \label{T-redcam}
\tabcolsep=8pt
\begin{tabular}{lcccl}
 \hline \noalign{\smallskip}
Lens   & $D$     & $R_{\rm c}$ & $d$    & Glass \\
         & (mm)   & (mm)  & (mm)  & \\
%  \noalign{\smallskip} 
\hline \noalign{\smallskip}
L1 & 200&		299.99    &	23.0  & S-FPL51-P2 \\
    & 200&		--2015.29~~~~&	0.75  & Air\\
L2 & 200&		375.84    &	25.5  & S-FPL53-P1\\
L3 & 200&		--558.96~~  &	7.0	& S-NBM51-P2\\
L4 & 200&		161.91    &	30.0	& S-FSL5\\
    & 200&	      AS 484.62 &	272.6& Air\\
L5 & 200&		360.93    &	35.0	& S-LAM2-P1\\
L6 & 200&		--299.99~~  &	10.0	& S-NBM51-P2\\
L7 & 200&		346.94    &	28.0	& SF57-P1\\
    & 200&		$\infty$     &		135.9& Air\\
L8 & 140&		--194.97~~ &		10.0	& S-TIH6-P1\\
L9 & 140&		--149.51~~ &		8.0	& Herasil\\
L10&100&	Toroidal	     &		12.0	& N-K5\\
    & 100& Toroidal AS     & 	6.5   & Vacuum\\	
% \noalign{\smallskip}
\hline
\end{tabular}
\end{table}

\subsubsection{Red camera}

The red-optimized camera consist of 10 lenses with an entrance aperture of 200\,mm (Fig.~\ref{F-camera}b). These are arranged in four groups of one lens (L1), three (L2--L4), three (L5--L7) and three (L8--L10)  lenses of various glasses and add up to~8 air-to-glass surfaces. A toroidal asphere on the last surface of L10 allows a mid-line lens separation to the CCD surface of 7.45\,mm and an edge-point separation of 2.7\,mm in all four CCD corners.  Table~\ref{T-redcam} summarizes the lens data. Its final Zemax merit function was 0.0037 and spots range from a diameter of 6 to 12\,$\mu$m for the central wavelength on axis to the reddest wavelength at the edge of an order. The respective EE80 energy resides between 11 to 16\,$\mu$m diameter. The measured focal length is 384.2$\pm$2.0\,mm.

\subsection{Field lens and dichroic coating}

An image of the \'echelle grating in the focus of the collimator appears spread out in wavelength with a full aperture of 400\,mm. In our case this is in vertical direction. A field lens collects the ``dispersed'' light over a large field of view and redirects it optimally to the transfer collimators. The lens assembly is a spherical  doublet of rectangular aperture 440$\times$65\,mm with a central total thickness of 60\,mm. Incidence angles are up to 15\degr . Its entrance lens L1 is made of N-BK7 with an outer radius of 1695\,mm convex and an inner radius of 3098\,mm concave. Both radii are off axis by 40\,mm. Central thickness is 30\,mm. The exit lens L2 is also made of N-BK7 with an outer radius of 3098\,mm convex but  an inner radius of 819\,mm concave. Its off-axis distance is also 40\,mm with a wedge angle of 88.56\degr . Its central thickness is also 30\,mm. L1 and L2 are glued together with NOA61. The entrance surface of L1 is AR coated for the entire wavelength range of the spectrograph with an average transmission of 98.8\,\% . The exit of L2 is also AR coated but optimized for the red-arm wavelength range.

%------------------------------- Fig   close-up Field lens design with ghosts
\begin{figure}
\includegraphics[angle=0,width=83mm,clip]{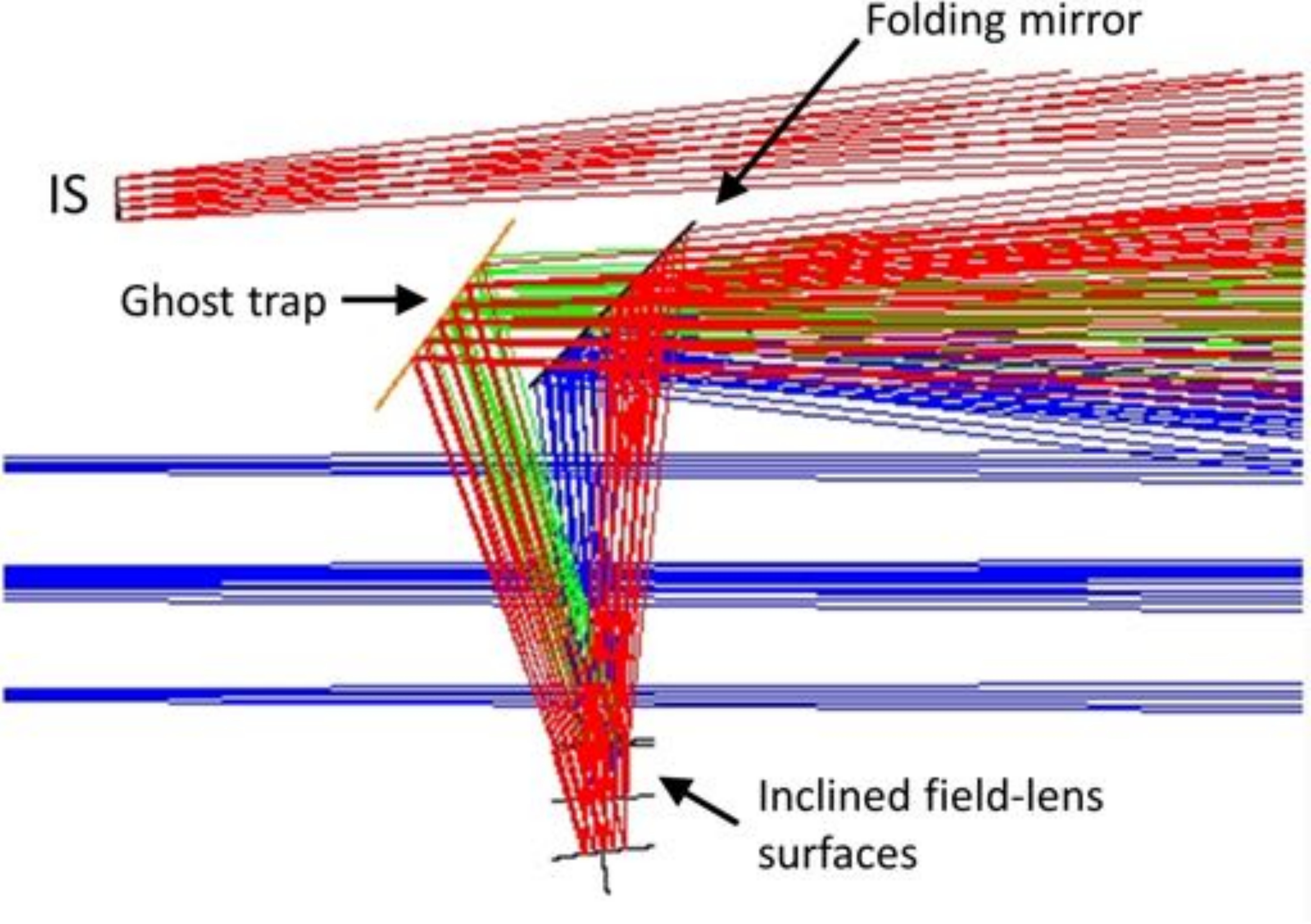}
\caption{Ghost trap near the double-pass folding mirror. The design of the field lens is such that ghosts from its internal reflections are diverted out of the beam and into a light trap. Compare to the overall design drawing in  Fig.~\ref{F-specdesign}. (IS Image Slicer).}\label{F-FLdesign}
\end{figure}

The L1-exit and L2-entrance surfaces have additionally a wedge angle of 2.45\degr\ in the cross-dispersion plane with respect to the collimator axis. This angle redirects all ghosts from the internal reflections within the field lens out of the field of view (Fig.~\ref{F-FLdesign}). The reflected blue-side ghosts miss the folding mirror for the double pass while the reflected red-side ghosts hit the folding mirror in double pass but then miss the blue transfer-collimator mirror. The transmitted ghosts, blue-side and red-side, miss the (red) folding mirror all together. Note that ghost strengths in the reflected beam are up to 3\,\%\ (depending on field position) while in the transmitted beam it is comparably negligible at the $<$\,0.1\,\%\ level throughout the field. This is possible because the field lenses are not in a pupil but are near the collimator focus. Therefore, field-lens ghosts do not significantly contribute to the stray-light budget of the instrument. Stray-light traps throughout the instrument generally keep it to a minimum.

The exit surface of the L1 doublet carries the dichroic coating for the light splitting to the blue and the red arm. Two sets of field lenses, each a doublet as described above, are available so that there is no blind wavelength region for the spectrograph. The difference between the two field lenses is just the cut-off wavelength of the dichroic coating, set to 508\,nm for the ``blue'' field lens (default unit)  and to 583.5\,nm for the ``red'' field lens. The blue field lens reflects the blue spectrum below 479\,nm and transmits the red spectrum above 537\,nm. Average transmission below 479\,nm is 1.1\,\%, transmission above 537\,nm is 97\,\%. The exit surface of L2 is AR coated for the wavelength range 537--907\,nm. The red field lens reflects the blue spectrum below 542\,nm and transmits the red spectrum above 625\,nm. Its L2 exit surface is AR coated for the wavelength range 625--907\,nm. The transmission below 542\,nm is 0.9\,\%, transmission above 625\,nm is 97\,\%.

Both field lenses are mounted side by side on a mechanical linear stage so that a quick exchange with micro-level repositioning accuracy is possible. The blue field lens is the default but when CD\,III is requested the change to the red field lens is automatically executed for complete wavelength coverage.

%--------------------------- F  slicer stack
\begin{figure}
\includegraphics[angle=0,width=83mm,clip]{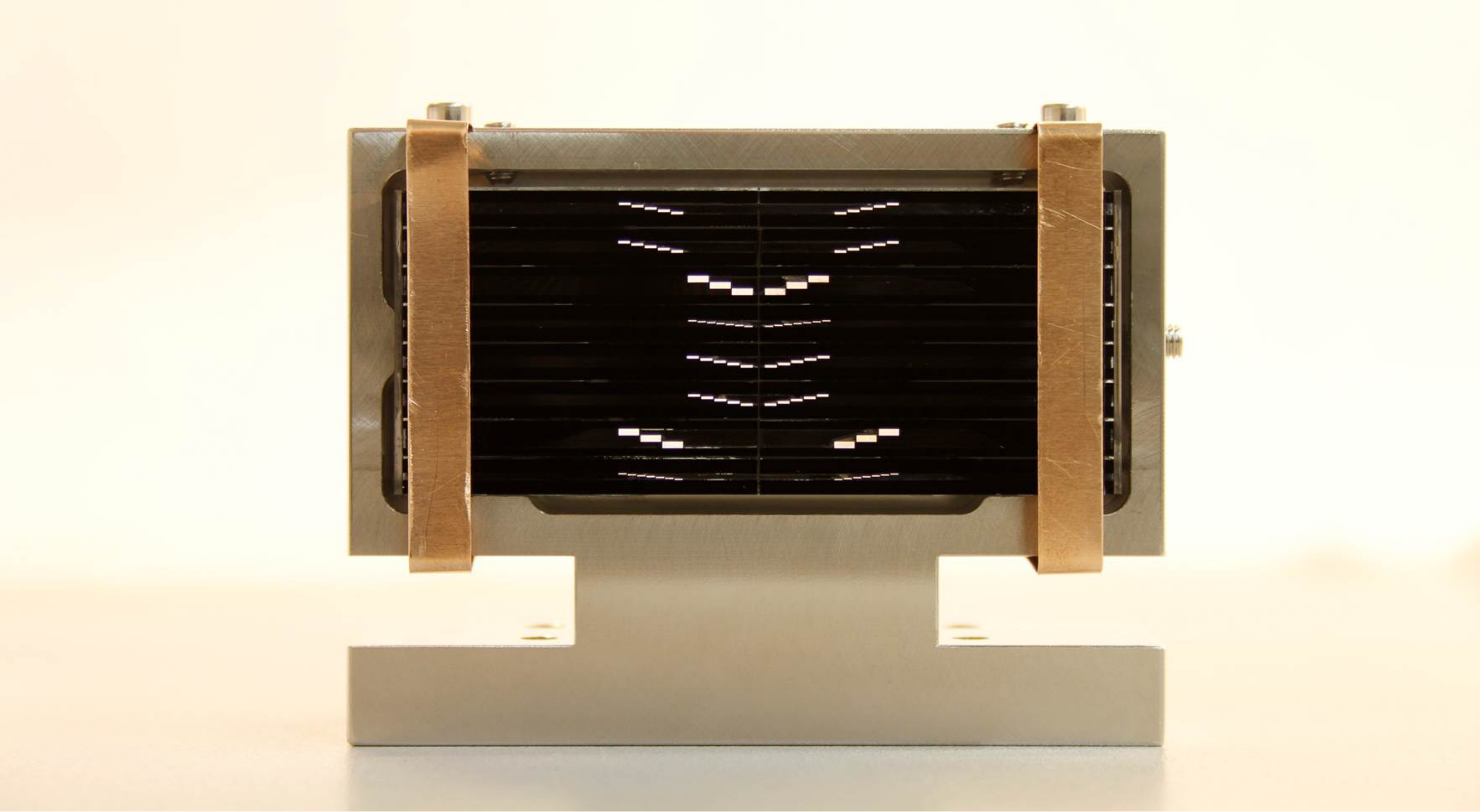}
\caption{View of the exit side of the PEPSI image-slicer stack. The figure shows image slicer \#1 with  approximately real dimensions. Its geometric center is aligned with the optical axis of the spectrograph. All image slicers are on the same Zerodur carrier. The stack remains fixed due to the large field of view of the Maksutov collimators. }\label{F-slicerstack}
\end{figure}

\subsection{Image slicers}

The three resolution modes of PEPSI are defined by the three different fibre-core diameters in combination with three image-slicer dimensions. Each fibre supports its own image slicer (IS). We adopted the so-called ``wave-guide'' IS concept. A single slicer consists of a stack of plane-parallel glass plates (slices) that are adhesively bonded together with high parallelism and where the bonding glue acts as the wave guide in total internal reflection. We used Epo-Tek OG134 ($n$ = 1.451) glue. The overall efficiency is set by the sampling of the diameter of the entrance beam with $n$ slices, i.e. by the width of the waveguide between the individual slices, and by the Fresnel losses at the two air-to-glass surfaces. For all slicers a waveguide thickness of 3--4\,$\mu$m was achieved. A detailed description of this new variant of image slicer was given by Beckert et al. (\cite{beckert}). The practical performance critically depends on the achieved mechanical production precision of tip and tilt of the individual slices with respect to each other and by the transformation of the fibre exit f-ratio to the (main) collimator f-ratio. Two sets of slicer stacks, each with 16 individual slicers, are available at PEPSI, one acts as a hot spare. Figure~\ref{F-slicerstack} shows the first of the two stacks in its test set-up.

%------------------------ Tabe 6   slicer dimensions
\begin{table*}
\caption{Image slicer dimensions in $\mu$m for each resolution/observing mode.} \label{T6}
\tabcolsep=10pt
\begin{tabular}{llllll}
 \hline \noalign{\smallskip}
         & $R$ = 270\,000 & $R$ = 120\,000 & $R$ = 43\,000  & $R$ = 120\,000 \\
         & \multicolumn{3}{c}{Integral Light} & Polarimetric \\
% \noalign{\smallskip} 
\hline \noalign{\smallskip}
Fiber core diameter $d$    & 100 & 200 & 300 & 200 \\
IS identification    & 1--4 & 5--8 & 9--12 & 13--16\\
Number of slices per IS & 7 & 5 & 3 & 5 \\
Slice thickness   & 70 & 175 & 500 & 175 \\
Wave-guide gap   & 3 & 3 & 3 & 3 \\
Reimaged fibre-core diameter & 463 & 926 & 1390 & 926 \\
Slicer entrance size & 637$\times$505 & 932$\times$887 & 1620$\times$1506 & 932$\times$887 \\
 & Underfilling & Overfilling & Underfilling & Overfilling\\
Slicer exit size & 637$\times$70 & 932$\times$175 & 1620$\times$500 & 932$\times$175 \\
Pitch between slices & 100 & 100 & 100 & 100 \\
Length of one slicer & 5059 & 5060 & 5060 & 5060 \\
Length Target+Sky slicer incl. pitch& 10518 & 10520 & 10520 & 10520 \\
Total length & 22000 & 22000 & 22000 & 22000\\
Geometric slicer loss (\%)  & 4.1 & 2.7 & 0.25 & 2.7 \\
Total efficiency (\%)  & 92.1 & 93.4 & 95.8 & 93.4 \\
% \noalign{\smallskip}
 \hline
\end{tabular}
\end{table*}

\begin{figure}[!t]
\center
\includegraphics[angle=0,width=83mm,clip]{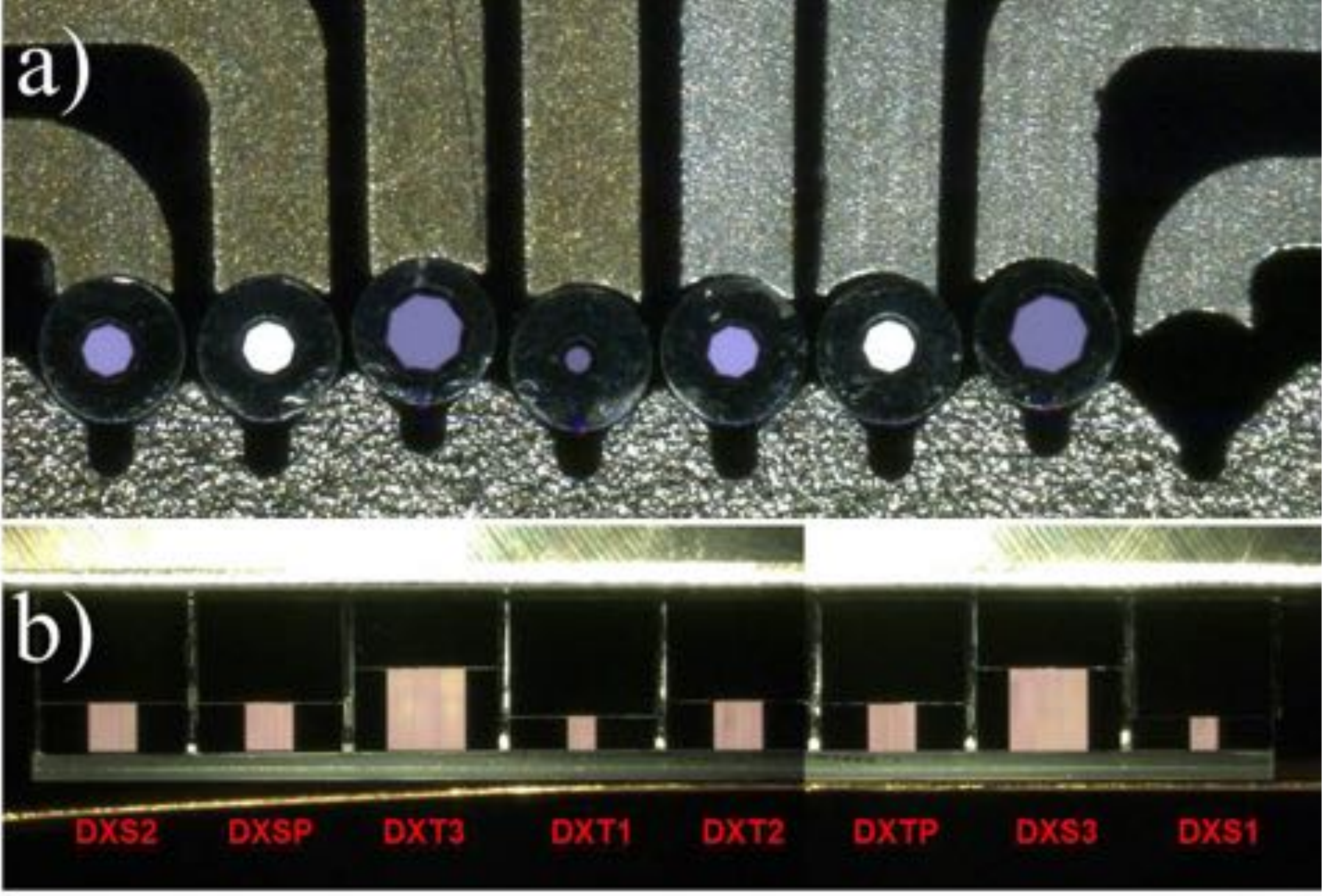}
\caption{Slicer interfaces for the DX side of the first of the two IS units. \emph{a}) Close-up of the fibre exits. Shown are seven of the eight octagonal fibre exits. One fibre is left away to show the leverage of one holding arm. The outer ferrule diameter is for all fibres the same (600 $\mu$m). \emph{b}) Close-up of the entrance surface of eight of the 16 image slicers. A transfer-optics connects the fibre exit with the slicer entrance. }\label{F-slicer}
\end{figure}

Table~\ref{T6} summarizes the slicer specifications and dimensions. The overall dimensions of a single stack are $N$ times the slice thickness plus the 3\,$\mu$m wave-guide gap in between. $N$ is 3, 5, and 7 for the 300 $\mu$m, 200 $\mu$m, and 100~$\mu$m fibres, respectively. Each slice is of different length and its distal end is polished under an angle of 45\degr\ and coated with protected silver. Thus, light exits the slicer at an angle of 90\degr\ with respect to the fibre entrance and with a spatial offset of the sliced light, called the pitch angle. Because all three resolution modes should fill the CCD real estate to its full extent, the slicer stack separation for all fibre cores in the focal plane of the collimator must be the same. The ultra-high resolution (UHR) mode must use the smallest of the three fibre cores, the 100 $\mu$m fibre with a 0.74\arcsec\  entrance aperture on the sky (the LBT-site's median seeing is 0.65\arcsec). When slicing this into 7 slices, we reach the practical limit due to the available space on the CCD detector (still with a sampling of 24\,pixels per slice though). Therefore, the 7-slice slicer sets the separation for the other slicers. Figure~\ref{F-slicer} shows the relative positioning of all 16 slicer stacks for slicer \#\,1. Slicer \#\,2 has a different slicer arrangement but is based on the same principle and will be described in a forthcoming paper. See Table~\ref{T6} for actual dimensions.

%--------------------------- F  slicer interfaces

Note again that every fibre entering the spectrograph collimator is equipped with its own image slicer. Two fibres per telescope are always ``active'', i.e. receive light. The total of four resolution/observing modes, three for integral light and one for polarimetric light, requires a total of 16 image slicers. Due to the wide field of view of the Maksutov main collimator, the position of these image slicers remain fixed and need not to be moved into position. Therefore, the entire image-slicer unit consists of a stable Zerodur carrier on which the 16 individual slicer stacks are assembled with respect to each other.  Two entrance and one exit window protects the slicers. The windows are broad-band AR coated for the entire wavelength range of the spectrograph. A retractable beam-splitting cube in front of the slicer-unit entrance can be used to monitor the adjusted state of the focused spots.

\begin{figure}
{\it a)}

\includegraphics[angle=0,width=83mm,clip]{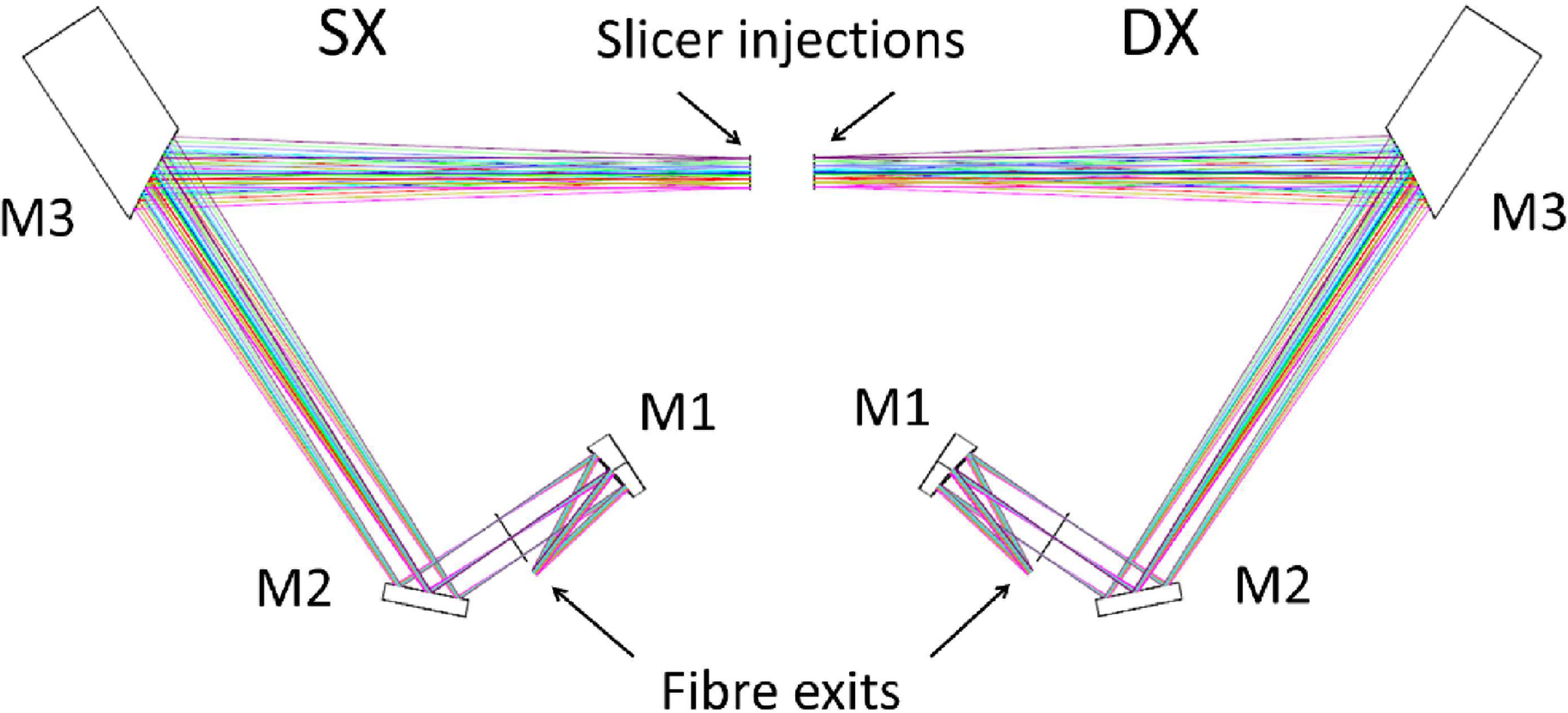}\\
{\it b)}

\includegraphics[angle=0,width=83mm,clip]{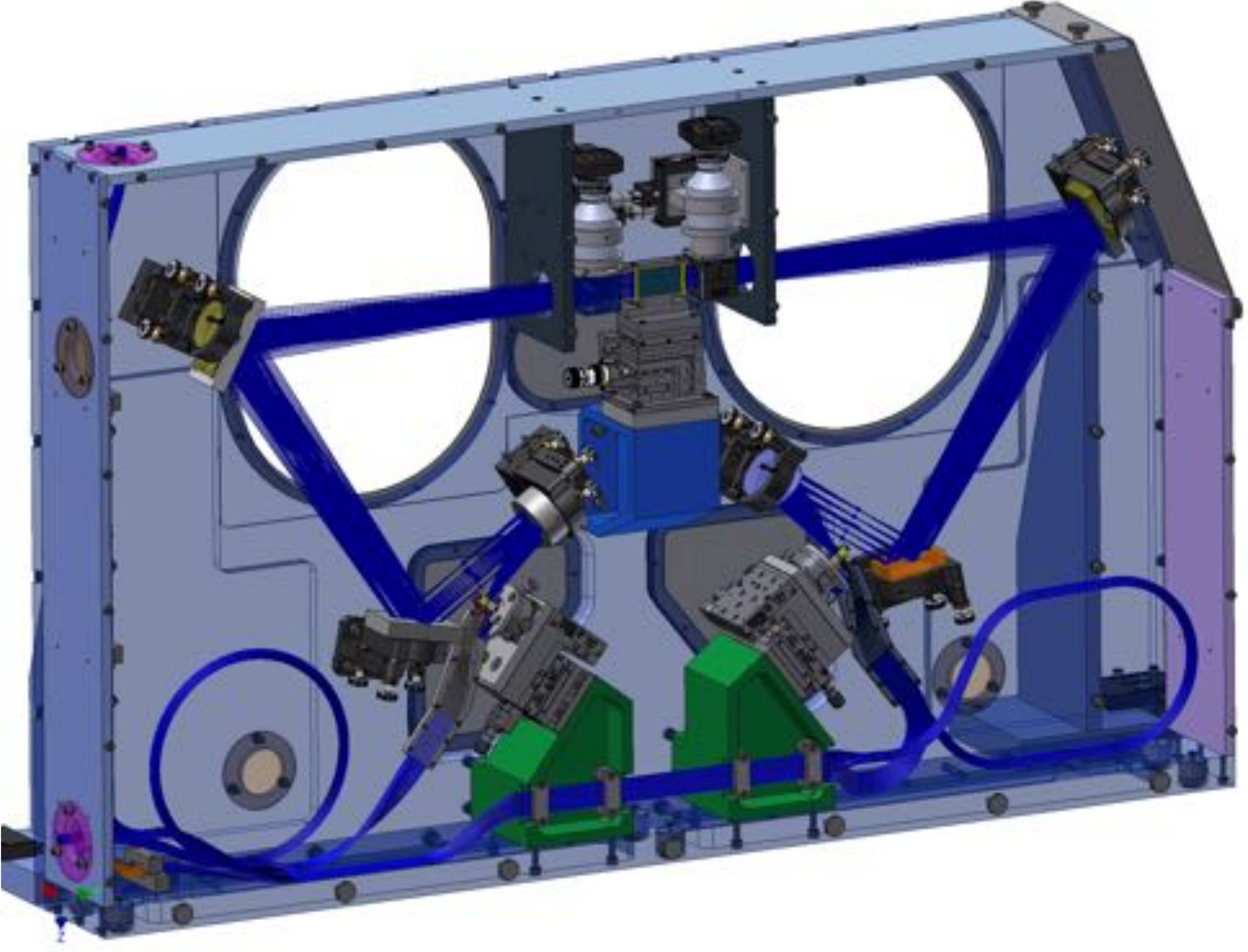}
\caption{Fiber-to-IS f-ratio transformation. \emph{a}) Zemax optical design. Two off-axis parabola (M1, M3) and one flat mirror (M2) convert the fibre-exit f/3 ratio to the main-collimator f/13.9 ratio. Eight fibres are injected for each LBT mirror (SX and DX). All mirrors are UV-enhanced silver coated and have an average combined reflectivity of 93\,\%. 
\emph{b}) Mechanical design. }\label{F-transf}
\end{figure}

Because light exits the 44\,m fibres at f/3 and because the image-slicer exit shall match the main collimator entrance f-ratio of f/13.9, we must transform the fibre exit from f/3 to f/13.9 before entering the image slicers (a magnification factor of 4.7). This is achieved with a three-mirror re-imaging system, one for each telescope fibre train (Fig.~\ref{F-transf}). Two of the three mirrors are off-axis parabolas and one is a flat. The critical mirror has an off-axis radius of 353\,mm at a size of just 73$\times$40\,mm (free aperture of 62$\times$32\,mm) and was manufactured by Carl Zeiss Jena. All surfaces are coated with an UV-enhanced silver coating with an average reflectivity of 97\,\%\  from 380--900\,nm. Its overall efficiency is thus close to 93\,\%\ for all wavelengths and free of wavelength-dependent aberrations. Field-dependent aberrations are still existent but are  minimized by placing the sky and/or the low- and medium-resolution fibres to the outermost locations of the image slicer.

The high magnification factor of 4.7, together with the fact that the IS shall make use of the entire CCD real estate, sets the geometric fibre separation to 647.5\,$\mu$m at the exit side. Because each fibre must have a ferrule for handling, this required a ferrule of outer diameter of just $\approx$\,600\,$\mu$m for any of the three (octagonal) fibre cores. All PEPSI fibres are equipped with such a custom-made stainless steel ferrule and arranged according to the IS entrance geometry shown in Fig.~\ref{F-slicer}a.

The full IS assembly has an exit window to the collimator of size 22.4$\times$22.0\,mm. The arrangement of the IS in vertical direction is uniform with a vertical pitch distance of 3.058\,mm with respect to the central slice in each IS. This separation sets the spectrum separation per \'echelle order on the CCD. The number of \'echelle orders per cross disperser has been fixed to 15 or 16. With a CCD real estate of 10.3k pixels square in total, i.e. 10\,560 pixels also in the cross-dispersion direction, it converts on average to 170\,pix per image slicer, i.e. per fibre. An individual IS splits this, again on average, into a spectrum height of 57\,pix/slice for the 3-slice IS, 34\,pix/slice for the 5-slice IS, and 24\,pix/slice for the 7-slice IS. Inter-slice space is at minimum one pixel for the ultra-high-resolution mode. The image scale on the CCD is given by the telescope image scale of 611\,$\mu$m/arcsec\ and the ratio of the  telescope-to-camera demagnification (both PEPSI arms have f/3 cameras). It amounts to 122\,$\mu$m/arcsec\ or, equivalently, to 13.6\, pix/arcsec\ or 0.074\arcsec/pix).

\subsection{Folding mirrors}

Two folding mirrors (FM) redirect the beam from the main collimator via the dichroic beam splitter to the two transfer collimators. While FM1 redirects the dispersed light from the entire wavelength range 383--907\,nm into the beam splitter, FM2 redirects just its exit light for the red arm from 537\,nm and longer. Both mirrors are $\lambda$/40\,rms in a central sub-pupil within the clear aperture of 380$\times$120\,mm (mirror size is 420$\times$160\,mm for both FMs).  The carrier is made of Zerodur and coated with a blue-enhanced Silflex VIS coating. Its reflectivity averages at 98\,\%\ at an incidence angle of 45\degr\ from 400--900\,nm. It shows a continuous drop to 95\,\% from 400\,nm to 380\,nm.

%--------------------------- F  slicer f/ratio transformation optics

\subsection{Shutters}

Both spectrograph arms are equipped with an own shutter. We employ two ``Bonn shutters" (Reif et al. \cite{reif}) with a free 200$\times$200\,mm aperture. Each shutter is made of two moving blades of a carbon fibre compound and run by two stepper motors. The blades can be counter- or co-moving depending upon exposure time. Four micro-controllers are involved; one for each shutter blade stepper motor, one for host communication and one for input signal filtering. Bonn Shutters are impact free, low acceleration devices. Instead of driving the shutter blades at high speed/acceleration the $<$1\,ms timing accuracy is achieved by a very precise motion control of both blades. The generation of every single stepper motor micro-step (8295 steps for the 200\,mm aperture) follows exactly a time table which is derived from a given velocity profile. Both blades are driven with identical velocity profiles of trapezoidal shape.

The exposure-time error is $\leq$\,300\,$\mu$s and the exposure non-uniformity is less than 1\,ms over the entire field of view. Minimum exposure time is thus 1\,ms. During long exposures the shutter blades are hold by the two motors which then dissipate 240\,mW in total. Due to the extreme environmental stability requirements on the optical bench all four shutter motors are capsuled and externally cooled via the central heat exchanger unit. Both shutter units are mechanically disconnected from the optical table by having them mounted on the inner spectrograph cover.

% ---------------- Table  efficiency of all optical elements
\begin{table*}
\caption{Telescope and spectrograph efficiency. Abbreviations: Al aluminum, Ag silver, AR antireflection, IMM immersed.}\label{T-effall}
\tabcolsep=15pt
\begin{tabular}{lllccc}
\hline  \noalign{\smallskip}
Optical element  & Material & Coating  & \multicolumn{3}{c}{Throughput/Reflectivity  (\%)} \\
                        &              &              & 390\,nm & 650\,nm & 900\,nm\\
% \noalign{\smallskip}
 \hline  \noalign{\smallskip}
M1 & Ohara    & Al & 90 & 89 & 88\\
M2 & Zerodur & Al & 91 & 90 & 89\\
M3 & Zerodur & Al & 91 & 90 & 89\\
% \noalign{\smallskip}
\hline  \noalign{\smallskip}
 LBT total & & & 74  & 72 & 70 \\
% \noalign{\smallskip}
 \hline  \noalign{\smallskip}
Collimator & FPL51 \& NSL36 & AR \& IMM & 98 & 98 & 98 \\
Prism       & N-BK7 & IMM & 99 & 99 & 99\\
ADC & N-Pk51 \& PBL6Y  & IMM & 97 & 98 & 98\\
Beamsplitter & N-BK7 & IMM & 98 & 98 & 98 \\
Camera & FPL51 \& NSL36 & IMM \& AR & 98 & 98 & 98 \\
Folding mirror$^1$ & Zerodur & Ag \& Al & 96 & 98 & 98 \\
% \noalign{\smallskip}
 \hline  \noalign{\smallskip}
 PFU total & & & 87 & 89 & 89 \\
% \noalign{\smallskip}
 \hline  \noalign{\smallskip}
Collimator & F-Si \& SF57 & AR & \dots & 99 & 99\\
Retarder$^2$ & PMMA \& glass & AR & \dots & 97 & 97\\
Foster beamsplitter & Ca(CO$_3$) \& N-BK7 & AR & \dots & 98 & 98 \\
ADC & F2 \& PSK53A &AR  & \dots  &96 &96 \\
Camera & S-FPL53 \& N-LAK5 & AR & \dots  &96 &96 \\
 \noalign{\smallskip}
% \hline  \noalign{\smallskip}
 Polarimeter total$^3$ & & & \dots & 87 & 87 \\
% \noalign{\smallskip}
 \hline  \noalign{\smallskip}
44\,m fibre + rod lens & Quartz  & \dots & $\langle 65\rangle$ & $\langle 77\rangle$ & $\langle 86\rangle$ \\
f-ratio optics & Zerodur & Ag & 86 & 94 & 94 \\
Slicer & Herasil & AR \& Ag & $\langle 92\rangle$ & $\langle 93\rangle$ & $\langle 93\rangle$\\
% \noalign{\smallskip}
\hline  \noalign{\smallskip}
 Fibres \& Slicer  total & & & 68 & 75 & 75 \\
% \noalign{\smallskip}
\hline  \noalign{\smallskip}
Main collimator$^4$ & Zerodur & Al & 95  &97  &97 \\
Main corrector$^4$ & N-BK7 & AR & 96 & 96 & 96\\
$R$4 grating & Zerodur & \dots & 67 & 67 & 67\\
Folding mirror & Zerodur & Ag & 96 & 98 & 98\\
Field lens & N-BK7 & AR &95  &95  &95 \\
Folding mirror & Zerodur & Ag & 96 & 98 & 98 \\
Transfer collimator  & Zerodur & Al & 95 & 97 & 97 \\
Transfer corrector  & N-BK7 & AR & 96 & 96 & 96 \\
Cross disperser & Various & AR  & 74 & 95 & 90\\
Cameras  & Various & AR & 77  & 94 & 90\\
CCD & Si & AR & 85 &  94 & 48 \\
% \noalign{\smallskip}
\hline  \noalign{\smallskip}
 Spectrograph total & & & 24 & 32 & 21\\
% \noalign{\smallskip}
 \hline  \noalign{\smallskip}
Grand total$^5$ & Integral light   & & 11 & 15 & 10\\
Grand total$^5$ & Polarized light & & \dots & 15 & 9\\
% \noalign{\smallskip}
 \hline
\end{tabular}

\vspace{1mm}Notes: $^1$\,Absent in the 100-$\mu$m mode; selectable Ag- or Al-coated version. \ $^2$\,Absent in linear polarization mode. \ $^3$\,Polarimetric wavelength coverage: 450--907\,nm. \ $^4$\,In double pass. \ $^5$\,Seeing losses are not included.

\end{table*}

% ------------------------------------------------------------------------------------------------------------------------------
\section{Spectrograph environment}\label{S-Environ}

\subsection{Environmental control}

%--------------------------- F  chamber
\begin{figure}
{\it a)}

\includegraphics[angle=0,width=83mm,clip]{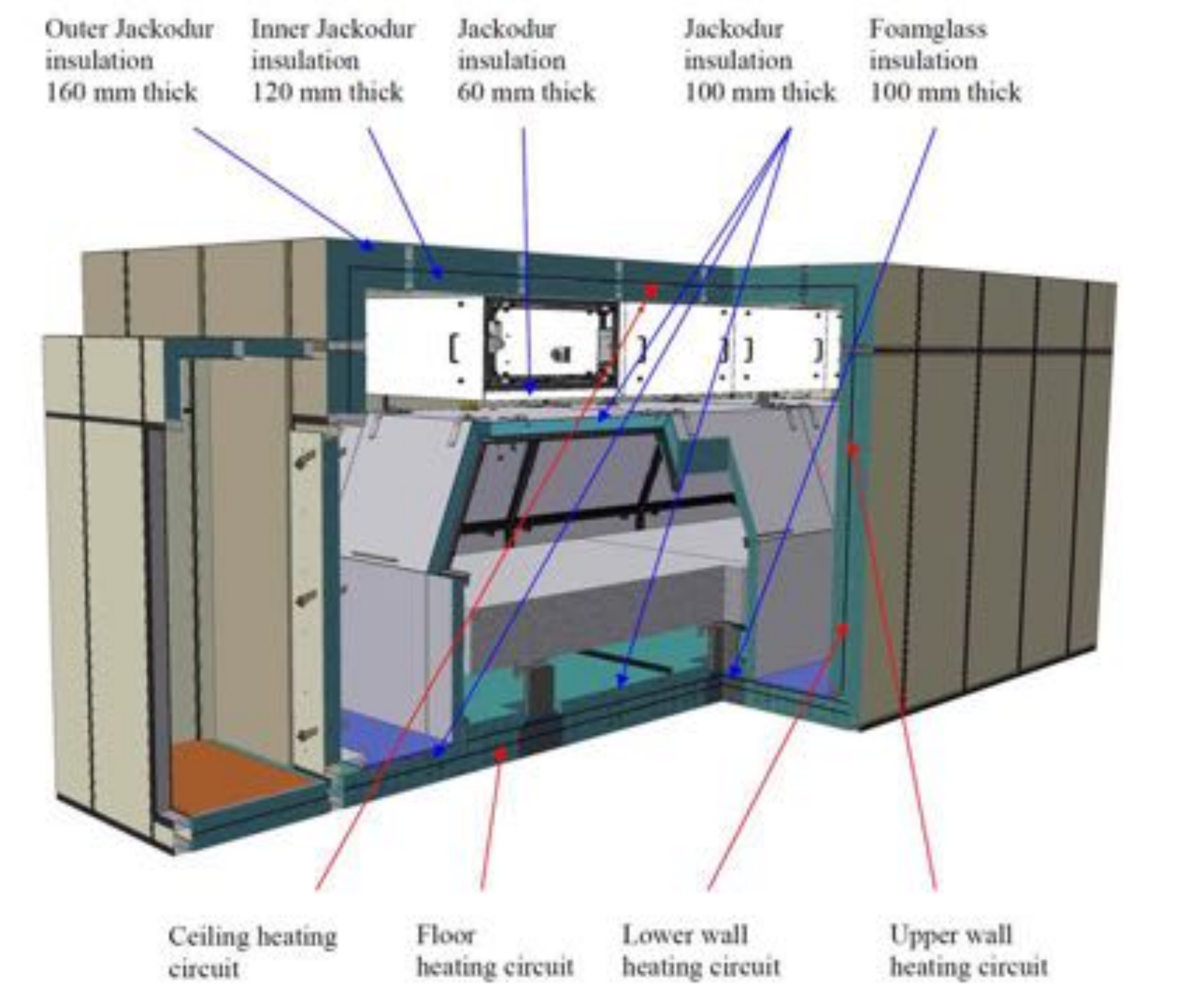}\\
{\it b)}

\includegraphics[angle=0,width=83mm,clip]{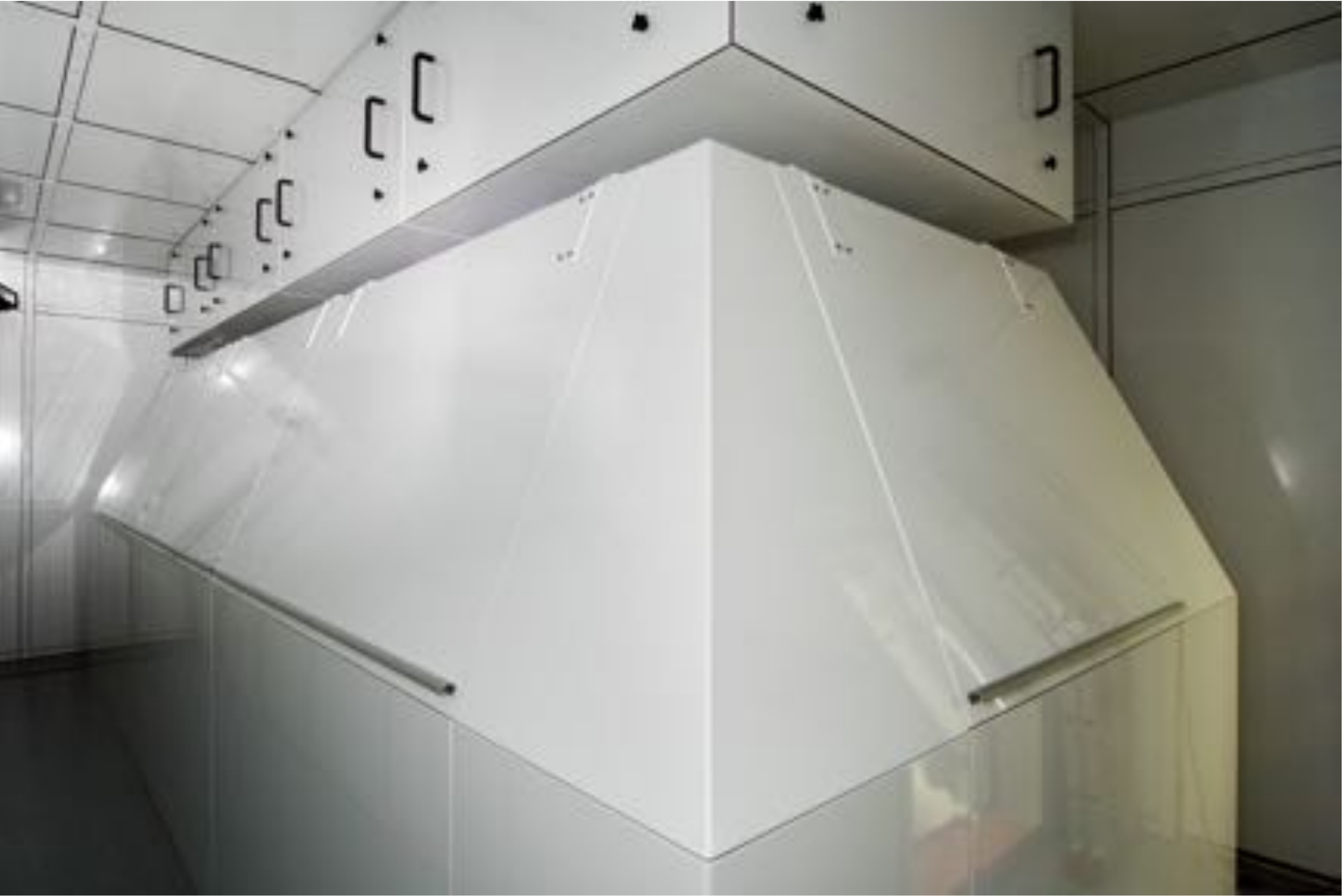}
\caption{The spectrograph environment. \emph{a}) Design of the chamber and cover components. \emph{b}) After  integration within the telescope pier at LBT. }\label{F-chamber}
\end{figure}

%--------------------------- F  chamber  p/T tests
\begin{figure}
{\it a)}

\includegraphics[angle=0,width=83mm,clip]{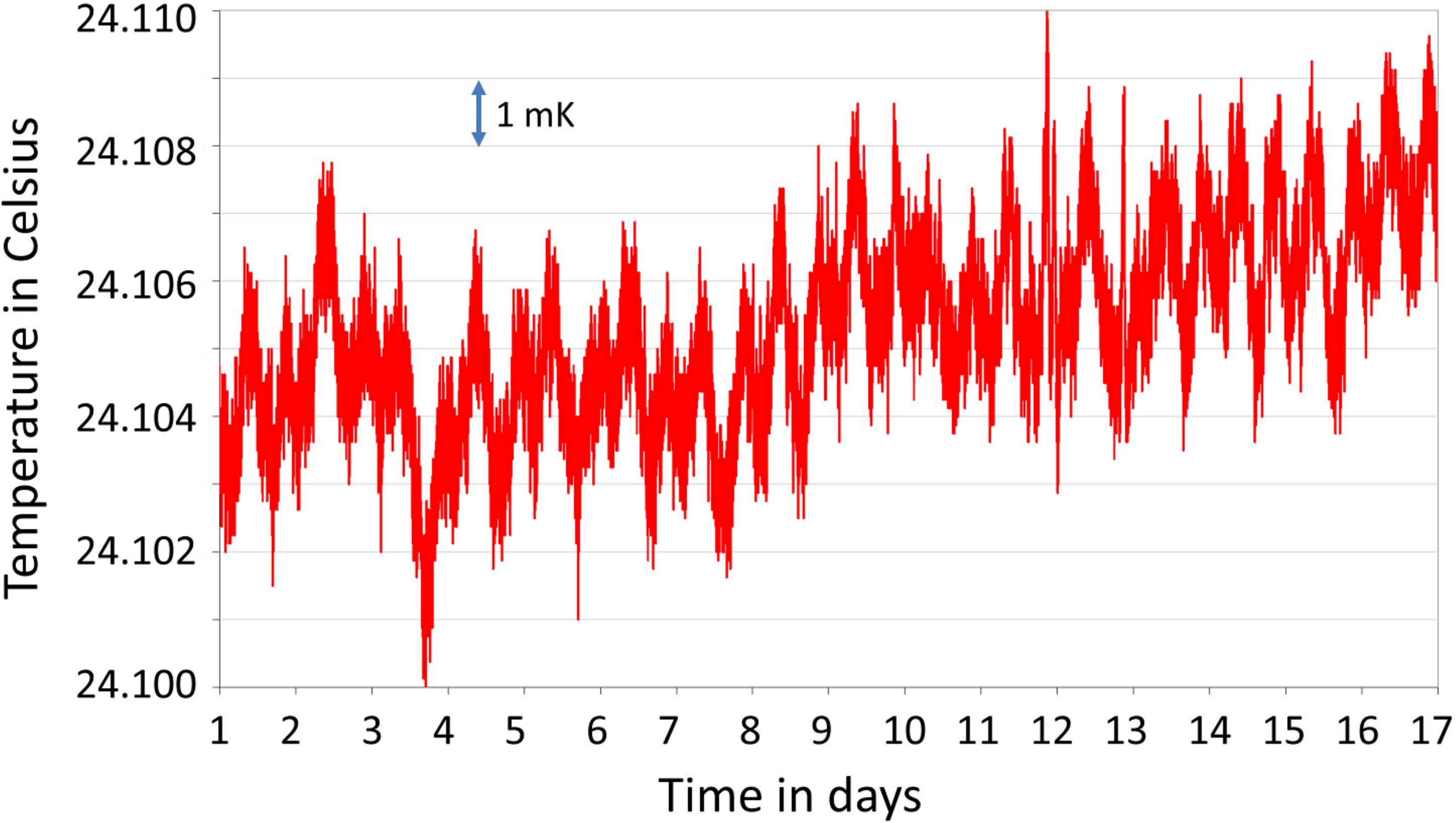}\\
{\it b)}

\includegraphics[angle=0,width=83mm,clip]{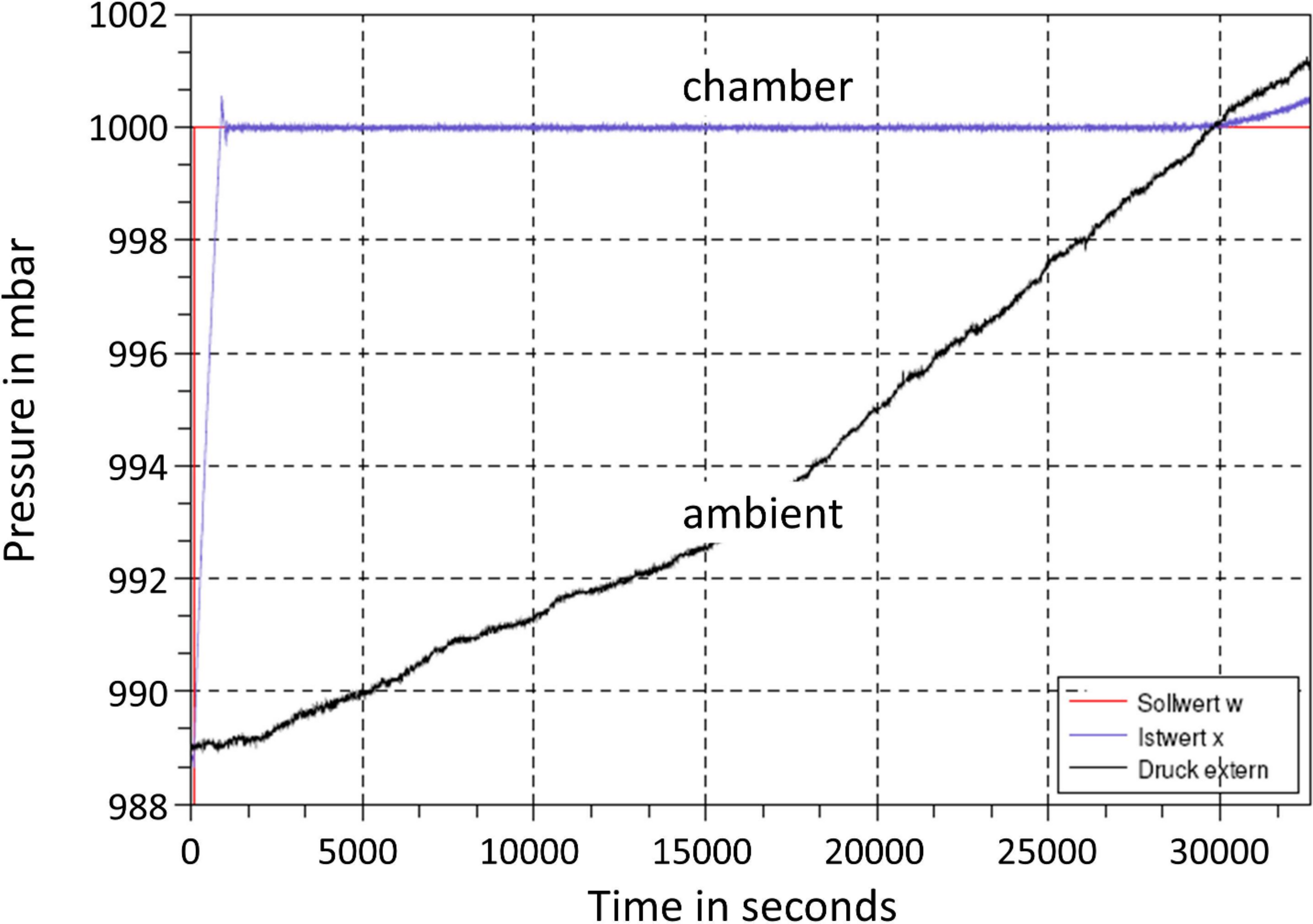}
\caption{Test results from the environmental stability system. \emph{a}) The temperature on the optical table for a 14-day test run. The rms is $\approx$\,1\,mK. Notice the day-night variations with a full amplitude of 3\,mK and a trend of about the same order. \emph{b}) Pressure test in the lab in Potsdam during an unstable weather period. The time axis ranges 9 hours. Outside barometric pressure varied from below 990\,mbar to over 1000\,mbar while the inside pressure remained constant to within 0.01\,mbar (the sensor  precision limit). Notice that the regulation limit was set to 1000\,mbar. }\label{F-pT}
\end{figure}

The spectrograph enclosure was already described in some detail in Strassmeier et al. (\cite{str:spie}). It is located in the observatory basement within the telescope pier on solid rock and surrounded by 1m-thick concrete walls. The ambient conditions at this location in the building are already quite stable. Nevertheless, barometric pressure may change by 25\,mbar p-v between summer and winter (from 675 to 700\,mbar). The air temperature changes by up to 8\,$\degr$C between January and August (from 12 to 20\,$\degr$C) and the relative humidity covers the range between 8\,\%\ and 52\,\%\ for the same period of time. The entire optical bench with its six support legs is enclosed by a passive thermal cover (see Fig.~\ref{F-chamber}). The cover itself is located in an actively-controlled pressure and temperature stabilized chamber. The chamber is a nested double structure in itself where no heat bridges between the inner and the outer structure exist. Each structure is made of a steel frame with a 10\,cm Jakodur insulation with a small air gap in between and an outer aluminum-sheet cover.  Areal heating foils are placed on the inside of all outer panels within the air gap and cover all walls, the floor, and the ceiling.

A total of 16 Pt-1000 sensors distributed within the chamber are read via a Linux PC with integrated PCI cards. A power transformer with an electric energy of 2\,kW is used to heat the foils. Solid-state relays are employed to distribute the power among four heating circuits according to the Pt-1000 signals. The control concept is to keep the air temperature as equal and as constant as possible to avoid vertical and horizontal temperature gradients in the air within the chamber (Materne 2005, Diploma thesis, Univ. Potsdam). The barometric pressure is measured with a high-precision GE~RPT410 sensor. The absolute accuracy of this sensor is given to 0.1\,mbar while its relative rms over time is found to be better than 0.01\,mbar. The control circuit is a closed loop and involves a dry air pump with an air-filter system that compensates the natural air losses from the chamber. Figure~\ref{F-pT} shows the results from two test runs of the chamber. The temperature on the optical bench is stable to within a few thousands of a degree and the pressure to better than 0.01\,mbar.

\subsection{Chamber interfaces}

\begin{figure}
{\it a)}

\includegraphics[angle=0,width=83mm,clip]{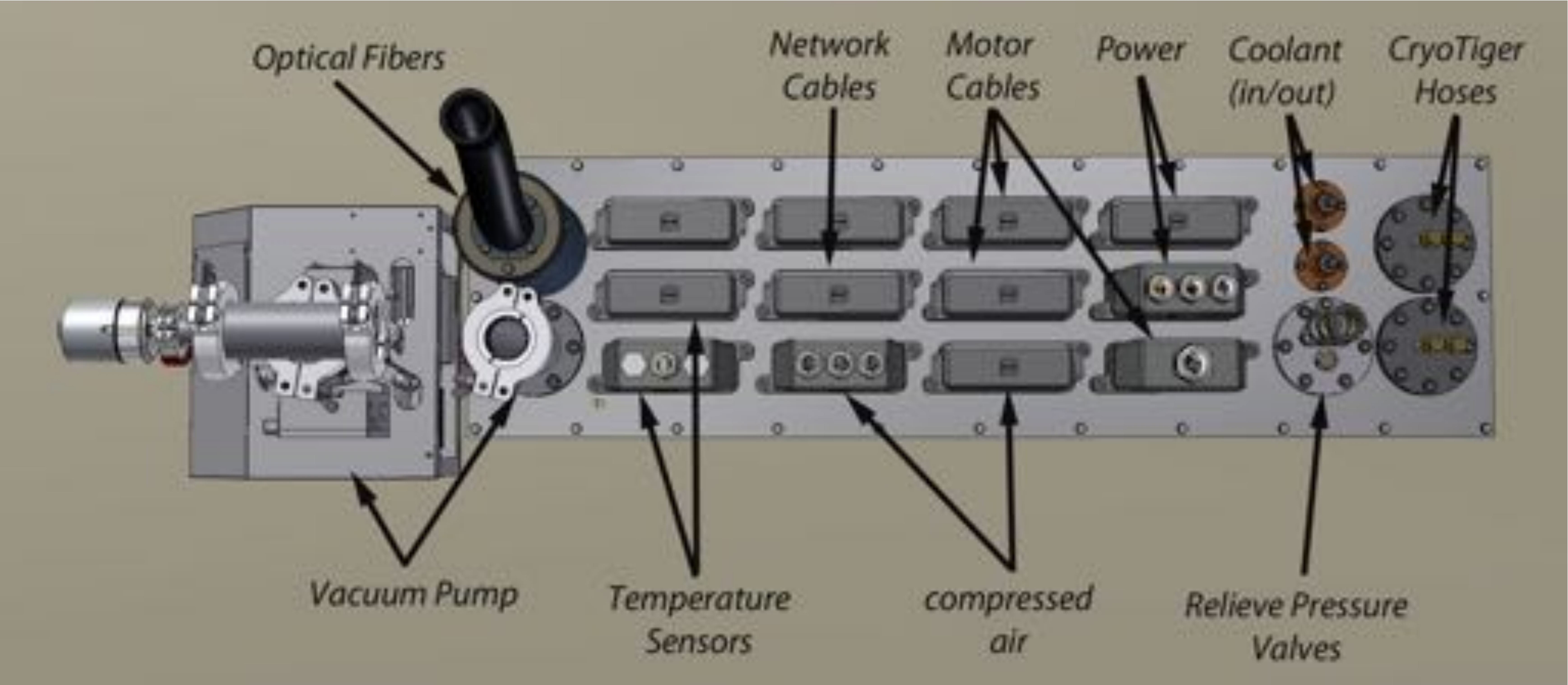}\\
{\it b)}

\includegraphics[angle=0,width=83mm,clip]{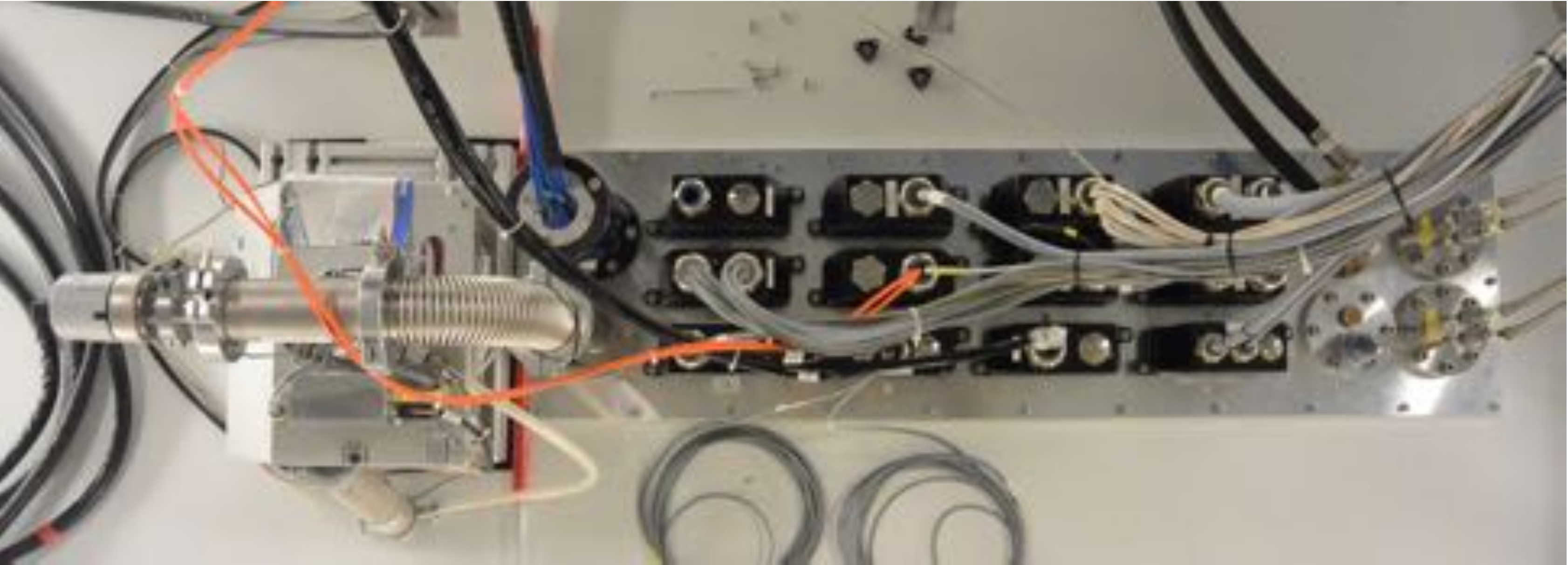}
\caption{The chamber interface to the electronic rack. \emph{a}) Schematic design. \emph{b}) Real unit. Media in/outlets include UPS power, direct fibres from the two CCD controllers, 32 ethernet fibres from the electronic racks, two water hoses from the cooling unit, four pressured air hoses from the clean-air system, and four coolant lines from the CryoTigers for CCD cooling. Additionally, a total of 26 science-light fibres enter the chamber. }\label{F-interfaces}
\end{figure}

The chamber itself can be entered through a pressure-save air-lock door and a preceding class-8 clean room\footnote{ISO standard 8 or class 100\,000.}. All media interfaces are located on the roof of the chamber on an interface plate and enter directly into the spectrograph cover through a supply and distribution box. Media in/outlets include 3\,kW 110V/60Hz UPS power, 2+2 fibre pairs for the two CCD controllers, 2$\times$16 ethernet fibres from the e-racks on the support bridge, two chilled water hoses from the cooling unit, four pressured air hoses for the optical bench suspension system and the air-pressure balance within the chamber, and four coolant lines from the two remotely-located CryoTigers for CCD cooling. Figure~\ref{F-interfaces} shows a picture of the interface plate.

%--------------------------- F  chamber interfaces

Additionally, a total of 26 fibres for science purposes enter the chamber through the same supply box as the other media. A special pressure-sealed connector was developed in house because fibre FRD is sensitive to any sort of external forces onto the fibres. The 3\,mm PVC jackets are glued into a pre-shaped 26-slot connector mask which then acts like a single fibre bundle on the inside of the chamber. It includes all pre-confectioned spare fibres as well. The fibres themselves remain loose within the Kevlar fabric inside the PVC jacket but the jacket exit regions, $\approx$10\,cm before the image slicer, are closed with a custom-made ferrule.

\subsection{Safety services}

The chamber door is a doubly interlocked safety mechanism that even if one opens the door during overpressure, the door remains locked and cannot push against a human in front of it. Once the pressure is equalized the second lock can be opened. The inside air pressure is measured continuously to very high precision and reported to the main control server where it is recorded and graphically displayed on the main control GUI of the observer. An OXY3690-MP oxygen sensor monitors the breathing air inside the chamber and its result is logged and additionally displayed at the chamber entrance.

% ------------------------------------------------------------------------------------------------------------------------------
\section{Calibration units}\label{S-Calib}

\subsection{Standard calibration}

\begin{figure}%[tbh]
\includegraphics[angle=0,width=83mm, clip]{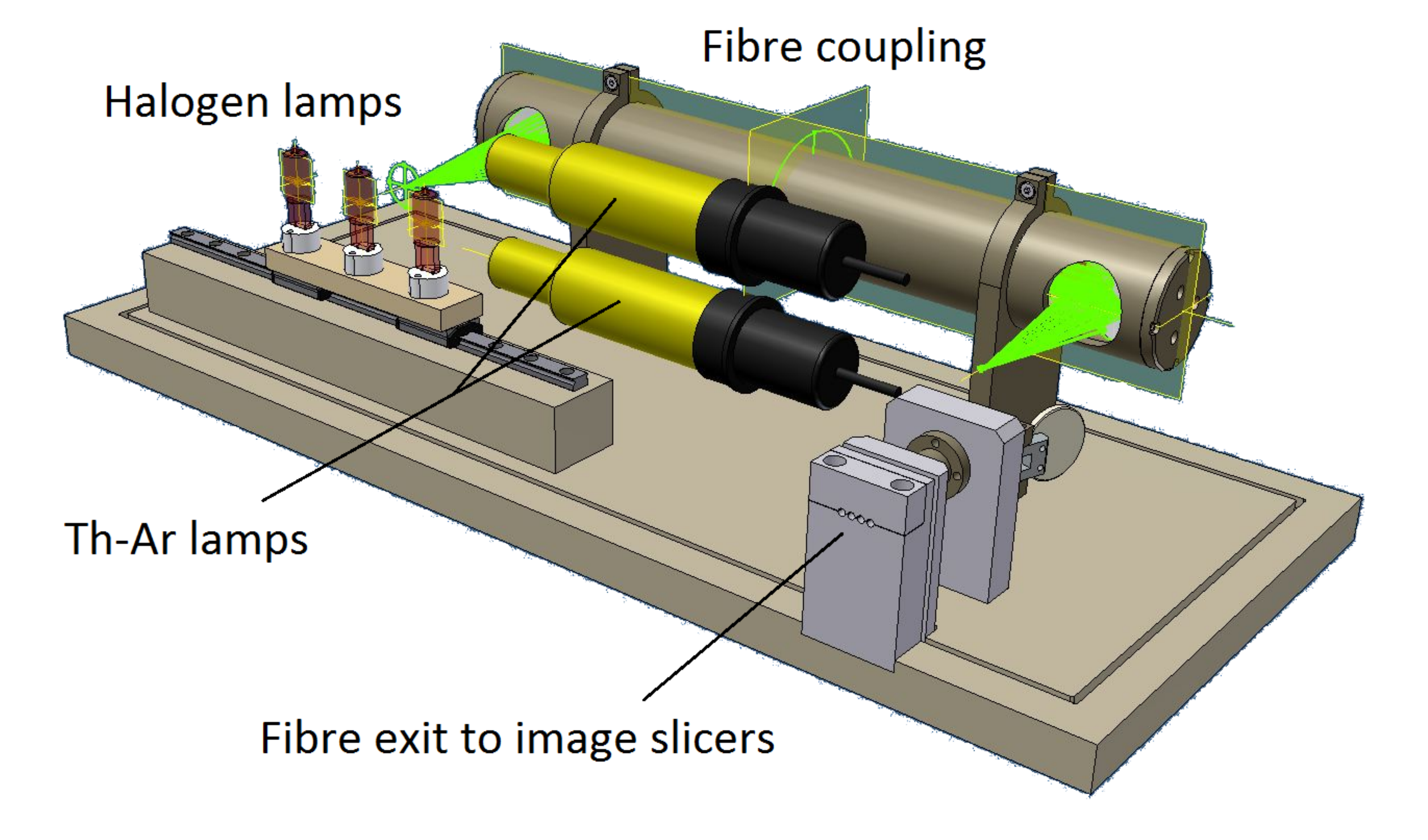}
\caption{Layout of the standard calibration unit. It provides either halogen light for flat fielding or Th-Ar light for wavelength calibration. Three 100\,W halogen lamps and two Th-Ar lamps are available. Fibres bring the chosen light to either the PFUs or the polarimeters and from there down the science fibres to the image slicers. }\label{F-calunit}
\end{figure}

The standard wavelength calibration source is a hollow-cathode Th-Ar lamp which provides an appreciable number of lines in the visible part of the spectrum. Its accuracy is limited by a number of factors such as line blending, uncertainty in the wavelength of the Th and Ar lines, poor wavelength coverage in the red, and uneven distribution of the line intensities (e.g. Lovis et al. \cite{lov:pep}). Moreover, at $R$~=~270\,000, the Th-Ar lines are mostly resolved and do not provide a unique and accurate reference wavelengths anymore although the principal stability  of the Th-Ar calibration can reach a level of 1\,\ms\ (Pepe 2008). However, many very strong lines in the red wavelength region are quickly saturated and prevent an even exposure level for the weaker lines. Both sources, Th-Ar and halogen, are located within the same unit (Fig.~\ref{F-calunit}).

\subsection{Fibre switches}

All four PEPSI foci receive calibration light via fibres. Each PFU has three dedicated 300\,$\mu$m fibres for its target-fibre calibration through an integration sphere and for its simultaneous wavelength calibration through a Fabry-P\'erot etalon. Additionally, the path for the VATT-light injection has its own calibration. The two polarimeter foci receive light through another pair of 300\,$\mu$m fibres.

%------------------------------ F     standard cal unit

A total of four fibre-switch units are required for the logistics. Their aim is to, firstly, bring the calibration light to the telescope foci and, secondly, inject it into the sky fibers (both for SX and DX). The four units are located within the e-racks on the PEPSI support bridge (see Fig.~\ref{F-fibrelogistics}). Each unit is made of a precision linear stage with an integrated DC motor and precision encoder and has four fiber ends in a fixed position (three sky fibres and one calibration-light fibre) and three fiber ends in a movable position (just the sky fibres). All fibres are terminated with a standard SMA plug.

\subsection{Light sources}

Two comparable standard calibration units exist (Fig.~\ref{F-calunit}), and can be internally exchanged if needed. One unit serves all standard telescope foci while the other serves just the Fabry-P\'erot unit on the optical bench.  Each unit contains three halogen lamps (two hot spares) and two Th-Ar lamps (one hot spare). We employ flat-field halogen lamps with four fibre exits and Th-Ar lamps with single fibre exits. The halogen sources are tunable 100\,W lamps from LOT-Oriel. The Th-Ar lamps are from Photron and Juniper. Cooling of the entire unit is by forced convection within the housing and by passive conduction of the housing to a heat exchanger.

Note that the four halogen-lamp fibres serve flat-field light from the same source to the two pairs of active target and sky fibres when the PFUs are in use. Similarly, they feed flat-field light to the two ordinary- and the two extraordinary-beam fibres when the polarimeters are in use. The same is the case for the standard Th-Ar calibration light.

%------------------------------ F      Fabry-Perot fringes
\begin{figure}
{\it a)}

\includegraphics[angle=0,width=83mm, clip]{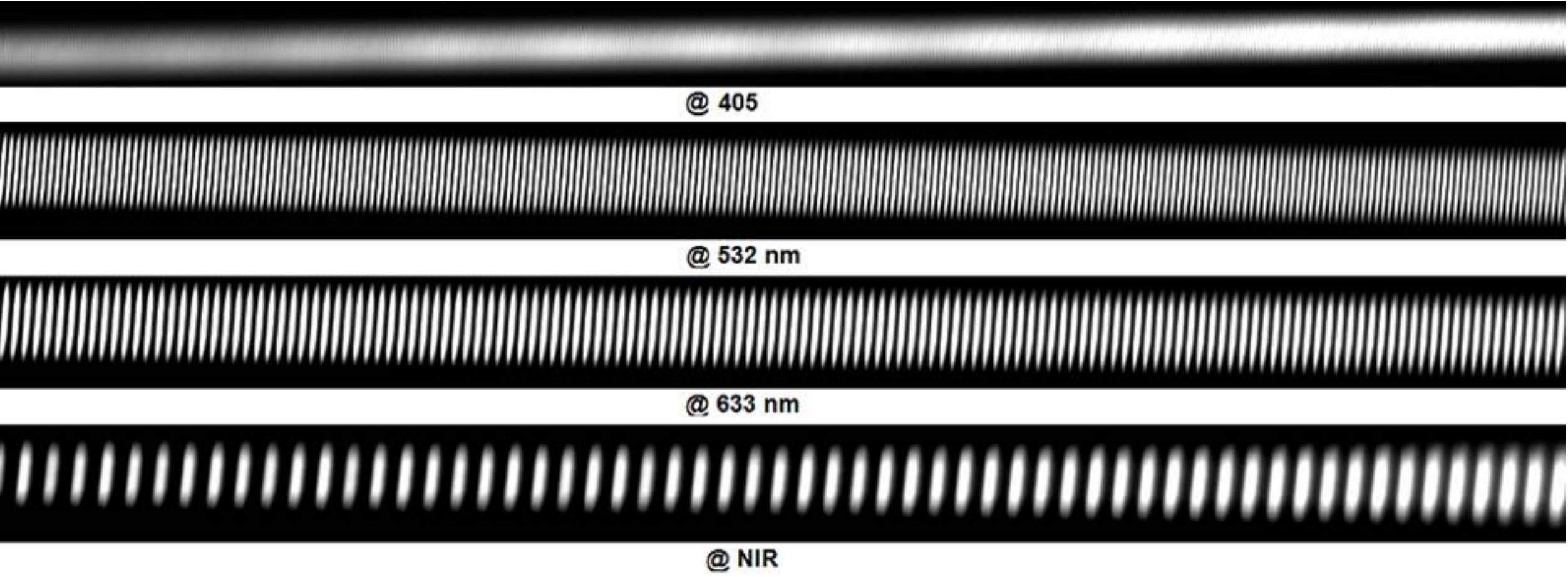}\\
{\it b)}

\includegraphics[angle=0,width=83mm, clip]{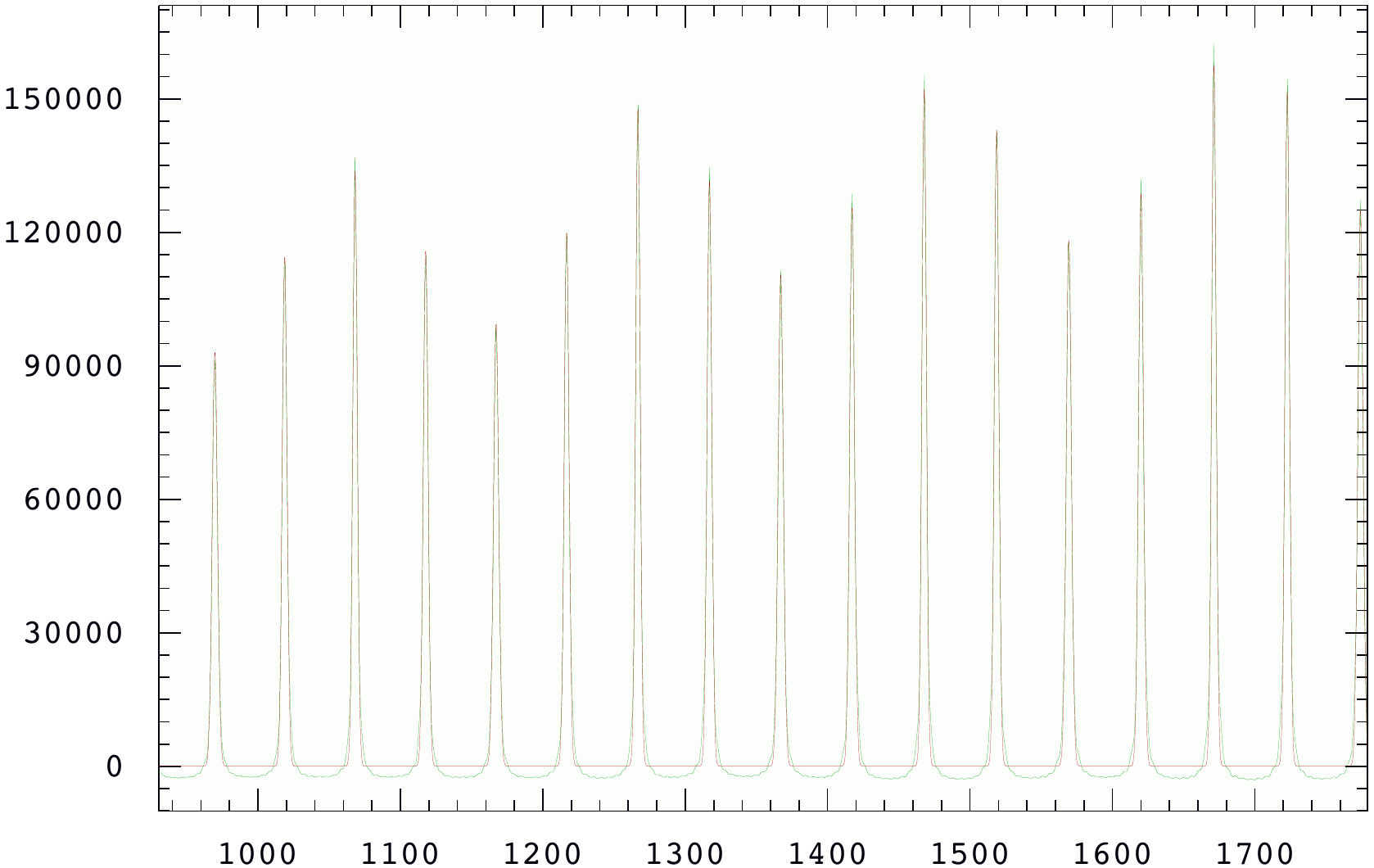}
\caption{Example of a set of Fabry-P\'erot fringes. \emph{a}) Two-dimß-ens\-ional fringe distribution for different wavelengths in low-res\-olution mode. \emph{b}) Extracted one-dimensional pattern (raw image) in ultra-high-resolution mode around 700\,nm. Only a small fraction of the 100 lines per free spectral range of one \'echelle order is shown. The $x$-axis is in pixels, the $y$-axis in ADUs. }\label{F-FPIfringes}
\vspace{-2.5mm}
\end{figure}

\subsection{PFU internal calibration sources}

A retractable integration sphere (``Ulbricht-Kugel'') per PFU allows coupling diffuse light into the target and the sky fibres. The sphere is either fed by its own Th-Ar lamp or by halogen light from the standard calibration unit via a \hbox{300-$\mu$m} fibre. The LOT-Oriel model is 50\,mm of inner diameter enhanced with a BaSO$_4$ coating and has a 14\,mm exit opening. Its reflectivity is 97.5\,\%\ across the 383--907\,nm range.

\subsection{Fabry-P\'erot wavelength calibration unit}

%------------------------------ F      Fabry-Perot optical design
\begin{figure}
{\it a)}

\includegraphics[angle=0,width=83mm, clip]{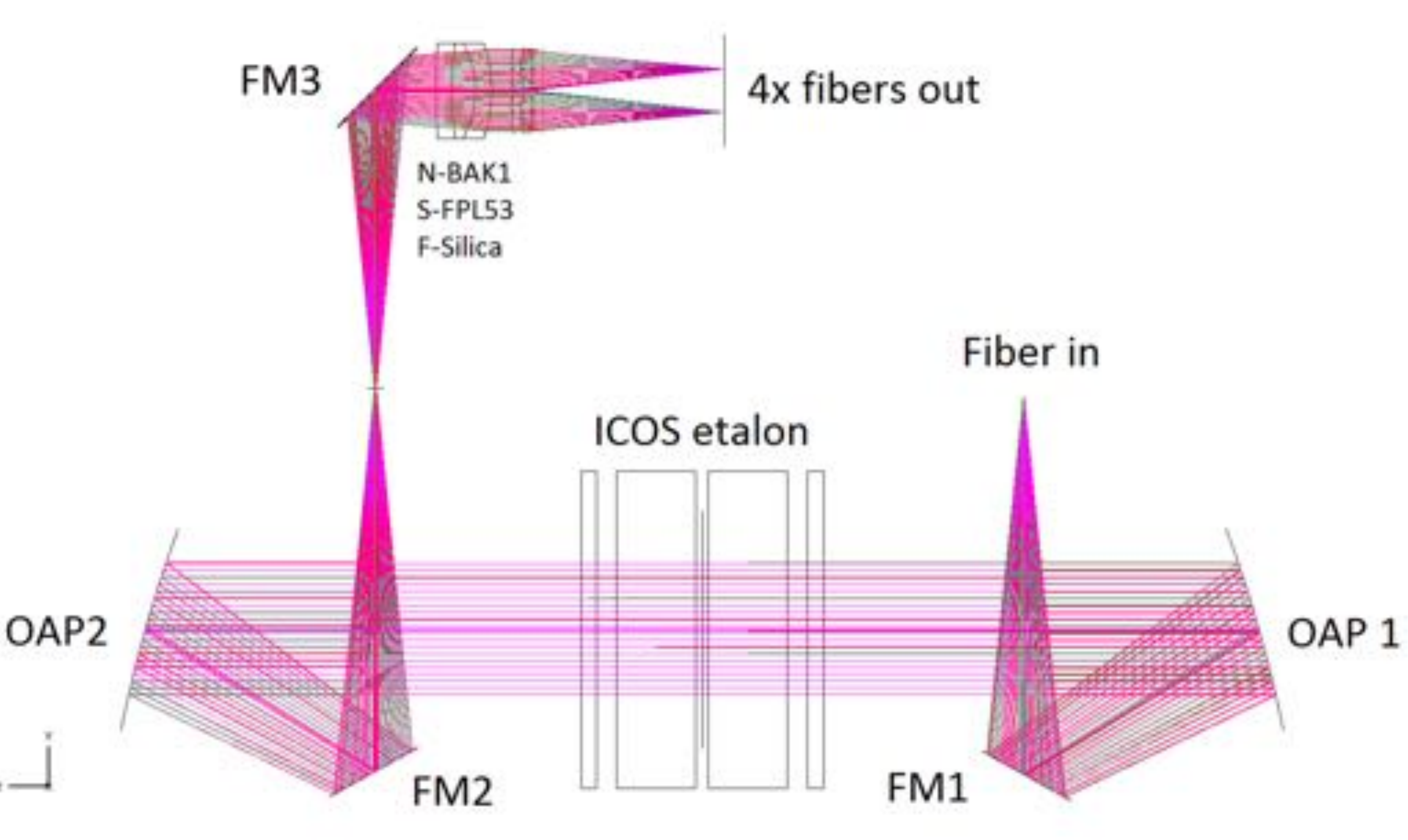}\\
{\it b)}

\includegraphics[angle=0,width=83mm, clip]{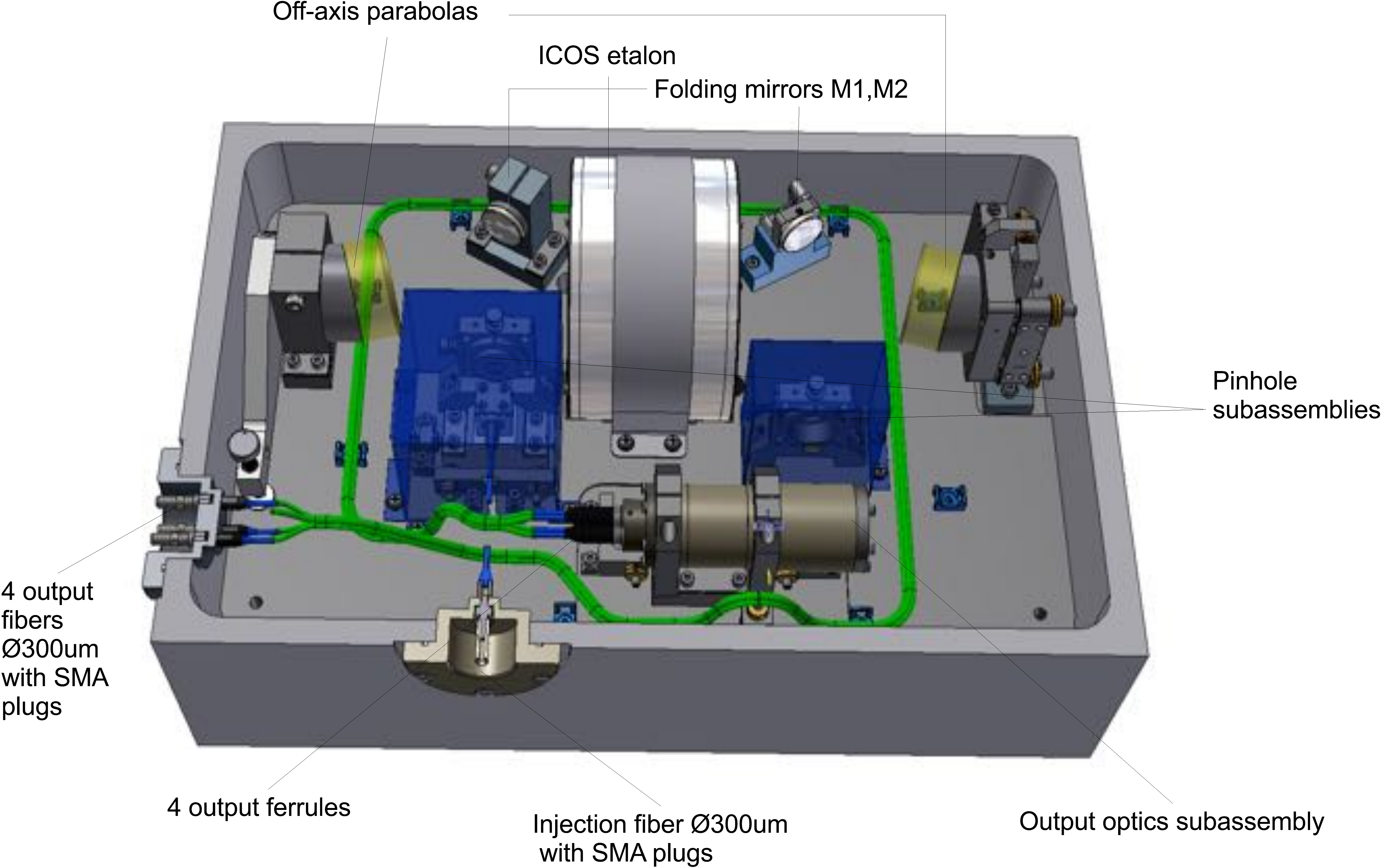}
\caption{The Fabry-P\'erot calibration unit. \emph{a}) Optical design. \emph{b}) Mechanical design. A single 300$\mu$m input fiber with an f/5 exit is converted to a parallel beam of diameter 30\,mm via an off-axis parabola (OAP1). It passes through the etalon and is then re-imaged with another off-axis parabola (OAP2) and split into four exit beams at f/10. These exit beams are picked up by four fibres, one fibre for each active image slicer, and connected to the central calibration unit.}\label{F-FPIdesign}
\end{figure}

In order to obtain high accuracy for radial velocity measurements, a very stable environment inside the spectrograph is required. The remaining changes of the air pressure and temperature within our chamber (see Sect.~\ref{S-Environ}) would still change the optical-path length in the spectrograph with a radial velocity shift of as high as 10\,\ms. A number of other factors also contribute to the long and short time scale instabilities, among are temperature variations of the CCD surface, guiding and centering errors of the star on the fiber, and modal noise within the fibers. To overcome most of these problems, a user mode is provided where a wavelength calibration spectrum is taken simultaneously with the target exposure. 

Such a mode is used in a number of spectrographs, most successfully in HARPS (Pepe et al. \cite{harps}). This allows to trace temporal variations in the spectrograph during target exposure, except due to guiding errors or any other factors specific to the target itself and the behavior of the calibration fibers. The main problem of a Th-Ar lamp for this simultaneous calibration mode is the large range of its emission-line strengths together with the small number of lines in the red wavelength region (where the brightest Ar lines are situated). The bright Ar lines create a strong and non-uniform scattering effect in the neighboring spectral orders occupied by the comparably weak target spectra. Therefore, we use the fringes from a sealed Fabry-P\'erot (FP) \'etalon instead.

Figure~\ref{F-FPIfringes}a shows the output of the PEPSI FP unit recorded at four different wavelength regions with a laboratory Czerny-Turner spectrograph\footnote{Schanne \& Sablowski;\\  http://spektroskopie.fg-vds.de/pdf/SSS\_manual.pdf}. Note that the spacing between individual fringes is constant but the plots have different wavelength coverages due to the wavelength-dependent dispersion of a Czerny-Turner spectrograph. Figure~\ref{F-FPIfringes}b shows a trace of the fringe profiles recorded with PEPSI in UHR mode and with CD-V for a 1800-pix section ($\approx$1.3\,nm) of an \'echelle order near 700\,nm. The slightly uneven amplitudes are partly due to fringing in the CCD coating and partly due to the pixel-to-pixel variations of the CCD (no flat-field division nor bias subtraction was done).

Figure~\ref{F-FPIdesign} shows the optical design of the full PEPSI FP unit. The unit is located on the optical bench of the spectrograph in its own stabilized vessel. A 12\,m 300\,$\mu$m fibre from the calibration rack provides halogen light, as well as Th-Ar light if desired, through the chamber and into the vessel. The etalon is a fixed-(air)-gap design from ICOS Optical Systems, Ltd. and is mounted in a sealed cell. Its cavity plates are made of fused silica with a clear aperture of 70\,mm (plate diameter of 88\,mm) and a flatness at 633\,nm of $\lambda$/180 ($\lambda$/150 after coating). The parallelism is $\lambda$/50. The spacer material is Zerodur with a length of 3.060\,mm resulting in a substrate wedge-angle uncertainty of 0$\pm$1 fringes at a tilt of 15\arcmin . The plate coating is a thin layer of silver enhanced with a layer of silica. Its reflectivity is measured to be between 72\,\%\ (at 400\,nm) and 93\,\%\ (at 900\,nm), its transmission between 18\,\%\ (400\,nm) and 4\,\%\ (900\,nm). The entrance surface is additionally AR coated. This coating was made as a single layer of MgF$_2$. This simple coating was chosen to match the thickness and internal stresses of the enhanced silver front surface coating.

The total finesse estimation is based upon the assumption that each instrumental defect has a Gaussian profile and
contributes to the total fringe width in a sum of squares. The total finesse for the high-resolution mode is mostly limited by the parallelism finesse and is in the range 8--17 for 400--900\,nm. The total finesse for the low-resolution and the medium-resolution mode is dominated by the spectrograph finesse and is in the range 5--12 and 2--4, respectively, always for the wavelength range 400--900\,nm. The thermal expansion of the Zerodur spacer causes a physical limit for the radial velocity stability of the transmitted peaks of 3~\cms/0.01\,K.

%------------------------------------------------------------------------------------------------------------------------------
\section{CCD detector system}\label{S-CCD}

\subsection{STA1600LN 10.3k$\times$10.3k CCDs}

%--------------------------- F  CCD
\begin{figure}
{\it a)}

\includegraphics[angle=0,width=83mm,clip]{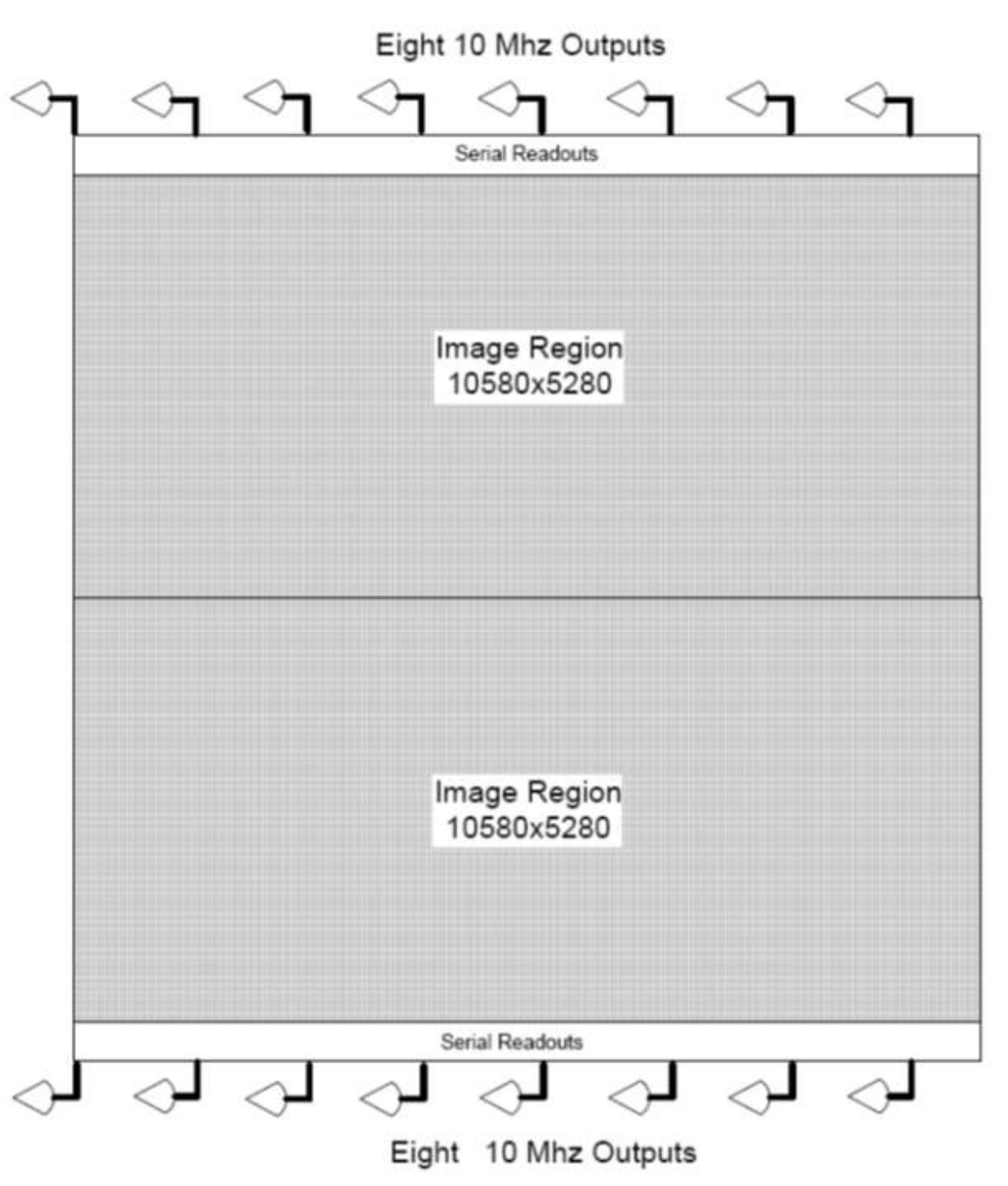}\\
{\it b)}

\includegraphics[angle=0,width=83mm,clip]{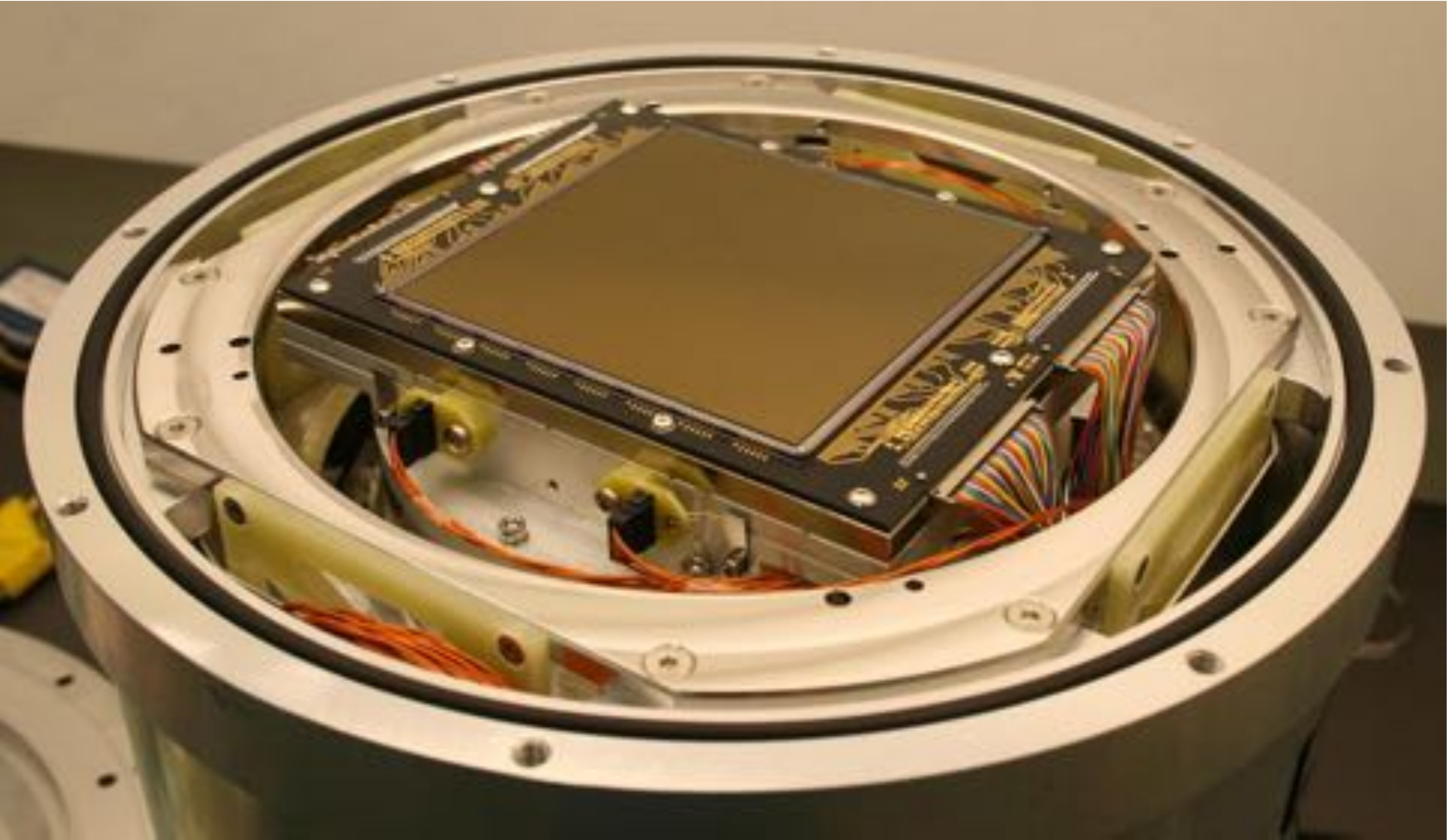}
\caption{PEPSI 10k CCDs. \emph{a}) STA1600 design of the photo-sensitive area. The full frame is divided into two halves, each with 10560 horizontal by 5280 vertical pixel and eight read-out amplifiers. \emph{b}) The PEPSI BLUE CCD during integration into the dewar.}\label{F-ccd}
\end{figure}

PEPSI runs one low-noise (LN) STA1600 Charge Coupled Device (CCD) per arm, dubbed PEPSI BLUE and PEPSI RED (Fig.~\ref{F-ccd}). The STA1600 device (Bredthauer et al. \cite{sta}) is a thinned, backside illuminated 10560$\times$10560-pixel solid-state CCD with a transparent polycrystalline silicon gate structure for creating electron hole pairs. Its full frame is organized in two halves each containing an array of 10560 horizontal by 5280 vertical photosites. Horizontal transport registers along the top and bottom permit simultaneous readout of both halves. The pixel spacing is 9$\mu$m$\times$9$\mu$m. For dark reference, each readout line is preceded by 8 dark pixels. This architecture provides video information as a single sequential readout of 5280 lines containing 1330 photosites. One output amplifier is located at the end of each horizontal register, thus, each device has in total 16 amplifiers. They are dual FET floating diffusion amplifiers with a reset MOSFET tied to the input gate. All CCDs were backside processed and characterized at The University of Arizona Imaging Technology Laboratory (ITL; Lesser \cite{itl}).

\begin{figure}
\center
\includegraphics[angle=0,width=83mm, clip]{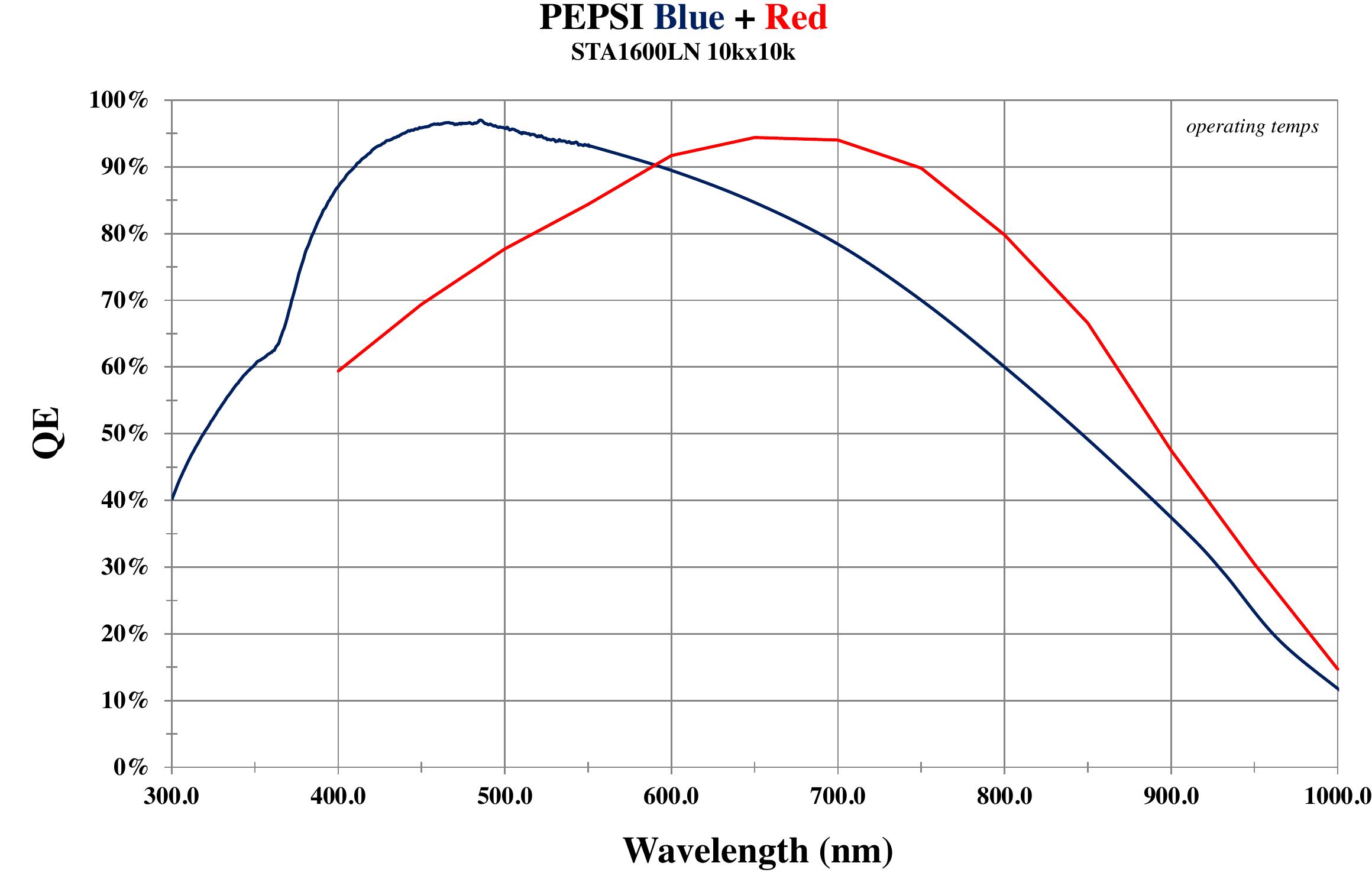}
\caption{Quantum efficiency (QE) curves for PEPSI BLUE (blue/darker line) and PEPSI RED (red/brighter line). The BLUE CCD sees wavelengths between 383 and 542\,nm, the RED CCD between 537 and 907\,nm. }\label{F-ccdqe}
\end{figure}

\begin{figure}
{\it a)} Flat field (BLUE) \hspace{15mm} {\it b)} Dark (BLUE)

\includegraphics[angle=0,width=40mm, clip]{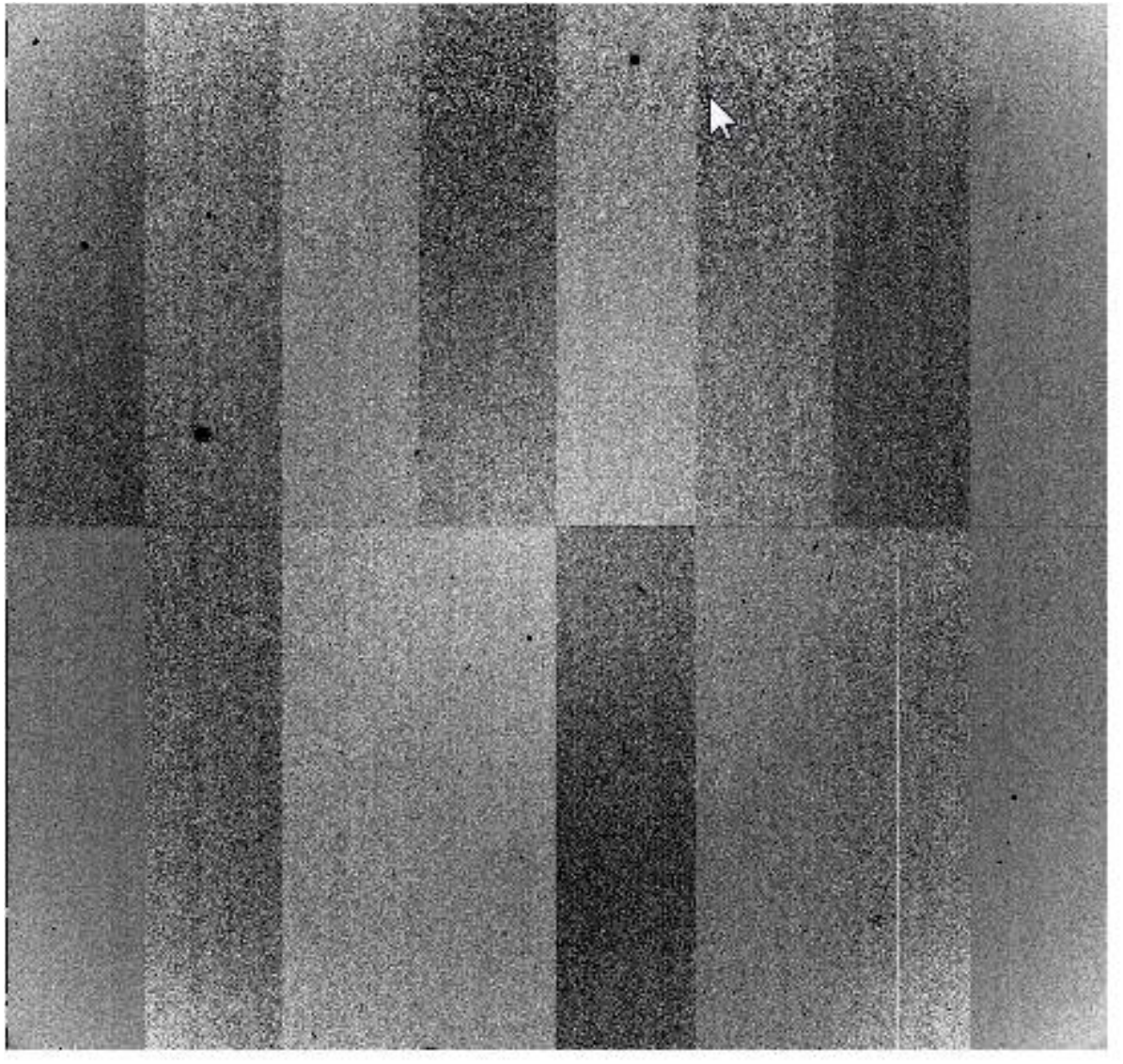}\hspace{1mm}
\includegraphics[angle=0,width=40mm, clip]{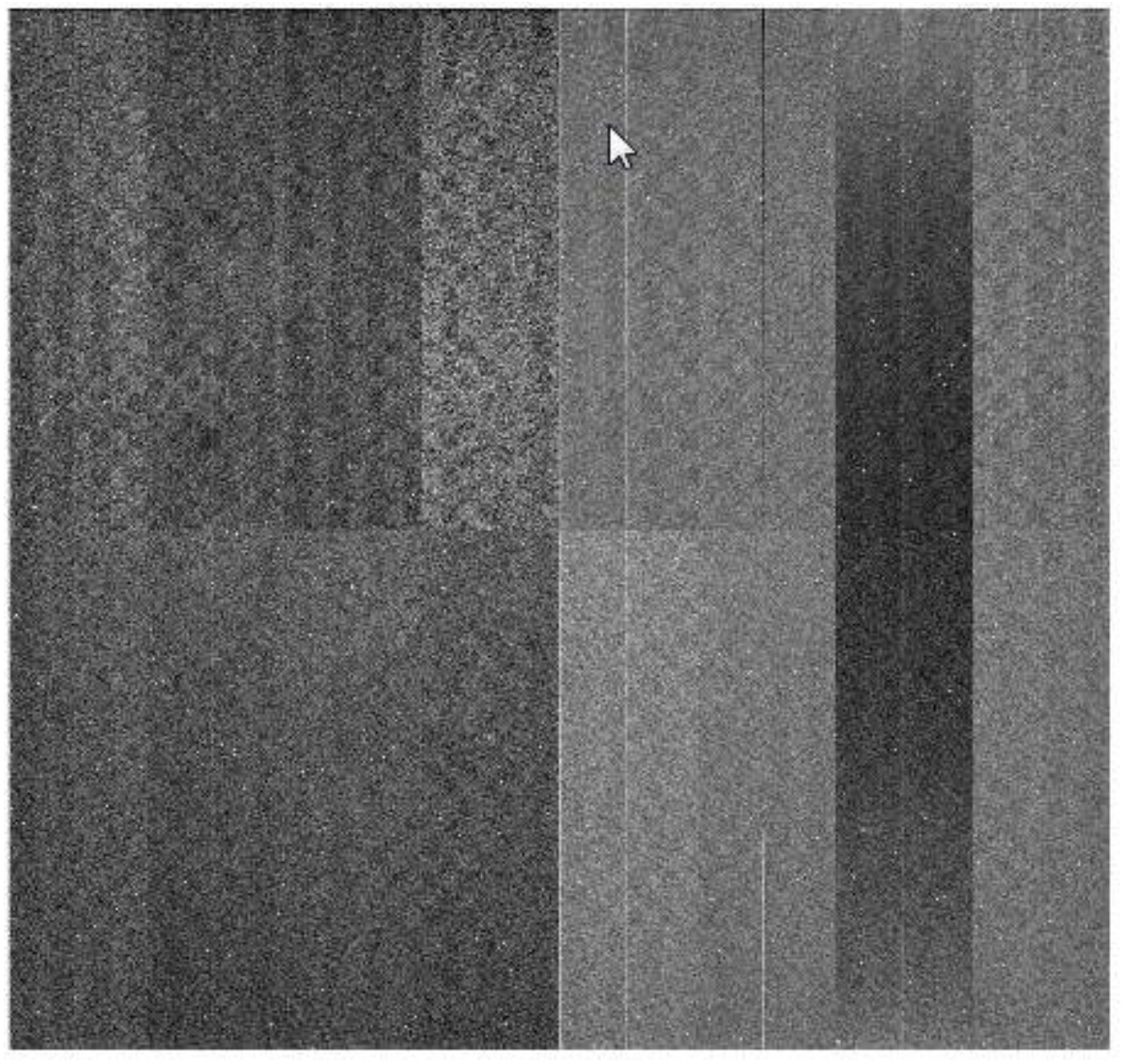}\\
{\it c)} Flat field (RED) \hspace{16mm} {\it d)} Dark (RED)

\includegraphics[angle=0,width=40mm, clip]{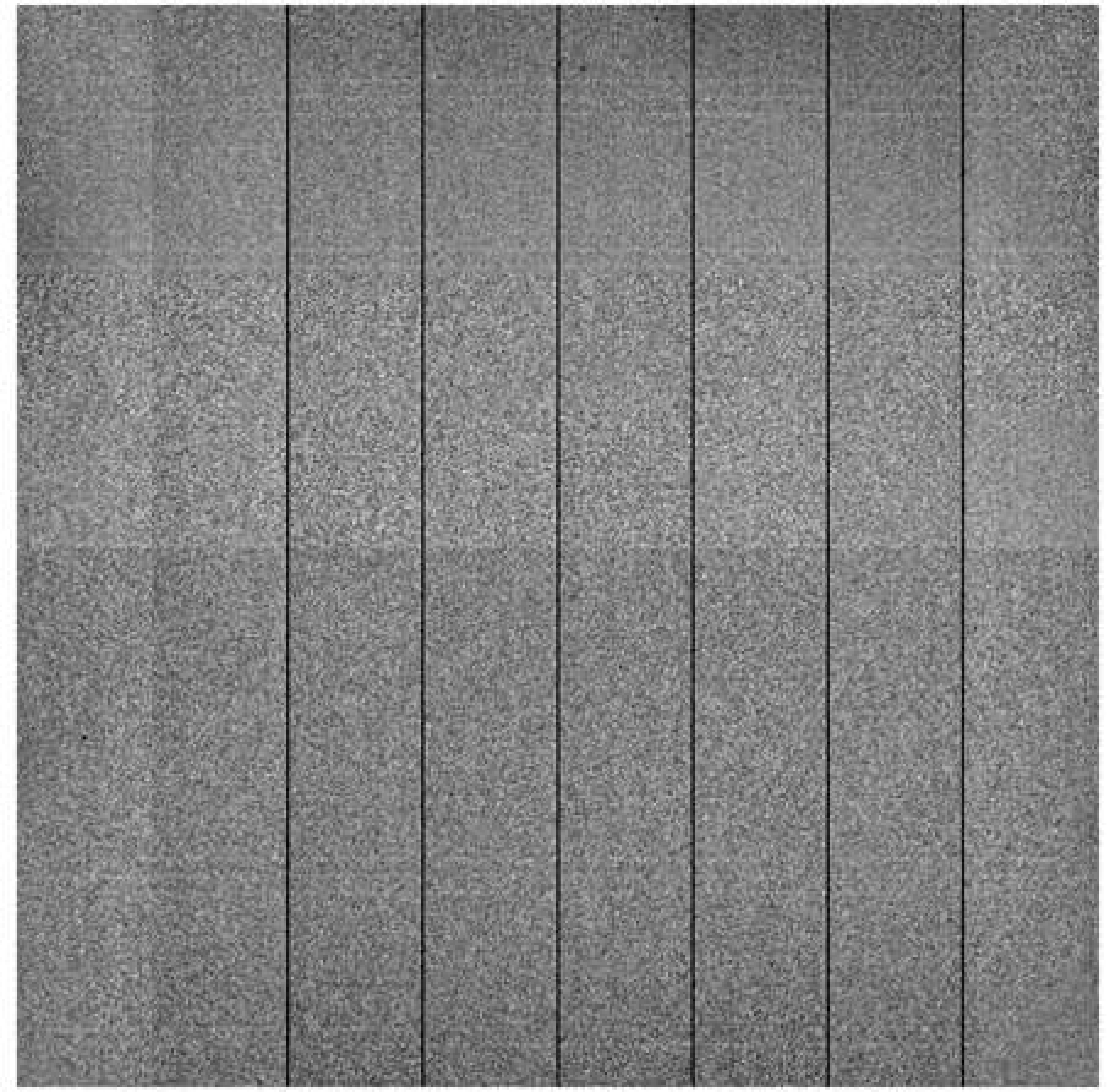}\hspace{1mm}
\includegraphics[angle=0,width=40mm, clip]{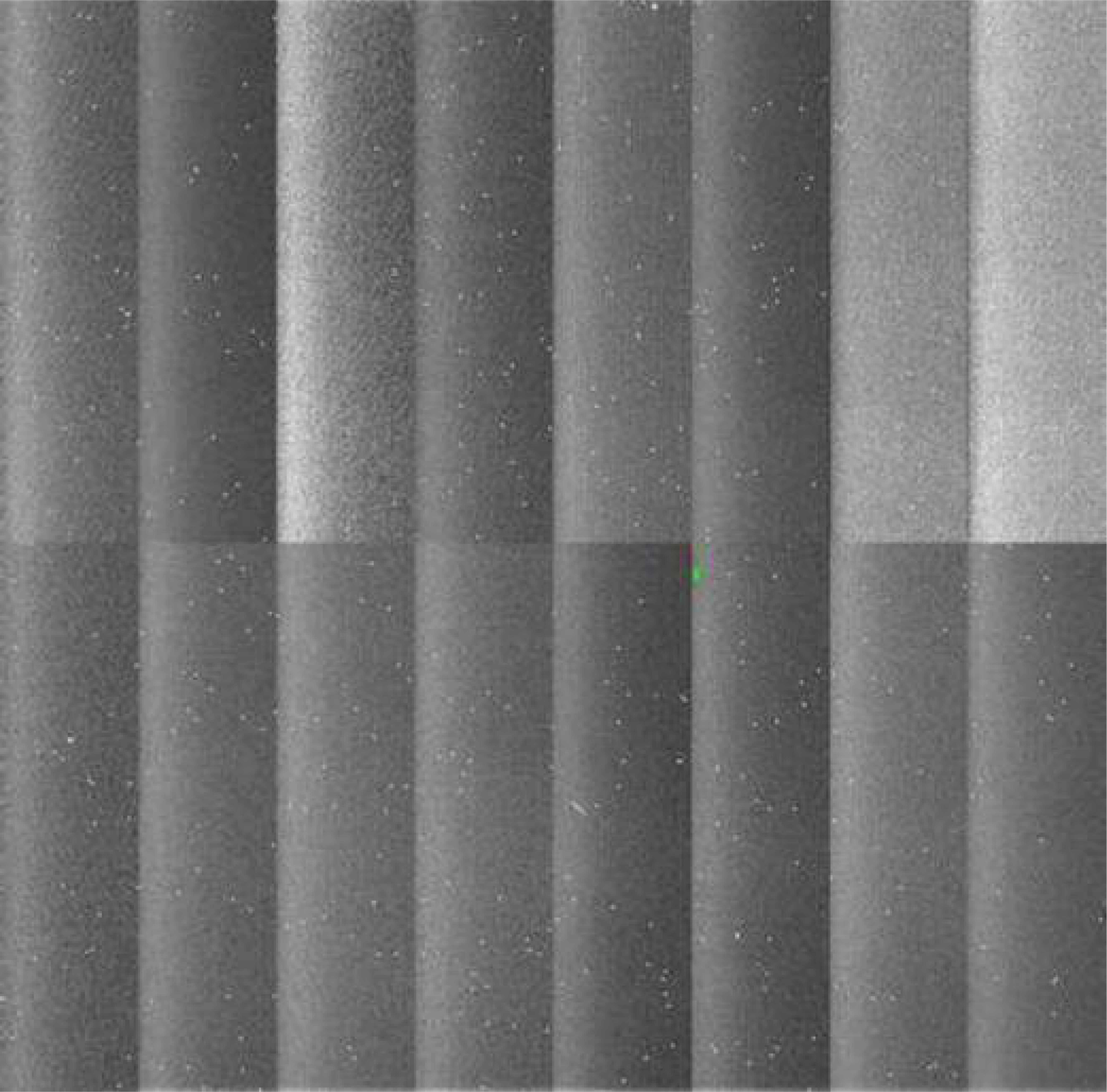}
\caption{Example flat fields and dark exposures. \emph{a}) 400-nm flat field for the BLUE CCD. \emph{b}) 10-min dark exposure for BLUE with amplifiers off; average dark read out is 1.1\,e$^-$/pix/hour. \emph{c}) 800-nm flat field for the RED CCD. \emph{d}) 10-min dark exposure for RED with amplifiers off; average dark read out is $<1.8$\ e$^-$/pix/hour.
}\label{F-ff}
\end{figure}

\vspace*{1mm}
\noindent
\emph{PEPSI BLUE.} (SN16536).  The total system gain and noise (unbinned) ranges between 4.4\,e$^-$ (gain 0.47~e$^-$/DN) and 5.9\,e$^-$ (gain 1.3). The median noise in the high-gain mode is 4.6\,e$^-$ and was  measured in the AIP dewar with the ARC-48D controller. The controller and internal wiring harness noise contributes about one to two electrons of noise. Total read-out time for optimal S/N is 60\,s through all 16 channels (at 125~kpix/s per amplifier). The fast read-out-mode (with the gain set to 1.3) takes 37\,s at 200~kpix/s per amplifier. Charge Transfer Efficiency (CTE) was measured with a Fe55 X-ray source in an ITL dewar and found to be excellent in all channels. Flat field is also excellent (Fig.~\ref{F-ff}). Dark current is measured to be 1.1~e$^-$ per pixel per hour at $-122\,\degr$C. No significant amplifier glows are seen when the OD voltages are turned off during integration. Full well is set by the 16-bit ADC limit to about 82\,000 electrons. There are a few dark and bright columns. One bright column is very strong at room temperature but mostly disappears below $-40\,\degr$C. The device is AR coated for optimal blue response and its peak QE was measured 96\,\%\ between 450--500\,nm (see Fig.~\ref{F-ccdqe}). Just recently, Lesser (\cite{itl_qe}) discussed the measurement procedure for absolute QE measurement and concluded that the measurements quoted here are better than 3\,\%\ accuracy over the 300--1100\,nm spectral region.

%------------------------------ F CCD QE

%\vspace*{1mm}
\noindent
\emph{PEPSI BLUE spare.} (SN13642). The device is our backup CCD and has a region of low QE mainly on the left side (image section 1). We believe this is related to dewar contamination while the device was installed in a previous system. Only 8 of the 16 amplifiers can be used for readout due to a serial short on one side of the device. Total read time for optimal S/N is then about 90 s through the 8 channels. The total system gain and noise (unbinned) ranges between 2.75\,e$^-$ (gain 0.898) and 3.80\,e$^-$ (gain 0.876). CTE was found to be excellent in all channels. Dark current is measured to be less than 3.0~e$^-$ per pixel per hour at $-122\,\degr$C. Full well is about 60\,000 electrons. The device is AR coated for optimal blue response and its peak QE was measured 93\,\%\ at 460\,nm (see again Lesser \cite{itl_qe} for a description of QE issues).

\noindent
\emph{PEPSI RED.} (SN15506). Flat field uniformity of this device is very good (Fig.~\ref{F-ff}). There are a few dark spots which are darker for bluer wavelengths. The internal CCD-bus structures can be seen in the redder images. All 16 channels are operable. The total system noise (unbinned) for the entire device is measured in the AIP dewar. Channel~1 always appears noisiest with 7.0\,e$^-$. The other channels range between 4.0\,e$^-$ (gain 1.475) and 5.9\,e$^-$ (gain 1.434). Again, the controller and internal wiring harness noise is included here and contributes about one electron of noise. Total read-out time for optimal S/N is 56\,s through all 16 channels (at 125~kpix/s per amplifier). The fast read-out-mode (with the gain set to 1.3) takes 37\,s at 200~kpix/s per amplifier. Horizontal and vertical CTE was measured with a Fe55 X-ray source in the ITL Bud dewar and found to be 1.0 as accurately as can be measured in all channels. Dark current is measured to be 1.8~e$^-$\,per\,pixel\,per hour at $-121\degr$C. We note a higher than expected background of charge events. They are not fixed from image to image and so are not hot pixels. Full well is about 85\,000 electrons. The device is AR coated for optimal red response and its peak QE was measured 94\,\%\ between 650--700\,nm (see Fig.~\ref{F-ccdqe}) but drops off to 47\,\%\ at the cutting wavelength of the spectrograph at 907\,nm.

%------------------------------ F flats and darks

\vspace*{1mm}
\noindent
\emph{PEPSI RED spare.} (SN18043).  Flat field uniformity is good in the red where the AR coating has been optimized. However, uniformity is not very good below 500\,nm due to features on the surface which appeared during backside etching. There are a significant number of bad columns (faintly bright or dark) in the device making it of lower grade than the PEPSI RED and BLUE CCDs. The electrical performance is about the same as for the other devices. All 16 channels are operable. CTE appears excellent from cosmic ray analysis. Dark current is ${\approx\!\rm 2~e^-}$\,per\,pixel\,per hour at the operating temperature of $-90\,\degr$C during testing. The median total system noise is about 4.5\,e$^-$ rms (range 3.7--4.9) with gain settings ranging between 0.48--0.60~e$^-$/DN at 135\,kHz read out.  No significant glows are seen when the OD voltages are turned off during integration. Full well is approximately at 65\,000\,e$^-$. Peak QE of 94\,\%\ is reached at 670\,nm and remains above 90\,\%\ for 600--750\,nm but drops to 85\,\%\ at 550\,nm and to 46\,\%\ at 907\,nm.

%%%  bis hierhin

\subsection{Dewars and cables}

The large CCD size of 95\,mm$\times$95\,mm required a special dewar. It was designed and manufactured at AIP. A diode protection circuit was designed and fabricated at ITL and installed in the dewar to reduce the possibility of damage from voltage spikes. Each 10k dewar has four vacuum connectors. Two of them are carrying the video signals (8 channels each). The third is connected to the clock signals and the fourth is for the temperature control. Therefore, the dewar requires a three-cable connection to the CCD controller (ARC with two eight-port video-channel cards; see next section) and one cable connection to the temperature controller (CryoCon Model~24C, from Cryogenic Control Systems, Inc.).

%------------------------------ F dewar
\begin{figure}
%\center
%\fbox{
\includegraphics[angle=0,width=83mm, clip]{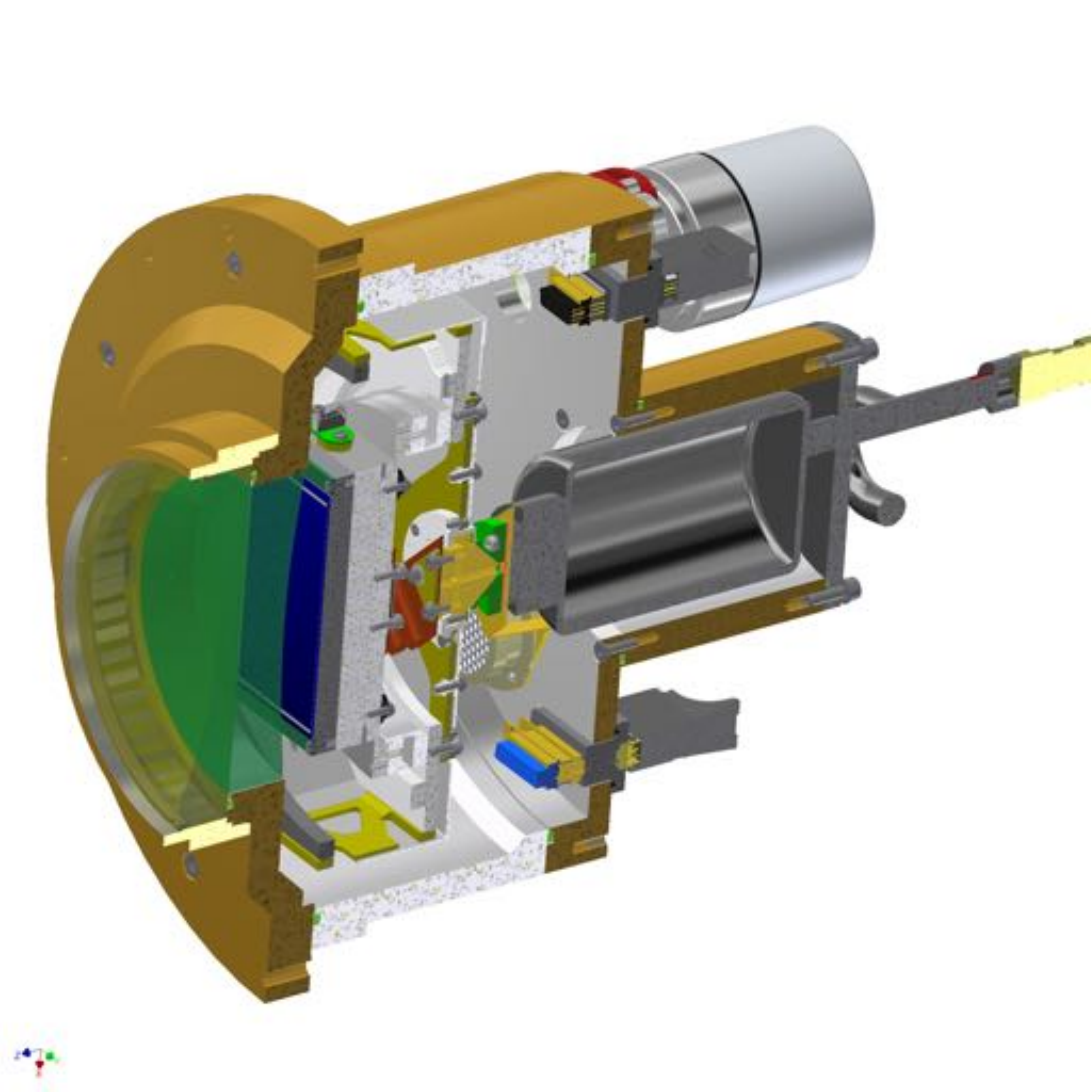}
%}
\caption{Cross section through one of the PEPSI dewars. Note that the last camera lens group also acts as the dewar window. }\label{F-dewar}
\end{figure}

The cables leading from the controller to the dewar are 1.5\,m in length.  They were made to be as small diameter and as flexible as possible.  All video, serial, and reset signals use small diameter coaxial cable (30\,AWG, silicon jacket, 95\,\%\ shield coverage). Bias voltage signals use groups of two pair shielded cables. The custom wiring inside the dewar brings the controller signals to the appropriate detector connectors through protection circuitry.  We measure about 1.5~DN noise (gain approximately 1~e$^-$/DN) for this system without the CCD installed and with a one microsecond correlated double-sample period.

The CCD sits with its Invar package on an aluminum cold plate. On two sides of the cold plate near the edges of the CCD are four PT-1000 temperature sensors. The heat bridge consists of some folded copper sheets in a perpendicular arrangement.  Around the connection of the copper sheets, located on the backside of the cold plate, four heating resistors are placed. On the cold head of the CryoTiger a copper plate is mounted that holds two getter pumps and a fifth temperature sensor of type PT-100, and also represents the other side of the heat bridge. Cooling is provided through two closed-cycle CryoTigers for an operations temperature of about $-120\,\degr$C. The CryoTigers run PT16 gas and are physically located outside of the spectrograph chamber on the PEPSI support bridge in a thermally capsuled rack. 

\subsection{ARC CCD controllers}

The two STA1600LN CCDs are run by two Astronomical Research Cameras, Inc. (ARC) controllers, each made up of two 8-channel video boards called ARC-48D. One \hbox{8-channel} CCD video board processes and digitizes the video outputs from half of the CCD and supplies DC bias voltages to the CCD. It contains eight identical processing circuits that can simultaneously process and digitize signals from eight video outputs. Eight 18-bit analog-to-digital (A/D) converters are provided on the board, along with a 16-bit data pathway to the backplane. The DC bias supply section of the board provides 32 separate low noise, digitally programmable voltages with a variety of voltage ranges suitable for direct connection to the CCD, and a video offset voltage for each of the eight video processors.

A 250\,MHz fiber optic timing board performs three main functions; communicating between the controller and the host computer, generating timing waveforms, and providing overall controller supervision. It has a duplex fiber optic link to a PCI-express interface board, the ARC-64, that operates at a bit rate of 250\,MHz, providing a sustained maximum image data transfer rate of 12.5\,Mpix/s. It controls the timing of the analog portions of the controller by writing 16-bit digital words as often as every 40\,ns over a backplane to other controller boards. It performs such supervisory functions as exposure timing, power sequencing, controlling a shutter, and synchronizing operations with signals external to the controller using the Motorola DSP\,56003 Digital Signal Processor, operating at an internal clock speed of 100\,MHz.

%------------------------------ F optical design
\begin{SCfigure*}
\hskip-2mm
\includegraphics[angle=0,width=137mm, clip]{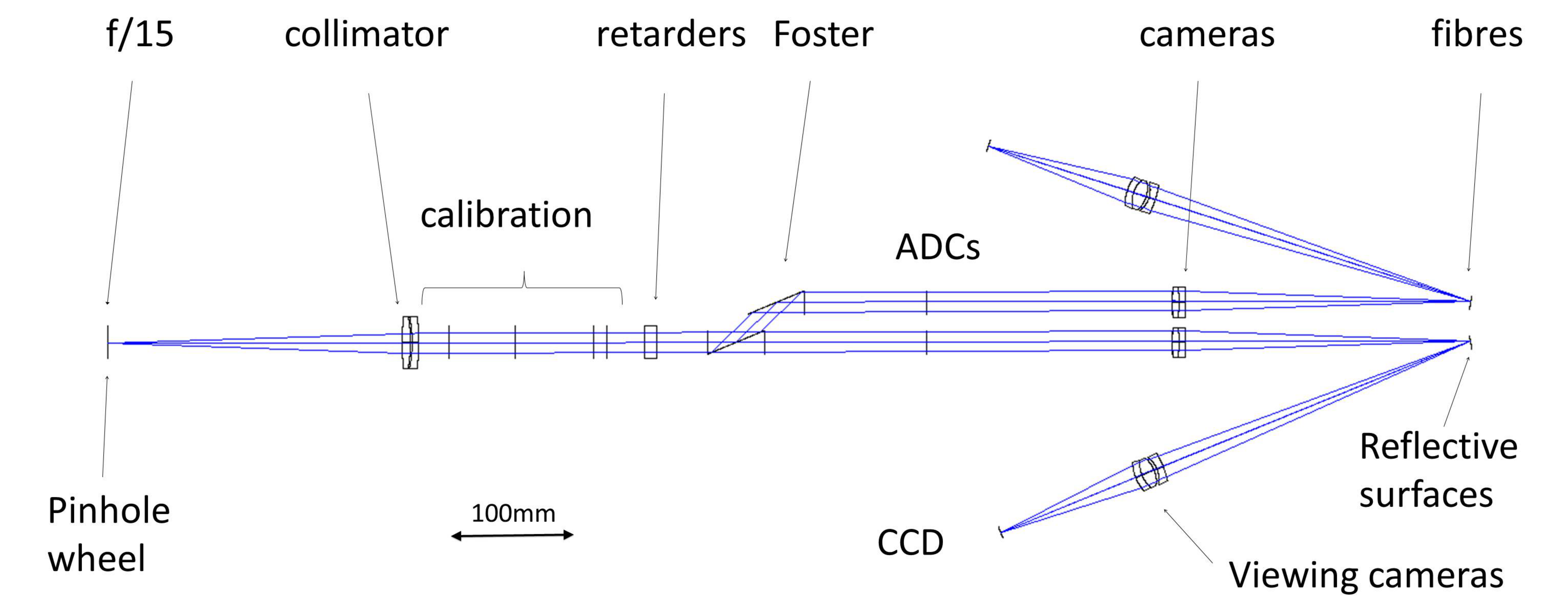}
\caption{Optical design of the PEPSI polarimeters. Light enters from the left and exits into the fibres on the right. Note that the fibres are positioned within central pinholes in the reflective surfaces. The individual optical components are marked. The calibration components and the retarders are revocable. }\label{F-pol_opt}
\end{SCfigure*}

% ------------------------------------------------------------------------------------------------------------------------------
\section{Polarimeters}\label{S-Pol}

The design concept and a description of the expected system characteristics were partly already presented by Ilyin et al. (\cite{ilya:etal}).  Its generalized second-order error propagation in the Mueller matrix and the expected error budget were given in Ilyin (\cite{ii}). We refer the reader to these two papers for further details. Here we focus on the technical description and the characteristics of the final system.

\subsection{Optical components}

With a Strehl ratio of 84\,\%\ the polarimeter has a diffraction limited performance. Figure~\ref{F-pol_opt} shows the full Zemax design. The two polarimeters are identical and can be used independently if needed. Both units are of a dual-beam design with a quarter-wave retarder (QWR) for circular polarization and a modified Foster prism beam-splitter as a linear polarizer. For linear polarization measurements the QWR is removed from the beam and the entire Foster unit is rotated. All polarization optics are located in a collimated beam. The entire unit is in a thermally controlled environment and kept constant at $+$20$\pm$0.5\,\degr C.

\noindent
\emph{Diaphragms.} Light enters the polarimeter through a pinhole wheel located in the f/15 focal plane of the telescope. The wheel has a series of diaphragms, adjustable from 5\arcsec\ for science observations up to 11\arcsec\ for alignment purposes, in particular the adjustment of the polarization-beam axes with the parallactic axis of the telescope. Furthermore, the diaphragms help to reduce stray light and minimize sky background polarization.

\noindent
\emph{Collimator.} A doublet (Table~\ref{T-Pol_lenses}) with a fused silica entrance lens and a SF-57 exit lens separated by 0.5\,mm by a low-expansion Dispal spacer generates a collimated beam of diameter 12\,mm. Both lenses are mounted in a Kovar tub that minimizes thermal expansion. The mount contacts of the lenses with the Kovar tub were manufactured with solderjet bumps in order to prevent any stress in the glass from gluing or mechanical torsion due a common ring strengthener. The residual birefringence peaks at 0.24~nm\,cm$^{-1}$. Both lenses are AR coated for the 400--900\,nm wavelength range.

\noindent
\emph{Retarders.} The super-achromatic QWRs are based on the Pancharatnam design and consist of five stretched acrylic PMMA (polymethylmethacrylat) zero-order retarder layers from ``Astropribor'' with their optical axes oriented at the specific angles (Samoylov et al. \cite{sam}). The anisotropic polymeric plates are laminated between two fine-annealed optical glass windows, each having a broadband anti-reflection coating with a reflectivity of $<$1\%. The glue and the glass used for the windows have a refractive index very close to that of PMMA ($n$ = 1.49). Therefore, no optical fringes at the level down to 0.025\,\%\ (1/4000) were detected on this type of retarder.

\noindent
\emph{Calibration optics.} The calibration optics are located on a rotary stage within the collimated beam and are interchangeable via two linear stages. The optics consist of a Glan-Thompson polarizing prism and two QWRs optimized for the red and the blue wavelength range, respectively. Calibration light is brought up to the telescope focus via a dedicated fibre and is injected into the polarimeter through a set of f/ratio-matching optics.

\begin{figure}
{\it a)}

\includegraphics[angle=0,width=83mm, clip]{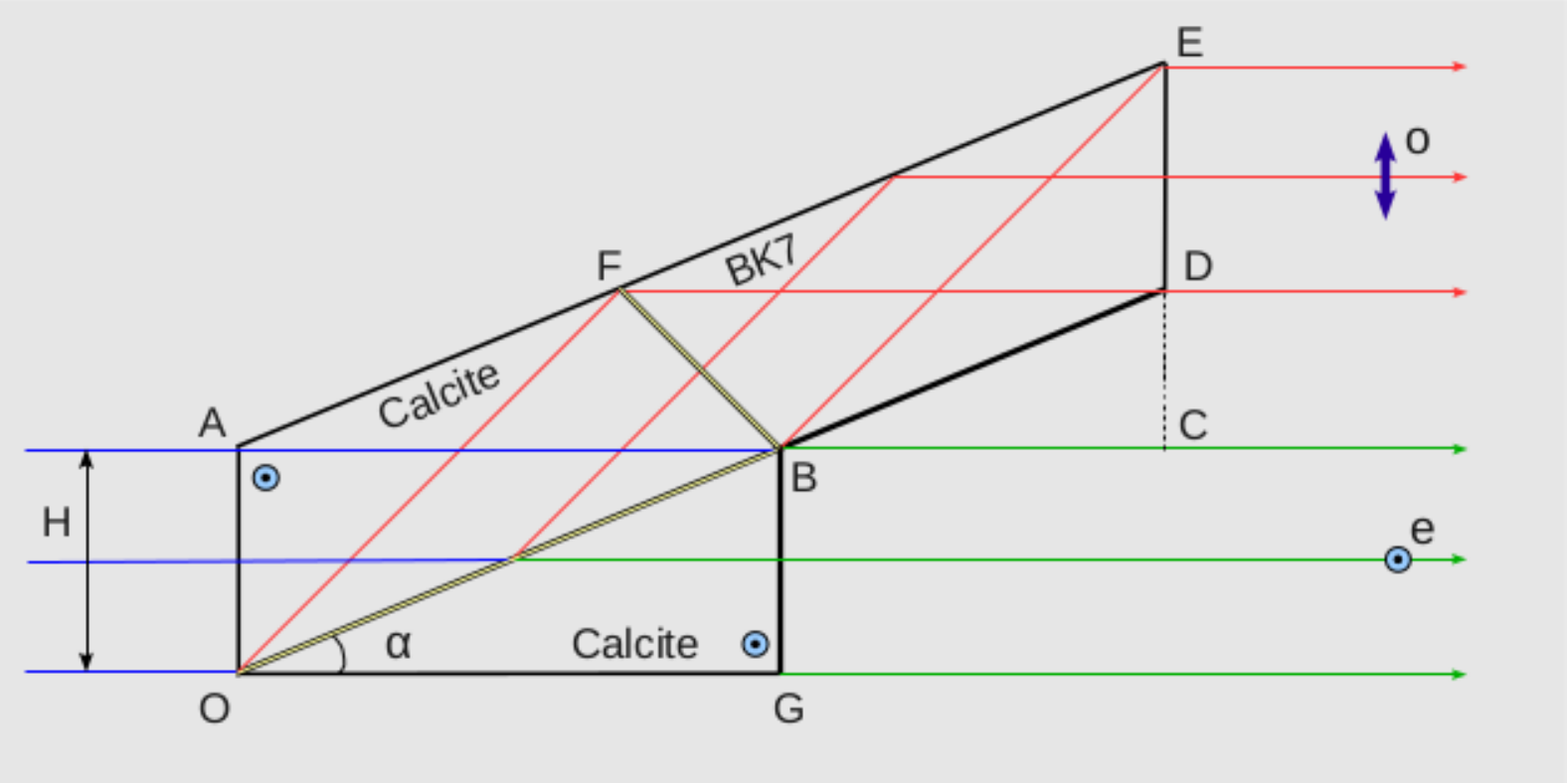}\\
{\it b)}

\includegraphics[angle=0,width=83mm, clip]{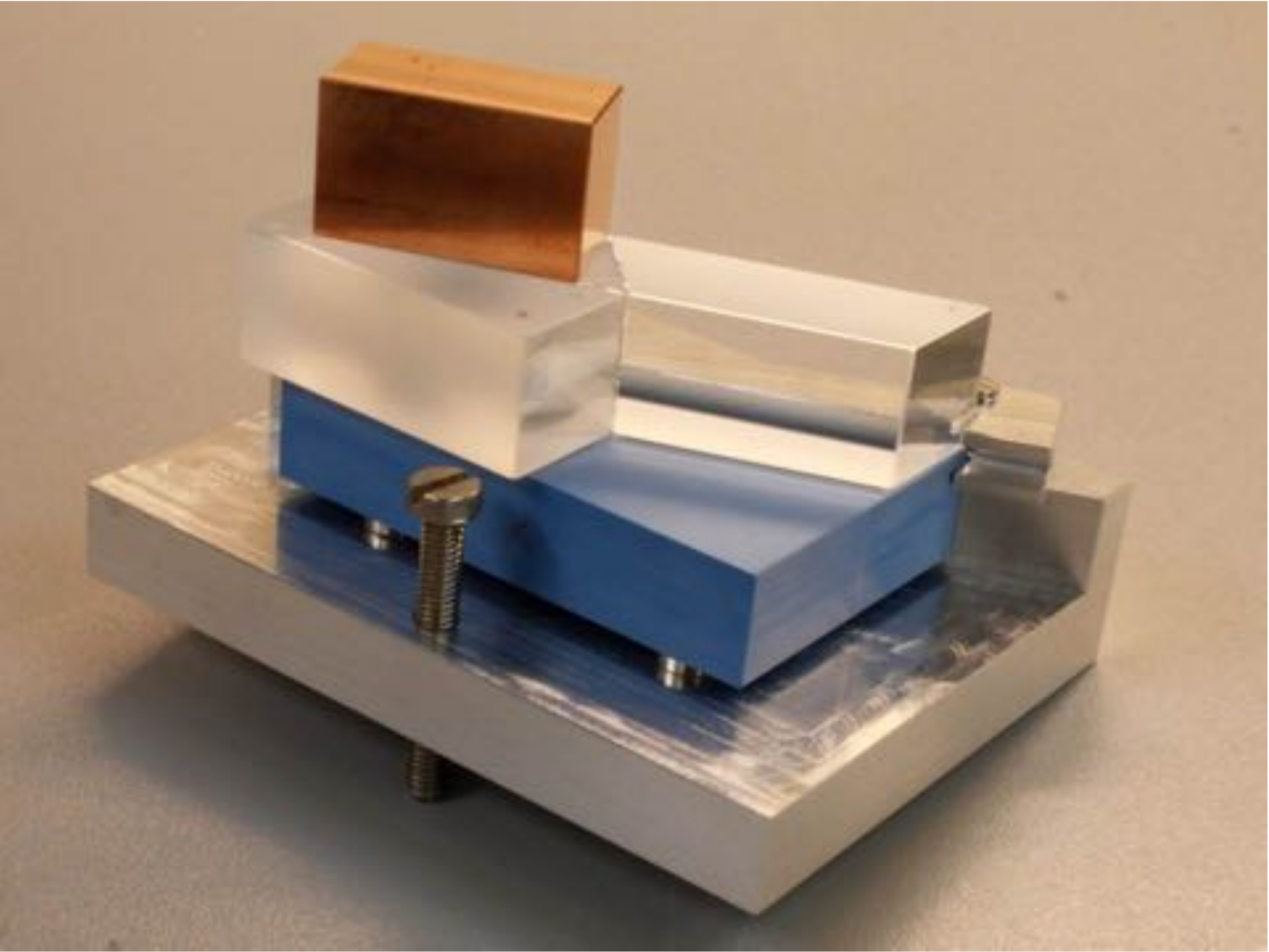}
\caption{Design of the Foster beam splitting unit. \emph{a}) Schematic design. \emph{b}) During assembly of the actual unit for the SX polarimeter.}\label{F-Foster}
\end{figure}

\noindent
\emph{Foster beam splitter.} A polarizing beam splitter is designed on the principle of total internal reflection of the ordinary beam between two calcite blocks. The extraordinary beam transmits straight through the two blocks while the ordinary beam is reflected and passes through the interface layer into the second calcite block where it is again totally reflected. A third prism made of N-BK7 is used to redirect the ordinary beam in the same direction as the straight-through extraordinary beam. Fig.~\ref{F-Foster} shows the conceptual layout and its realization. Each calcite block measures 14$\times$14\,mm for the cross section. One is shaped as a prism, the other as a rhomboid. The unit is mounted on a stress-free holder made of a low-thermal expansion material within a thermo-stabilized environment to prevent stress-induced birefringence. We found a solution for the optimal adhesive bonding of the two calcite blocks by inserting an acrylic matching liquid (Ilyin et al. \cite{ilya:etal}). This minimizes the influence on the polarization state of the transmitted and reflected light down to a level of 0.2--0.02\,\%\ from the blue to the red wavelength edges.

% ------------------------------ Table Linsen Polarimeter
\begin{table}
\caption{Polarimeter collimator and camera lens data. $D$ diameter, $R_{\rm c}$ radius of curvature, $d$ mid-lens  thickness.} \label{T-Pol_lenses}
\tabcolsep=8pt
\begin{tabular}{llccl}
 \hline \noalign{\smallskip}
 Surface   & ~~~$D$    & $R_{\rm c}$   & $d$     & Glass \\
 Number   & (mm)   & (mm)  & (mm)   & \\
%\noalign{\smallskip} 
\hline \noalign{\smallskip}					
Collimator & & & & \\	
1	&31.5 &+145.355~~	&5	&F-Silica	\\
2	&31.5 &--85.155	&~~0.5 &Air	\\
3	&31.5 &--75.949	&4	&SF57	\\
4	&31.5 &--104.591     &\dots	&Air	\\[2pt]					
Camera & & & & \\	
1	&24.0 &+56.615	      &4	&S-FPL53	\\
2	&24.0 &--80.953	 &4	&N-LAK8	\\
3	&24.0 &--587.592~~     &\dots	&	Air\\
%\noalign{\smallskip}
\hline
\end{tabular}
\end{table}

%------------------------------ F Foster unit

\noindent
\emph{ADCs.} Two atmospheric dispersion correctors (ADCs) are located behind the Foster unit within the collimated beam. Each one consists of two counter-rotating prisms immersed into a fluid with suitable refractive index. The relative angle of rotation between the prisms depends on the zenith distance of the target and the overall orientation is thus defined by the telescope AGW de-rotator angle.

\noindent
\emph{Camera optics.} A doublet (Table~\ref{T-Pol_lenses}) for both polarimetric beams re-images the entrance diaphragm onto the fibre entrance microlenses. The doublet consists of a S-FPL53 entrance lens and a N-LAK5 exit lens. Both lenses are broad-band AR coated.

\noindent
\emph{Fibre heads.} The fibre heads consist of of a tilted reflective surface with a central pinhole containing a microlens. The two science fibers, one for the o-beam and one for the ao-beam, have a separation of 23.9\,mm. The mirrors have tilt angles of 9\degr\ and 11\degr\  for the ordinary and the  extraordinary beam, respectively. The pinhole has the same size as the micro-lenses, 0.9\,mm (1.5\arcsec ).

\noindent
\emph{Fibre viewers.} Two f/8 transfer optics re-image each fibre entrance to a Basler GigE 1600$\times$1200 CCD with 4.4\,$\mu$m pixels video camera. The field of view is 29.3\arcsec , chosen in order to obtain accurate centering of the polarized beams on the fibers as well as for possible position corrections during long exposures. Exposure times range from 1\,$\mu$s to 102\,s depending on the target magnitude range of 0\fm --18\fm .

%------------------------------ F inner & outer structure
\begin{figure}
{\it a)}

\includegraphics[angle=0,width=83mm, clip]{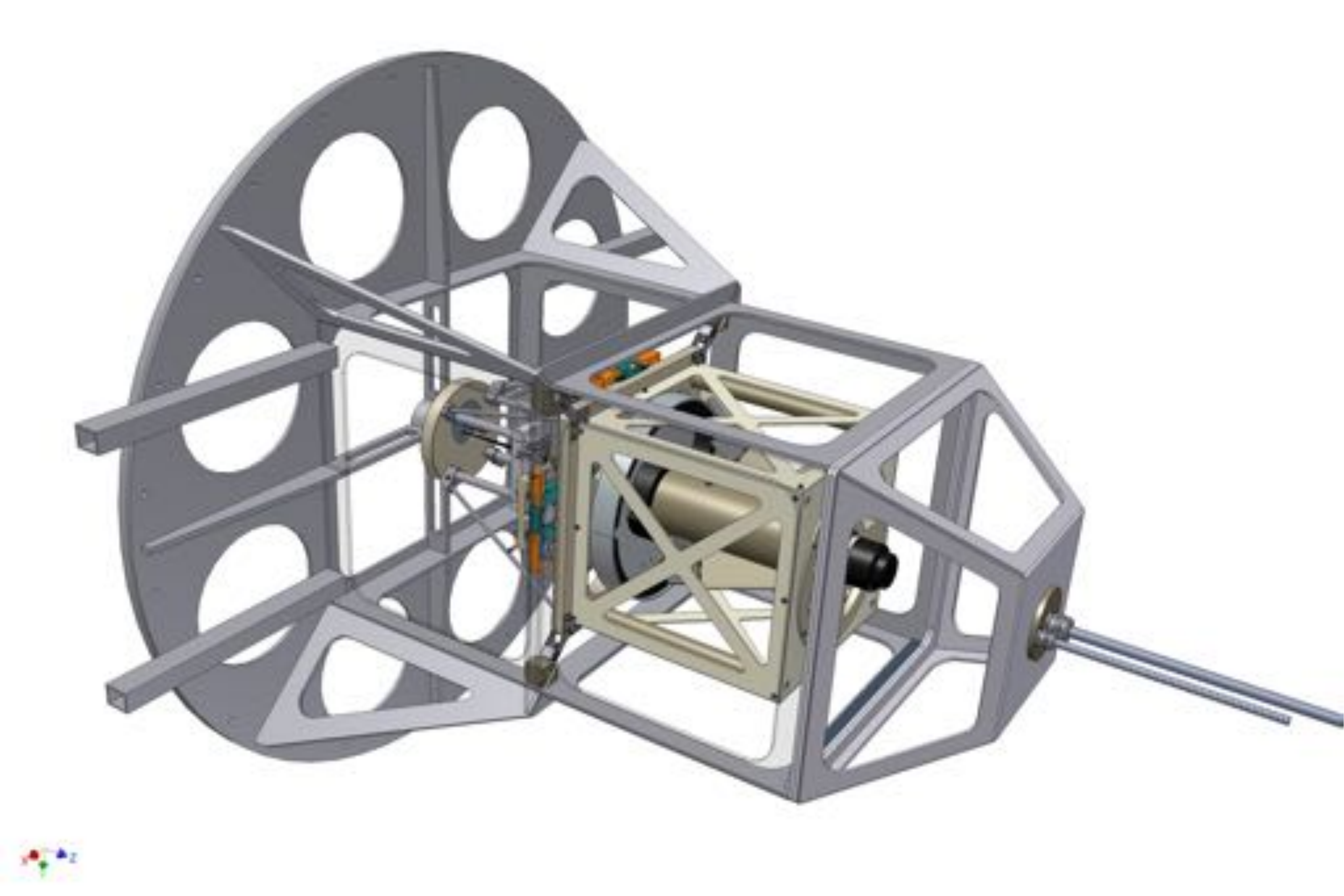}\\
{\it b)}

\includegraphics[angle=0,width=83mm, clip]{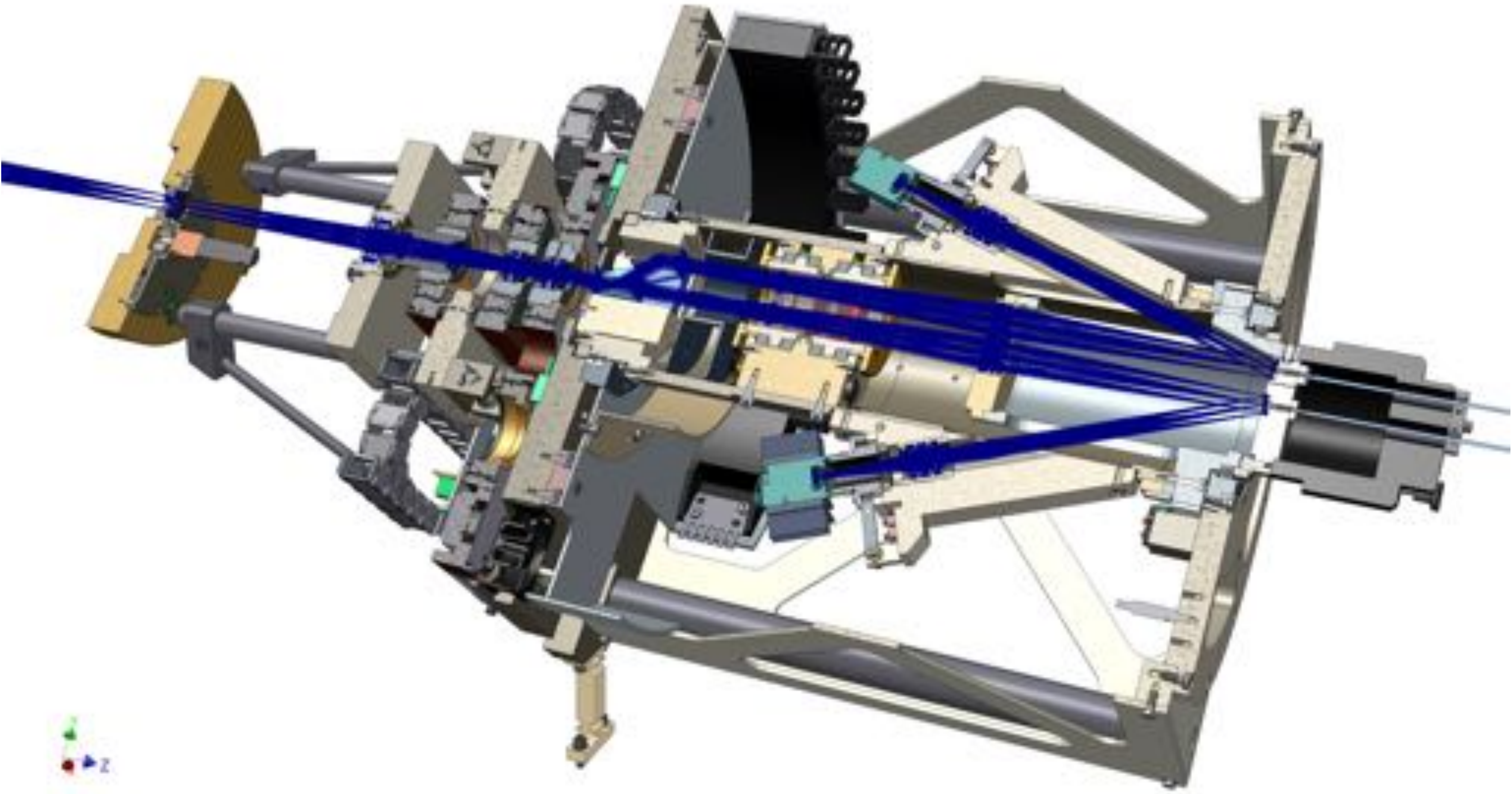}
\caption{Mechanical design of the polarimeter structure. \emph{a}) Outer structure. The circular flange on the left side has a diameter of 1.3\,m and is fitted to the AGW structure shown in the following figure Fig.~\ref{F-polagw}. \emph{b}) Inner structure. Light from the telescope enters from left and the fibres to the spectrograph are shown on the right. }\label{F-pol_struc}
\end{figure}

\subsection{Mechanical structure}

The mechanical system consists of an inner structure which holds all optical components, an interface plate which defines the interface to the telescope AGW unit, and an outer structure main frame which holds the inner structure, the interface plate, the electronics boxes, and the thermal cover. The inner structure at the entrance side is supported by a flexure-type diaphragm which fixes the translational degrees of freedom (DoFs) but releases the rotational DoFs and, at the Foster distance, by an additional fixture. The outer structure is linked to the inner structure via four flexure arms with a square cross section of 15\,mm made of Alimex ACP5080. The interface flange and the outer structure are made of St37 steel, whereas the inner structure is of the same aluminum alloy as the flexure arms.

The full system was analyzed for a gravity vector oriented in its $Y$ and $Z$ directions (cf. Fig.~\ref{F-pol_struc}) and for a thermal load with a temperature gradient of ${\Delta T=35}$\,K set by the minimum environmental temperature of --15\,\degr C and the inside +20\degr C . Without thermal load, the resulting maximum elongation along the $Z$ direction at zenith-pointing position is ${\Delta Z=105}$\,$\mu$m, the highest sag at horizontal position is ${\Delta Y=52}$\,$\mu$m. In case of horizon pointing, and with the maximum thermal gradient displacement in $Y$ direction, the fiber pinhole sags with respect to the entrance diaphragm by 312\,$\mu$m, or equivalent by 0.51\arcsec .  Even this is within the tolerance. Highest Von-Mises stress relative to the proper weight is about 6.9\,MPa, the one associated to the thermal load is 72.9\,MPa.

A modal analysis was run to determine the first 10 eigenfrequencies. Even if dynamic resonances do not constitute an issue in our case, the modes give a deeper insight into the overall kinematic and structural behavior. The lowest eigenfrequency was found at 32.6\,Hz due to bending of the entire inner structure around the $X$-axis, the highest was at 102.9\,Hz due to bending of the external plates around the $Y$-axis.

%------------------------------ F AGw 3 & 4
\begin{figure}
{\it a)}

\includegraphics[angle=0,width=83mm, clip]{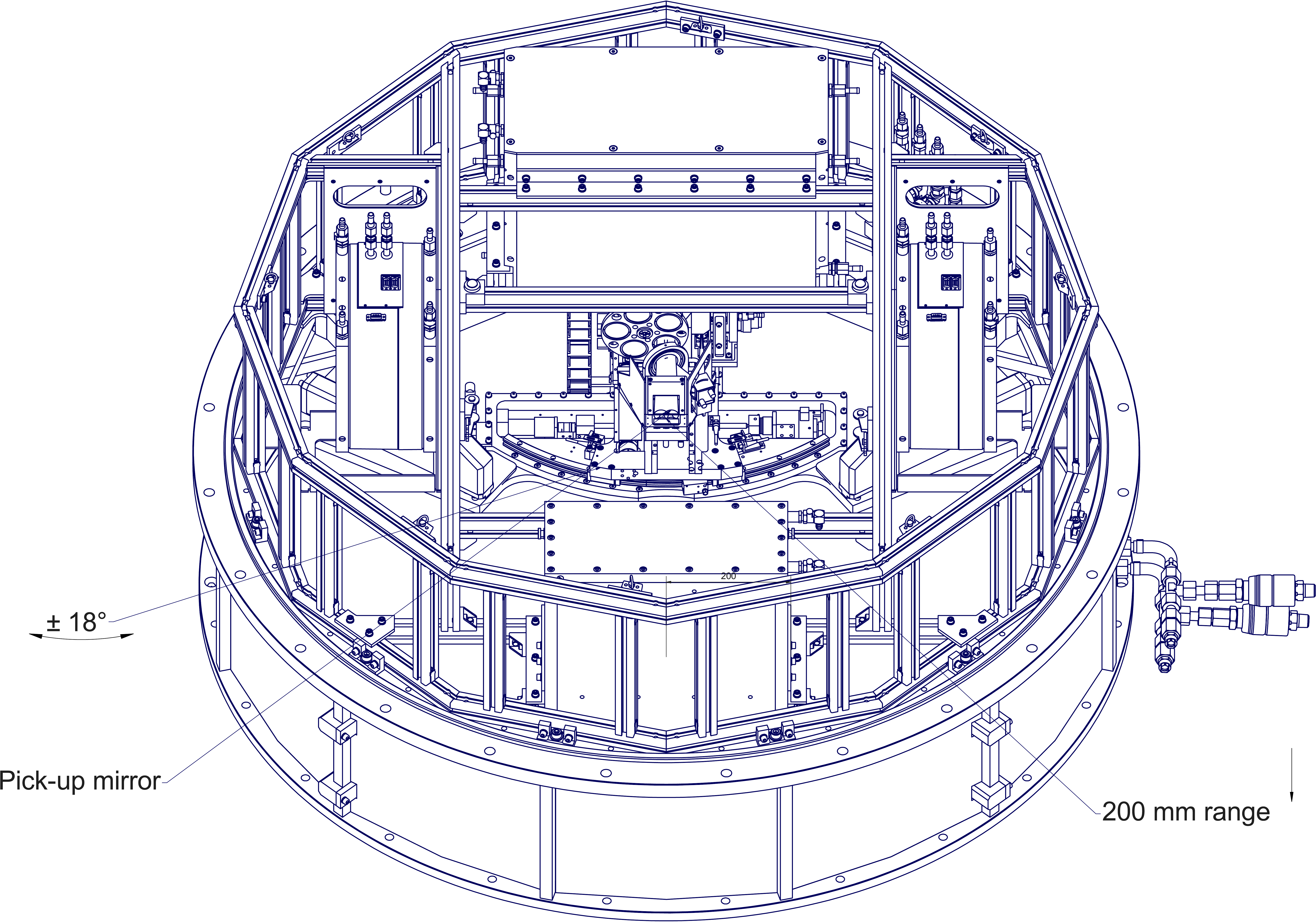}\\
{\it b)}

\includegraphics[angle=0,width=83mm, clip]{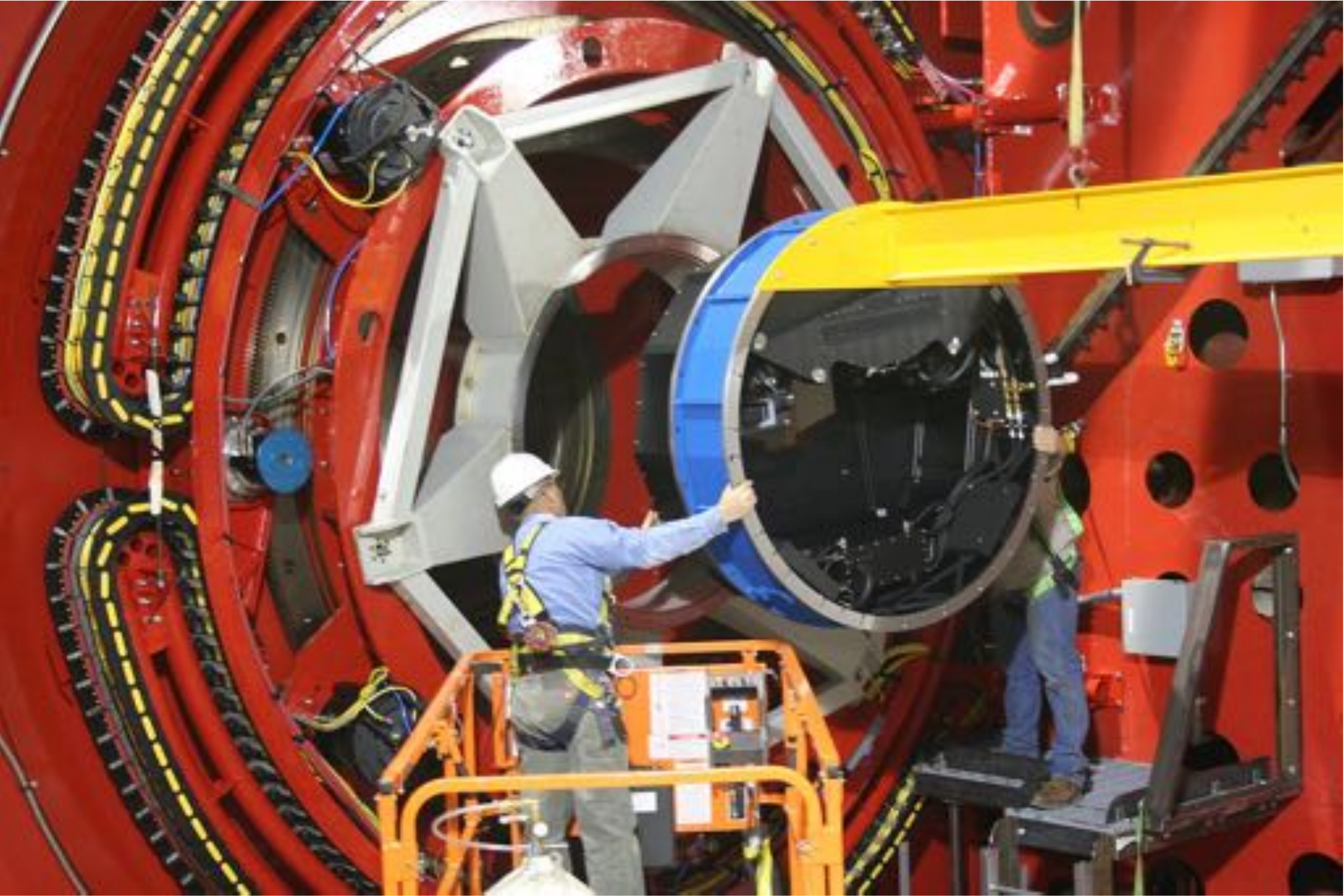}
\caption{Polarimeter AGW units. Shown is one of the two units. \emph{a}) Mechanical design. For scale, the diameter of the entire unit is 1.3\,m. \emph{b}) Unit \#3 during test installation at the telescope focus. In the background one can see the 3-m diameter flange already connected to the image derotator of the telescope. }
\label{F-polagw}
\end{figure}

\subsection{Thermal design}

A thermal gradient has been applied to the optical model in Zemax, assigning a temperature of 21\,\degr C to the four surfaces of the collimator doublet, 19\,\degr C to the Foster prism, the ADC, and the camera lenses, while keeping all  other optical elements at 20\,\degr C. The associated aberration is a defocus of 7\,$\mu$m, or 0.01\arcsec , which is negligible considering the size of the fiber pinhole. The related spot diagram and the chromatic aberration with respect to the normalized entrance pupil were shown recently in DiVarano et al. (\cite{igor}). We conclude that having a thermo-stabilized internal volume with a constant temperature of 20\,\degr C$\pm$1\,\degr C fulfills the performance goals of the instrument.

A less trivial thermal situation is within the two electronic boxes which together dissipate 300\,W. These boxes are flanged to the outer structure and are actively cooled by a liquid cooling circuit. They have a nested cover similar to the PFUs. The inner box is protected against the outside by a Jakodur cladding of 40\,mm thickness. Two identical heat exchangers cool the air in two regions located in the gap between the internal insulation panels and the cover plates of the outer structure. The heat exchangers were dimensioned proportional the boundary condition of a constant temperature of 20\,\degr C for the inner structure and a minimum outer temperature of --15\,\degr C with a heat transfer coefficient of 10\,W\,mK$^{-1}$ for the Jakodur cladding. Their cooling capacity compensates for the internal heating of about 75\,W for both cooling circuits. This power is recycled from the electronics cabinets.

\subsection{AGW units}

For each of the two polarimetric focal stations, we have one acquisition, guiding and wavefront sensing (AGW-) unit. Figure~\ref{F-polagw}a shows a design drawing of one of the two, otherwise identical, units. The 27\arcsec\ by 27\arcsec\ field of view can be moved in the telescope focal plane from the optical axis to about 5\arcmin\ off-axis giving an unvignetted patrol field of about 20 square arc\,minutes.  This allows complete sky coverage with guide stars down to about Sloan ${r'=18}$\,mag. Mechanically the movement is realized with a rotational motion ($\pm18^\circ$) as well as a radial motion (200\,mm full range) with the
rotation axis offset by 612.5\,mm from the optical axis. In addition, a focus stage with a range of 45\,mm allows to compensate for the telescope field curvature. These stages were  manufactured by Feinmess Dresden, Germany. The stages and the filter wheel in the guiding arm are all controlled by a Delta~Tau UMAC system with in-house software.

The on-sky acquisition and guiding is performed with an e2v CCD57-10 frame transfer CCD. The CCD camera is located after a dichroic, which reflects the light red-ward of 720\,nm to the wavefront sensing arm, and a focal reducer which gives a plate scale of 0.054\,\arcsec/pix. The wavefront sensing is done with the same kind of CCD but in this case a small 15$\times$15 aperture lenslet array has been glued on top of the photon-sensitive CCD surface. A pupil image is formed  on the lenslet array with a single lens, and a Shack-Hartmann pattern is registered with the CCD. A pinhole in the focal plane in front of the last lens limits the field of view of each lenslet to 3\arcsec\ diameter which is imaged on to 12 by 12 pixels on the wavefront sensing CCD. The University of Arizona Imaging Technology Lab has glued the lenslet arrays to the CCDs and delivered the camera-control system based on Magellan controllers.

The units are electronically identical to the PFU AGW units used at the bent Gregorian focal stations of the telescope (Fig.~\ref{F-agw1}) and were thus seamlessly integrated with the telescope control system.
On-sky performance of the units have shown image FWHM of less than 0.35\arcsec\ in the Sloan $r'$ band on occasion, and the guider camera detects about 10\,000 photons/s for a ${r'=18}$-mag guide star.

%------------------------------ F  polarimeter storage
\begin{figure}
\includegraphics[angle=0,width=83mm, clip]{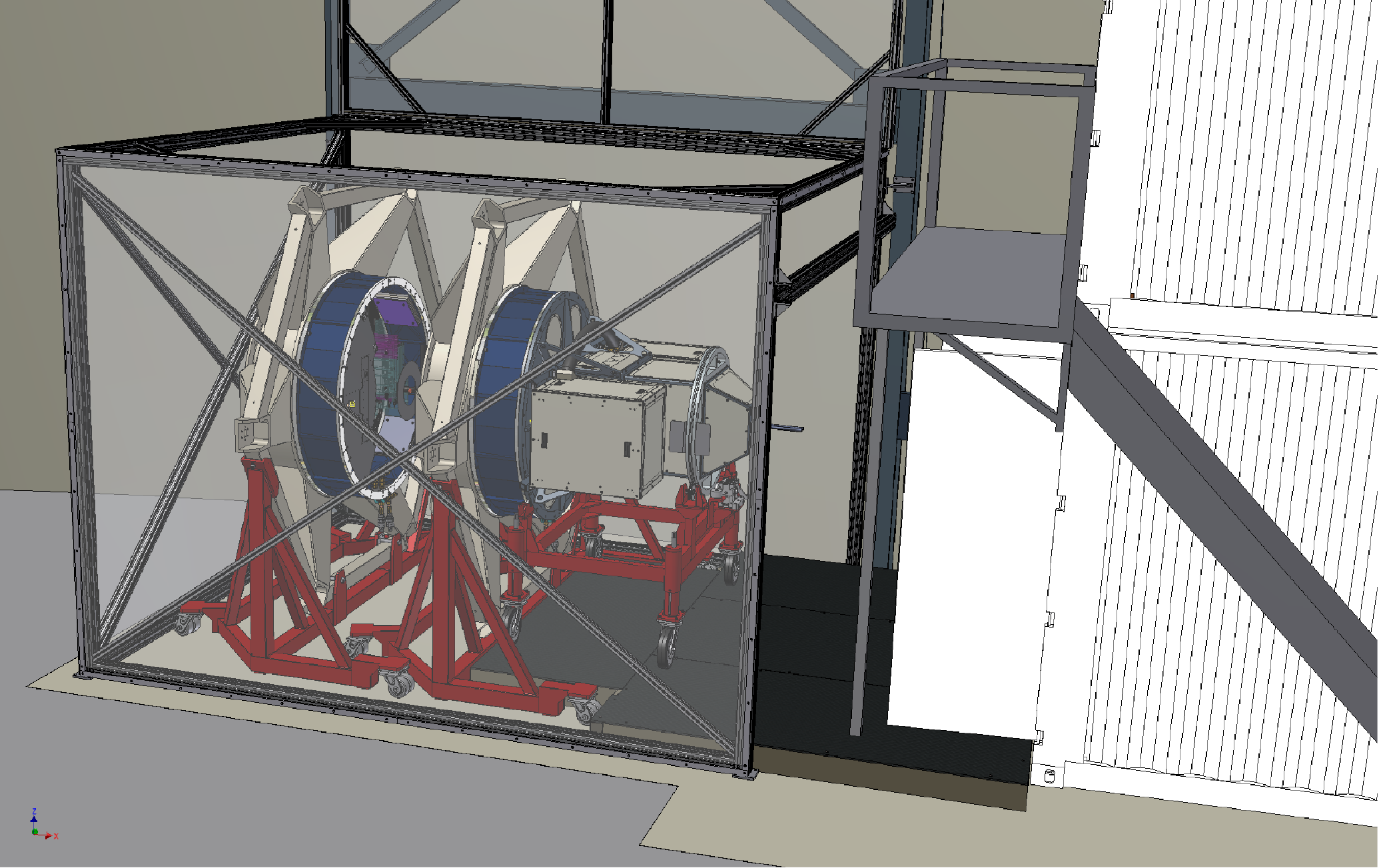}
\caption{Storage of the two polarimeters. A converted container serves as a clean room (bottom right container). The two 3\,m flanges and the two AGWs are pre-mounted on a steel rack in front of the clean room. Before observing, the polarimeters are mounted to the AGWs and moved to the telescope by crane. }\label{Pol-storage}
\end{figure}

\subsection{Storage and handling}

Polarimeter optics are very fragile and particularly prone to dust and dirt, and proper storage and handling are a prerequisite for survival and performance. Because the polarimeters could not be mounted on the telescopes permanently, we store them in a dedicated clean room when they are off the telescopes (Fig.~\ref{Pol-storage}). Only the fibre heads remain in a safe box on the telescope. A standard 6m-shipping container was converted into a low-class clean room within the lower third floor in the LBTO building. It is pre-entered through a plastic assembly tent. The tent is also used for the storage of the two 3m-diameter telescope flanges that connect the polarimeter AGWs to the telescope image derotators. The AGWs remain pre-mounted to these flanges but could also serve other instruments. The polarimeters are attached to the AGWs only shortly before lifting to the telescope.

Within the clean environment the polarimeters are mounted on a trolly on which they can be rotated around their main optical axis. This allows easier functional testing before an observing run.

% ------------------------------------------------------------------------------------------------------------------------------
\section{PEPSI control and data-reduction systems}\label{S-Control}

\begin{figure}
\includegraphics[angle=0,width=83mm, clip]{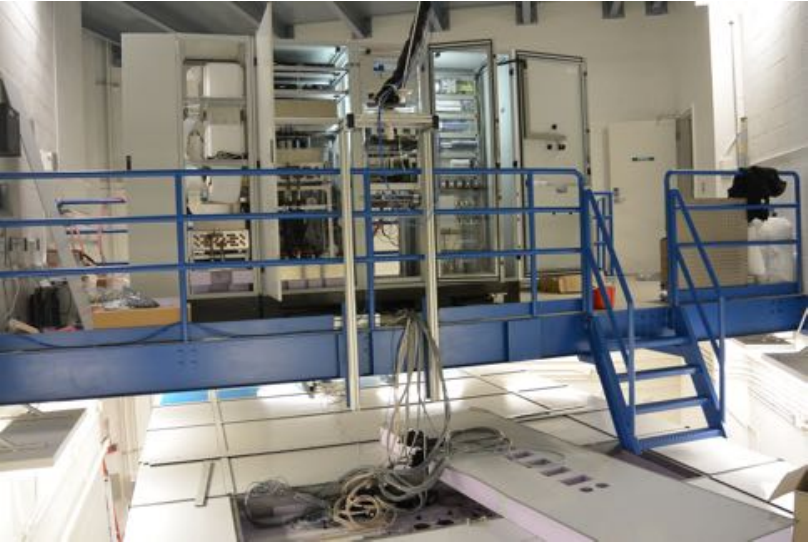}
\caption{PEPSI main electronics rack on the support bridge (blue). The top of the chamber is visible in the lower part of the picture. }\label{F-Rack}
\end{figure}

\begin{figure*}
\center
\includegraphics[angle=0,width=169mm, clip]{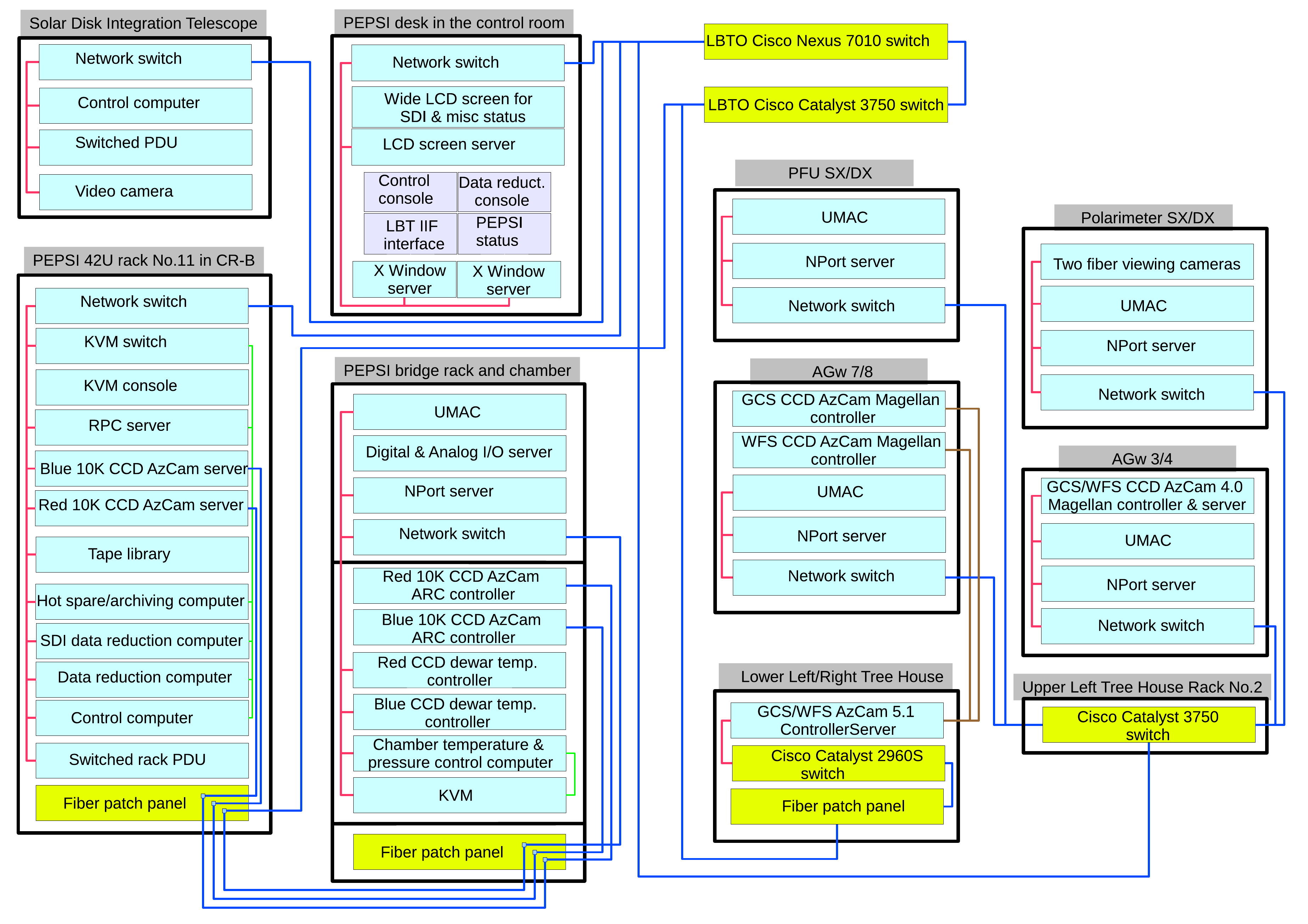}
\caption{Overview of the PEPSI network components. The yellow boxes are LBTO provided and mark the respective interface. The ``Tree Houses'' are the walkable electronic cabinets on the telescope. See text. }\label{F-control}
\end{figure*}

The entire suite of network switches, compute servers, and X-window clients is shown in Fig.~\ref{F-control} for an overview. Almost all of the control hardware is located in four 2m-racks  (Fig.~\ref{F-Rack}) on the PEPSI support bridge.

%------------------------------ F  electronics rack

\subsection{Control interface}

The operation during observations is done from the LBT control room. PEPSI is a software client of the LBT telescope control system (TSC) which is executing control commands from the client. The communication is performed with XML scripts via Remote Procedure Calls (RPC) and is completely hidden from the user. The communication protocol is implemented in a C/C++ library, the so-called Instrument Interface (IIF), and consists of a number of functions including pointing the telescope, selecting a guide star, fine offsetting of the telescope mirrors, etc.. The PEPSI control system is written in C++ under a Red-Hat/FedoraCore Linux distribution with a gcc v4 compiler.

The main components of the control interface are the spectrograph status and command widget, the quick look and analysis facility, a number of table widgets, and various status windows. All the information and parameters are held in database tables, like the objects to be observed, the spectral settings with the specific parameters like exposure times for calibration spectra, or the list of FITS images acquired. 

The table of objects is SIMBAD-enabled, which means that the coordinates (and other parameters) of any new target can be picked via the client server or typed by hand. The real-time status of the spectrograph settings is displayed on the control console widget window. Although the settings can be changed manually by clicking on the respective button, its main purpose is to show the status information of the spectrograph. 

%------------------------------ F  network components

\subsection{Network components}

PEPSI uses a variety of network switches due to historical reasons and physical-size constraints. An overview is shown in Fig.~\ref{F-control}. Details are subject to change.

\subsection{Computer server and storage system}

This system provides storage and computing power to deal with the large CCD image-data arrays. One 10k-CCD image with 10\,560$\times$10\,560 pixels with two bytes per pixel comprises a raw image of 223\,MB. One exposure with the complete six wavelength settings is thus a 1.34-GB image. An image in floating-point format with variances is 892\,MB for image processing. Several such images must be located in memory at the same time. The system that provides the acquisition and processing chain of the raw CCD images consists of the following servers.
\begin{itemize}
\item A control computer for hardware control of a number of distributed low-level devices of the spectrograph subsystems. It provides access to the LBT control system and the SDI telescope, displays status and initiates the data reduction queue. It works over the local network and is not accessible from outside.
\item The RPC server is a gateway for the control computer to access and issue low-level commands to control all the hardware components of the spectrograph located in the chamber, on the support bridge, the two PFUs, and the two polarimeters.
\item Two data-reduction computers are dedicated to the data obtained during day-time with the SDI telescope and to the data obtained from night-time with the LBT or the VATT, respectively.
\item A hot spare computer can replace any of the other without a change of any hardware or cables. It also serves as an off-load archiving computer.
\end{itemize}

The data system currently provides a shared disk storage of 32\,TB (as of early 2015), so that each server can independently and at any time access the same files without corrupting the file system. The large capacity storage array provide sufficient transmission for time-critical applications dealing with very large data arrays. The system also provides a magnetic tape library accessible by each server at any time by request.

\subsection{AzCam CCD-data acquisition system}\label{AzC}

AzCam is a CCD image acquisition and analysis system developed at the University of Arizona Imaging Technology Laboratory.  It is used to acquire images for the PEPSI RED and BLUE 10k cameras as well as for the AGW-cameras. The AzCam system consists of three main components: 1) the main server named AzCamServer, 2) a camera server which interacts with the controller hardware named ControllerServer, and 3) various client applications such as AzCamTool and AzCamConsole.

\noindent
\emph{AzCamServer} is the core program which coordinates all AzCam activities.  It communicates over internet sockets with control clients, hardware servers, logging processes, etc.. AzCamServer processes commands and interacts with the various software components as necessary, receiving image data, creating FITS image files, and sending image files to remote data processes. When necessary, AzCamServer also connects to other instrument and telescope servers for status information and control. AzCamServer is written in Python with C++ components.

\noindent
\emph{ControllerServer} communicates with AzCamServer over a socket and directly commands the controller hardware (a PCI-Express card for PEPSI) via drivers.  For the PEPSI systems, the ControllerServer process runs on the same machine as AzCamServer. ControllerServer is written entirely in C++.

AzCamServer supports several different clients which are used for control and analysis.  For local testing and debugging of the PEPSI CCD cameras the user GUI is AzCamTool, which is a LabVIEW application.  This client allows users to select image file names, binning, DSP code, and many other parameters.  For normal PEPSI observations, the PEPSI software communicates directly to AzCamServer via a command parser which is specific to the PEPSI command syntax.  There is also a python console client named AzCamConsole which allows python command syntax and general scripting.  It also supports a wide variety of debug tools.

\subsection{CCD image processing with S4S}

Our ``Software for Stellar Spectroscopy'' (S4S) is a generic package in C++ under Linux and currently consists of about 80\,000 lines of source code which are partially based on the 4A package (Ilyin \cite{ii2000}). S4S exploits the many advantages of object-oriented and template-classes programming. An extensive Numerical Template Library (NTL) is specifically tailored to the needs of the data reduction and analysis tools. Basic approximation tools include smoothing, circular, and polar splines with adaptive regularization parameter to minimize the curvature or the second derivative of the residuals, orthogonal polynomial series, and Chebyshev polynomials in one, two, and three dimensions. Non-linear least-squares include multi-profile fitting, multi-harmonic fit to periodic signal (also circular splines), and an orbital solution to radial velocities. NTL also implements a special class of numeric iterators which allow to use it as a vector language (like in Java or IDL). An extensive matrix template class incorporates a variety of different types of 2D matrices and a 3D cube matrix necessary for solving least-squares fitting problems with a number of decomposition template classes.

The FITS class supports all possible extensions and data compressions. The data image and vector are the prime container classes to store and operate observational data, their variances, masks, and various arguments (pixels, wavelength, or time). The Graphical Toolkit is designed upon the native Linux X11 library. The Communication Bus is a middleware class to communicate between different network nodes and exchange C++ objects between them.

\subsection{Data-reduction toolkit}

\begin{figure}
\includegraphics[angle=0,width=83mm, clip]{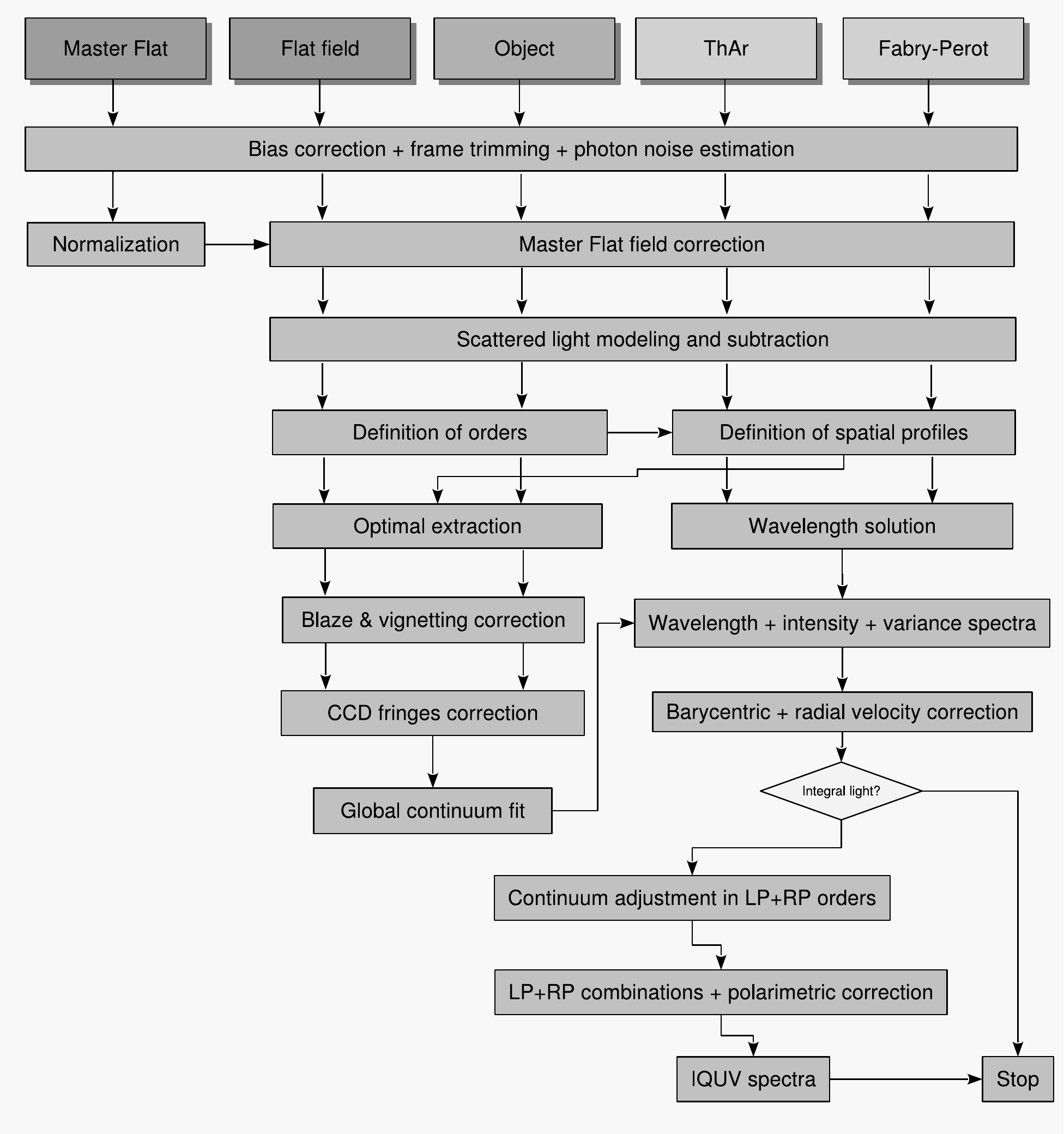}
\caption{Flow chart of the PEPSI $IQUV$ data-reduction pipeline. ``Object'' frames are any of the four Stokes-component exposures. }\label{F-datared}
\end{figure}

The Image Processing and Reduction Toolkit (IPRT) within S4S is based upon adaptive selection of parameters by using statistical inference and robust estimators. The standard components include bias overscan detection and subtraction, scattered light surface extraction and subtraction, definition of \'echelle orders in integral light or in polarimetric light, weighted extraction of spectral orders, wavelength calibration, and a self-consistent continuum fit to the full 2D image of extracted orders. Figure~\ref{F-datared} shows the flow chart for the detailed reduction steps.

%------------------------------ F Data reduction

%------------------------------------------------------------------------------------------------------------------------------
\section{VATT fibre feed}\label{S-VATT}

\subsection{The VATT telescope}

The Vatican Advanced Technology Telescope (VATT) is a 1.83\,m alt-az telescope located approximately 350\,m away from the LBT building. It is jointly operated by the Vatican Observatory Foundation with the University of Arizona. Its optics consist of a f/1.0 honey-combed borosilicate primary mirror with a 0.38-m f/0.9 Zerodur concave secondary mirror making up a f/9.0 Gregorian focus on the rear side of the primary mirror. The image scale is 12.52~\arcsec/mm in a vignetting-free field-of-view of 15\arcmin.

%------------------------------ F VATT injection
\begin{figure}
{\it a)}

\includegraphics[angle=0,width=83mm, clip]{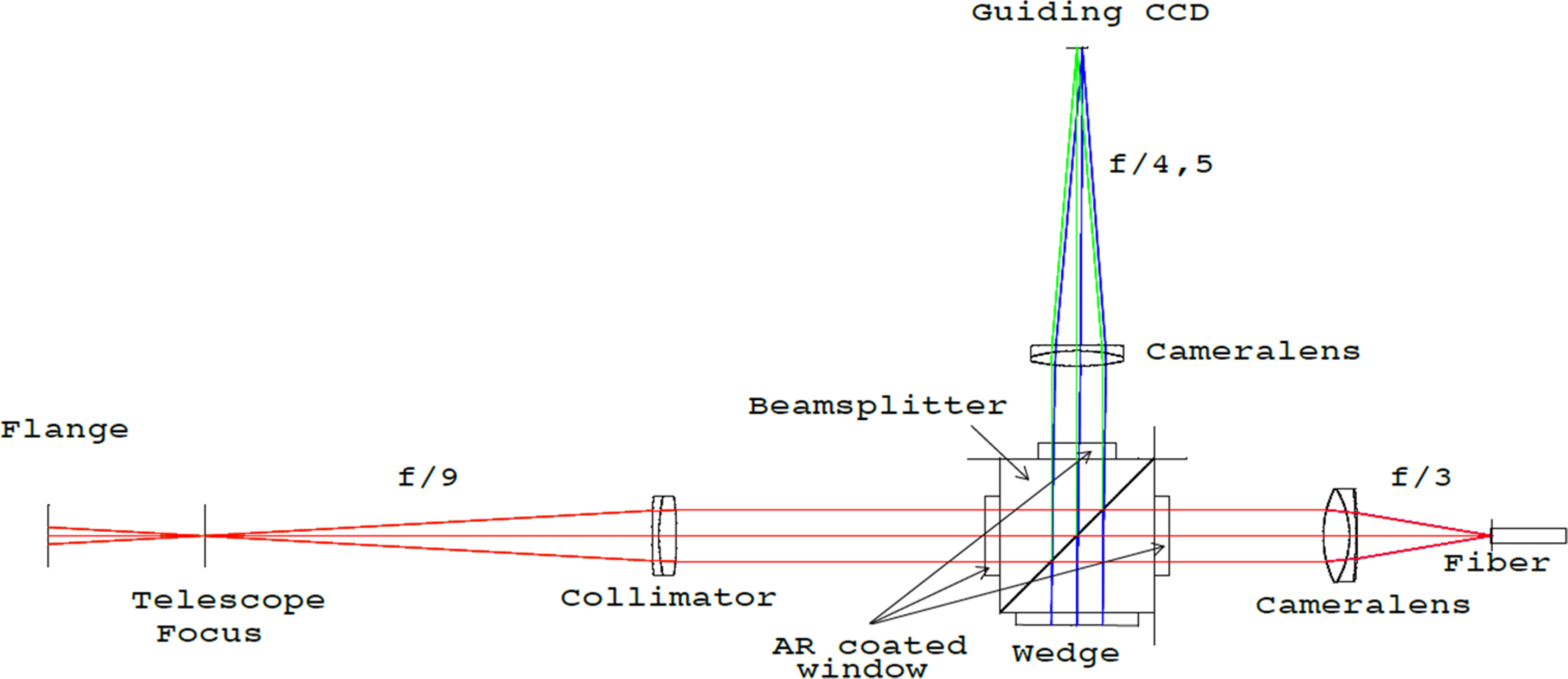}\\
{\it b)}

\begin{center}
\includegraphics[angle=0,width=50mm, clip]{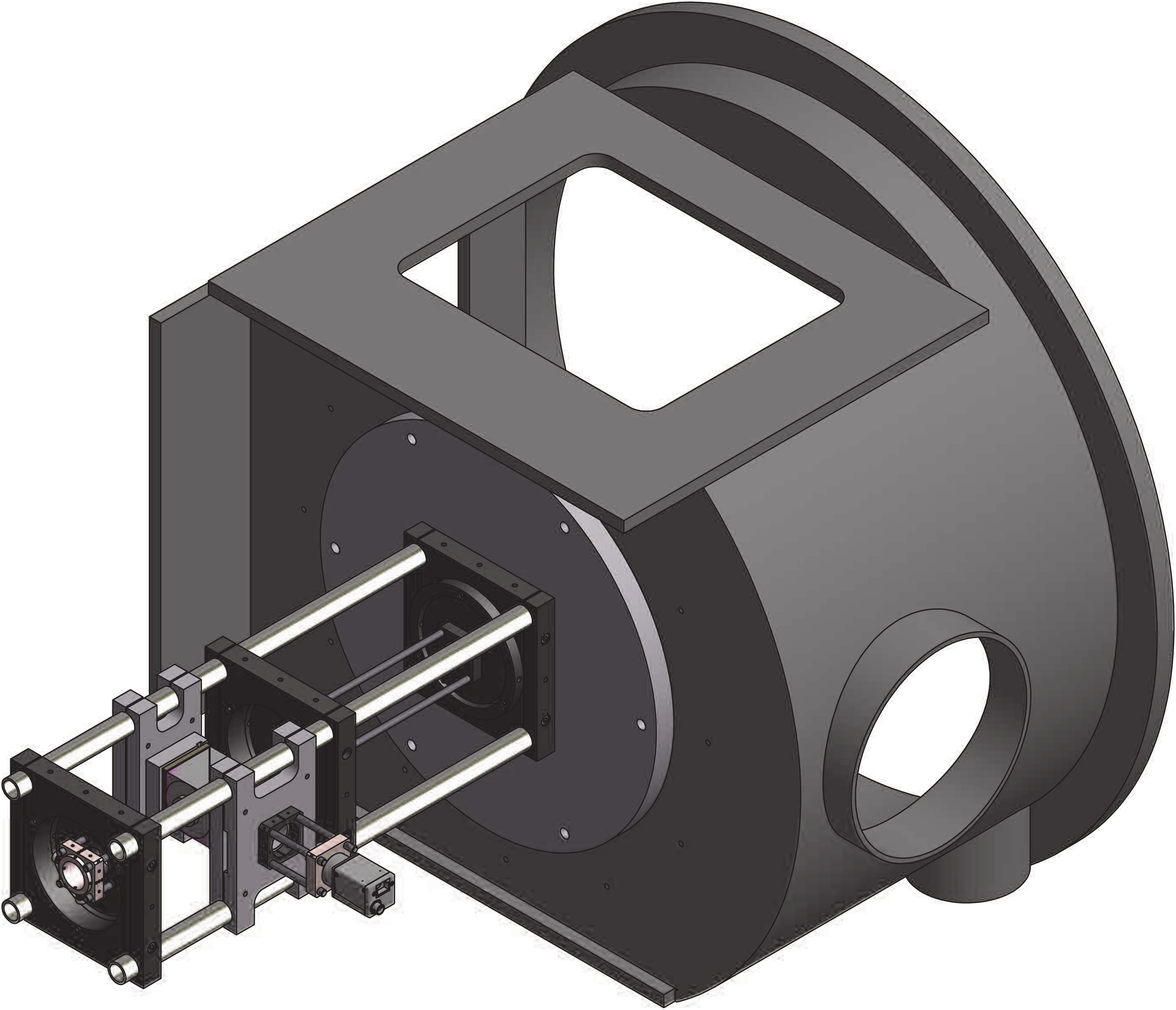}\\
\end{center}
{\it c)}

\includegraphics[angle=0,width=82mm, clip]{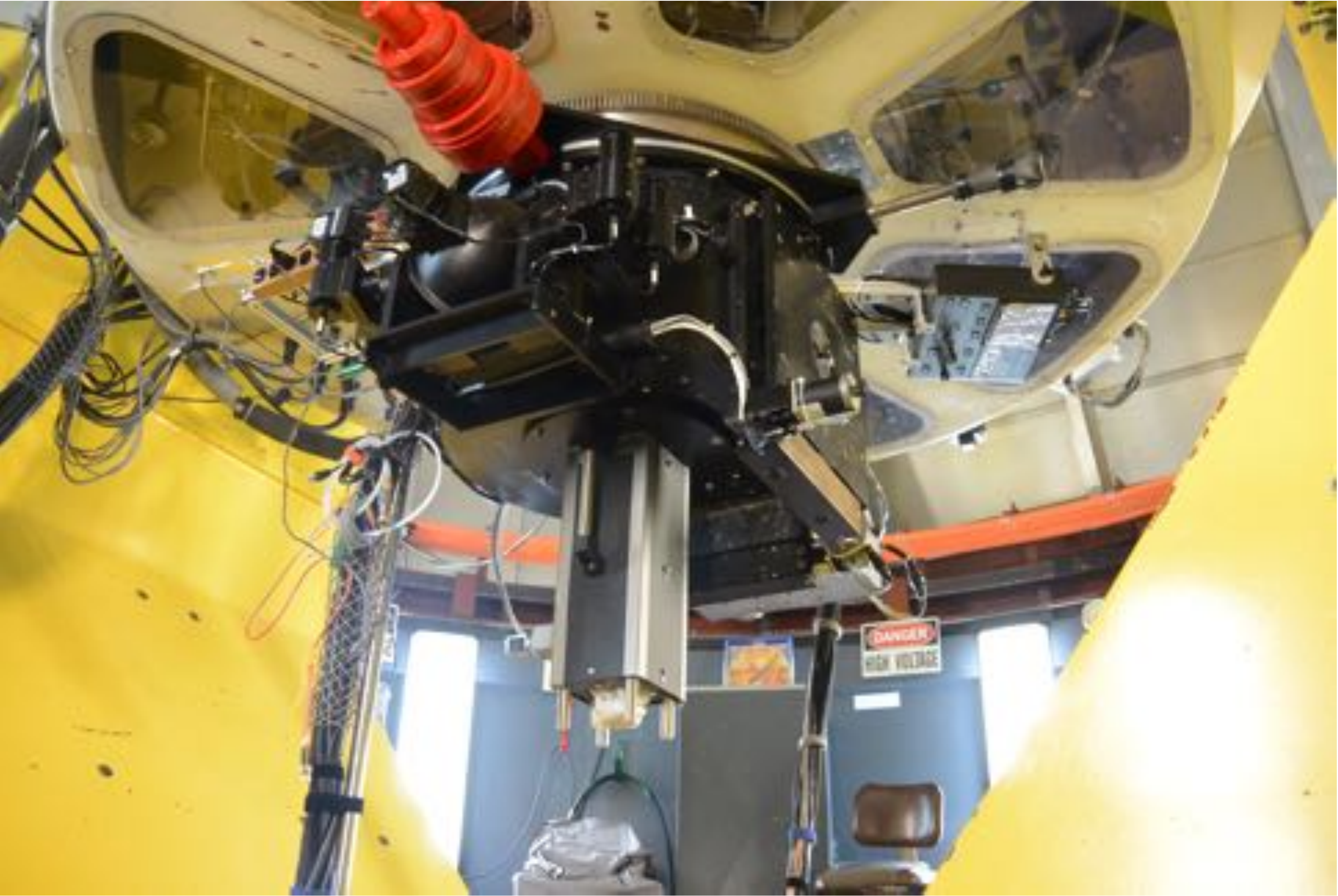}
\caption{Fibre injection unit (FIU) and acquisition \& guiding (A\&G) unit for the VATT. \emph{a}) Optical design of the FIU. The telescope f/9 focus is on the left side, the fibre pinhole on the very right. Indicated are the collimator doublet (Col1), the camera (Cam), the beamsplitter cube (BS) with its backside wedge, and the A\&G CCD. \emph{b}) Mechanical design of the FIU and A\&G plus its telescope flange. \emph{c}) Real unit mounted on the telescope. }\label{F-VATT}
\end{figure}

\subsection{Fibre connection to PEPSI}

A set of three PolyMicro FBP200240280400 fibres connect the VATT telescope with the PEPSI spectrograph. The fibre bundle runs in an existing underground duct along the same path as the communication uplink of the LBTO. Total fibre length is 453\,m, 29 of which are inside the VATT building, 35\,m are from the LBT-duct outlet to the PEPSI bridge, 39\,m from the PEPSI bridge to PFU-SX, and the remaining 350\,m are inside the duct system between the two buildings. We chose 200-$\mu$m fibres to match the medium-resolution mode (120\,000) of PEPSI. A trial cable production using under-water cabling techniques failed in 2013. Therefore, we decided to add an extra polyamid buffer of outer diameter 400\,$\mu$m and put the fibres into a classical protective tubing system named ``HelaWrap''. For protection against moisture, an additional sleeve was wrapped around before pulling the fibre bundle. Both ends of the three fibers are equipped with SMA-905 connectors.

%------------------------------ F VATT fibre transmission
\begin{figure}
\includegraphics[angle=0,width=83mm, clip]{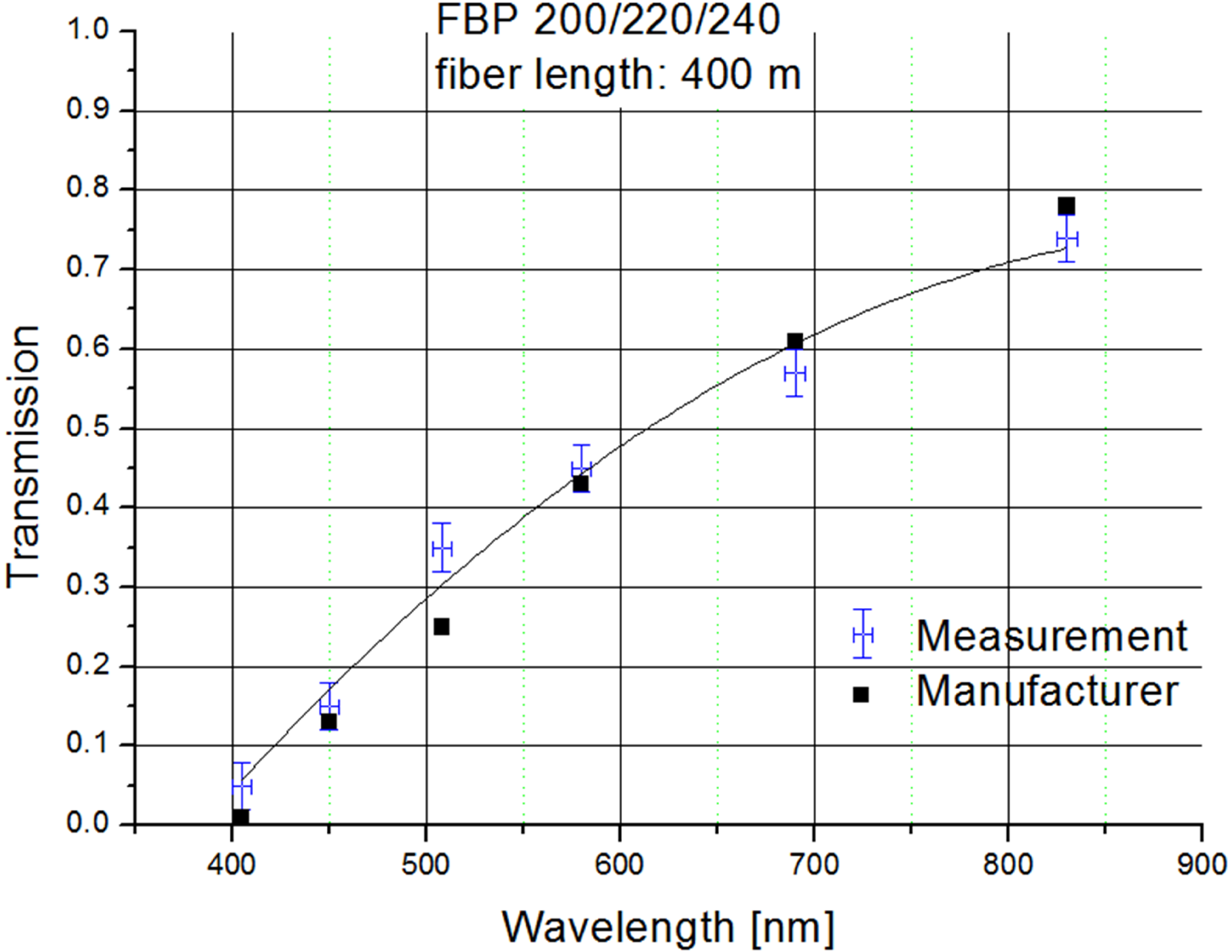}
\caption{Transmission measures for the VATT-PEPSI fibre connection. Our own measures are shown as error bars, the extrapolated manufacturer values are overplotted as black squares. The wavelength bins equal to the central wavelengths of the six cross dispersers.}\label{F-VATT400}
\end{figure}

From the FBP data sheet for 50-m fibre lengths, we expect the transmissions at the central wavelengths of CD-V (685\,nm) and CD-III  (508\,nm) to be 93\,\%\ and 82\,\%, respectively. This converts to approximately 52\,\%\ and 17\,\%\ for the 453-m pieces. However, actual throughput of one of the fibres was measured to be significantly better, 57\,\%\ and 35\,\%\ for  685\,nm and 508\,nm, respectively. Our measurements are shown in Fig.~\ref{F-VATT400} for the central wavelengths of the six cross dispersers. Transmission peaks for the reddest cross disperser at 73\,\% . However, the transmission at the very blue cross disperser  is on average just 5\,\% .

\subsection{Fibre injection unit}

The VATT fibre injection unit (FIU) is special for PEPSI because it is ``single-eyed''. All other observing modes of PEPSI create four spectra per \'echelle order on each CCD. E.g., the LBT/PFU mode consist of two on-axis fibers for the target spectra from the two LBT mirrors and two off-axis fibers for the two sky or calibration positions. The SDI mode consists also of two target fibres plus two simultaneous calibration-light fibres. The VATT observing mode provides only a single target spectrum through one fiber but is also accompanied by one (simultaneous) Fabry-P\'erot spectrum through one of the PFU sky fibers.  No sky-background light is available in the VATT mode.

The telescope focus is f/9. No ADC nor field corrector are installed. Optimal fiber injection is achieved at f/3 in air with an achromatic camera as shown in Fig.~\ref{F-VATT}a. An aperture stop of diameter 200\,$\mu$m (7.5\arcsec ) is laid out as a pinhole in a reflective, thin, stainless-steel plate. The back-reflected light passes again through the beam splitter and is redirected to the guider CCD by a reflective wedge on one side of the beam splitter. It creates an image of the pinhole on the detector that is separated by $\approx$\,38\arcsec\ from the directly reflected beam. Guiding is done with the directly reflected light from the beamsplitter. Note that the beamsplitter itself is within a collimated beam of the telescope and redirects 1\%\ of the incoming light to the guider CCD. The FIU's mechanical interface is the flange shared with the VATT science 4k-CCD camera.

Guider camera is a Basler GigE 1600$\times$1200 CCD with 4.4\,$\mu$m pixels. It is uncooled. Typical exposure time is 100\,ms with a read-out time of the full frame of less than 10\,ms. The camera re-imaging optics magnifies the telescope image scale by two times, which makes the scaling factor on
the guider camera to be 36.3 pix/\arcsec . It provides a field-of-view for the guider of 44\arcsec$\times$33\arcsec .

\subsection{Operation}

The VATT is currently still a manual telescope requiring a night-time telescope operator. Attempts to automate the steady-state operations part are underway and remote-control observations from Tucson already possible. At the time of writing, we treat the FIU as a VATT visitor instrument and mount and dismount it whenever telescope time is available.

\subsection{Data acquisition and storage}

VATT-PEPSI data share the same interface as any other PEPSI mode and are administered by the PEPSI control server. Data are physically located in the LBT computer room B which is part of the LBT storage area network.

%------------------------------------------------------------------------------------------------------------------------------
\section{Solar-disk-integration (SDI) telescope}\label{S-SDI}

Disk-unresolved solar light is fed to the spectrograph from two small Sun-as-a-star telescopes via a fibre connection to the calibration unit. Its aim is, firstly, to provide an external comparison source for the instrumental profile and monitor its long-term stability and, secondly, to monitor the Sun for measuring line-bisector variations over an entire magnetic-activity cycle, among others.

%------------------------------ F SDI
\begin{figure}
{\it a)}

\includegraphics[angle=0,width=83mm, clip]{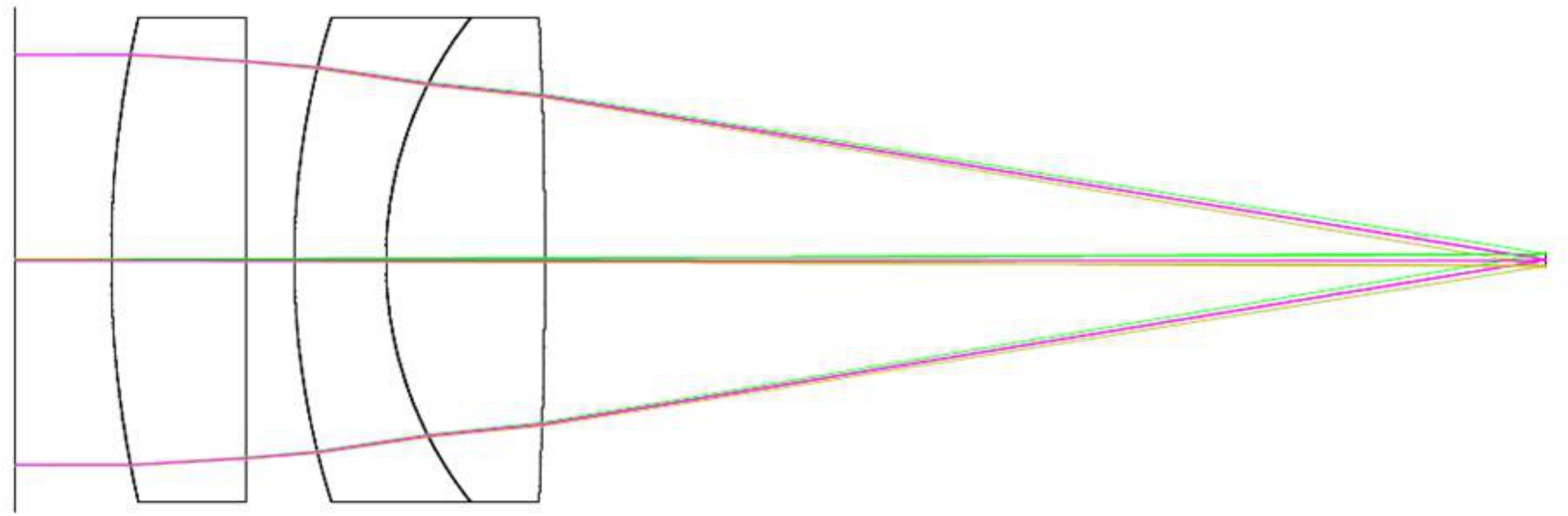}\\
{\it b) \hspace{50mm} c)}

%\begin{center}
\includegraphics[angle=0,width=45mm, clip]{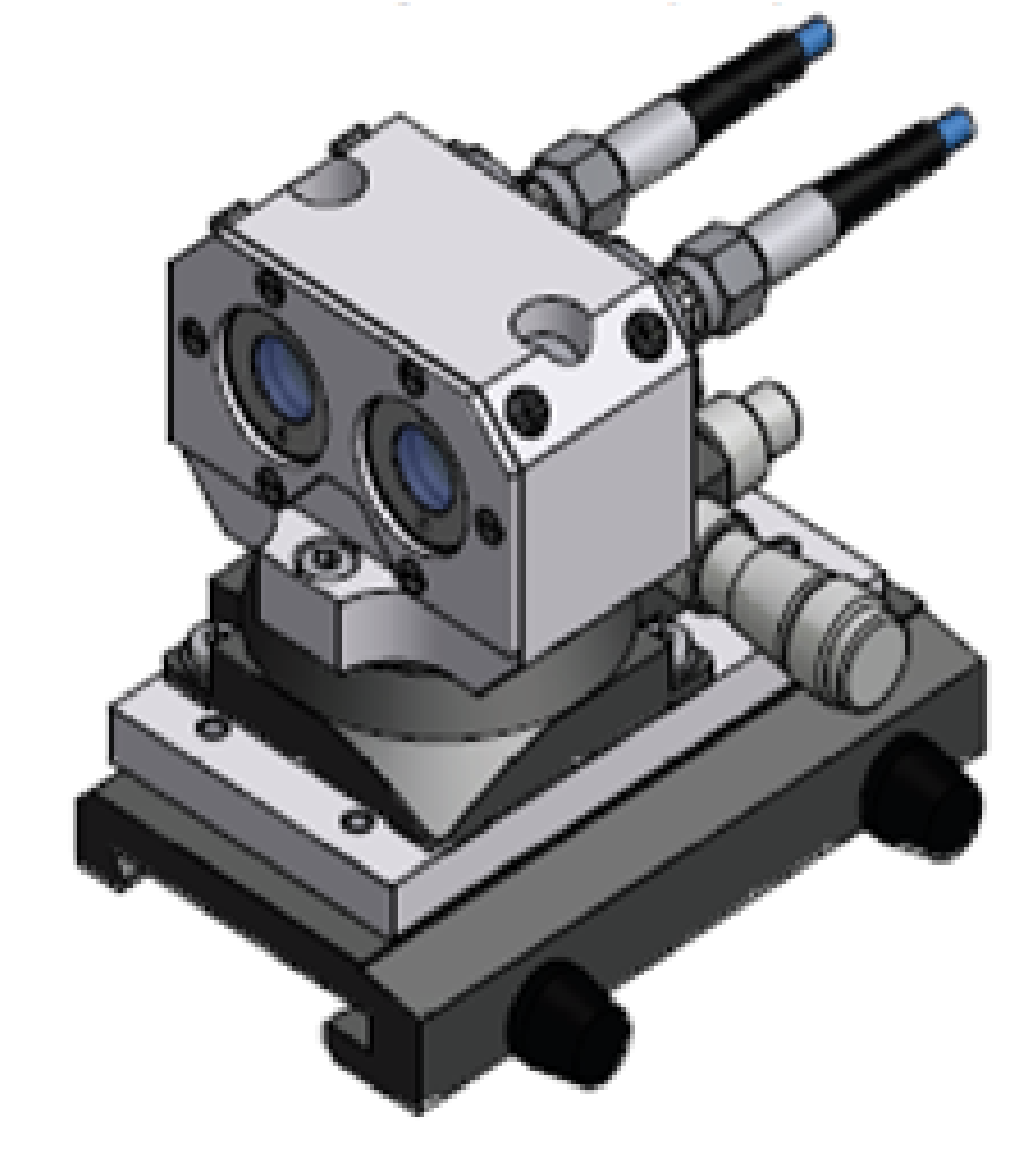}\hspace{10mm}
\includegraphics[angle=0,width=25mm, clip]{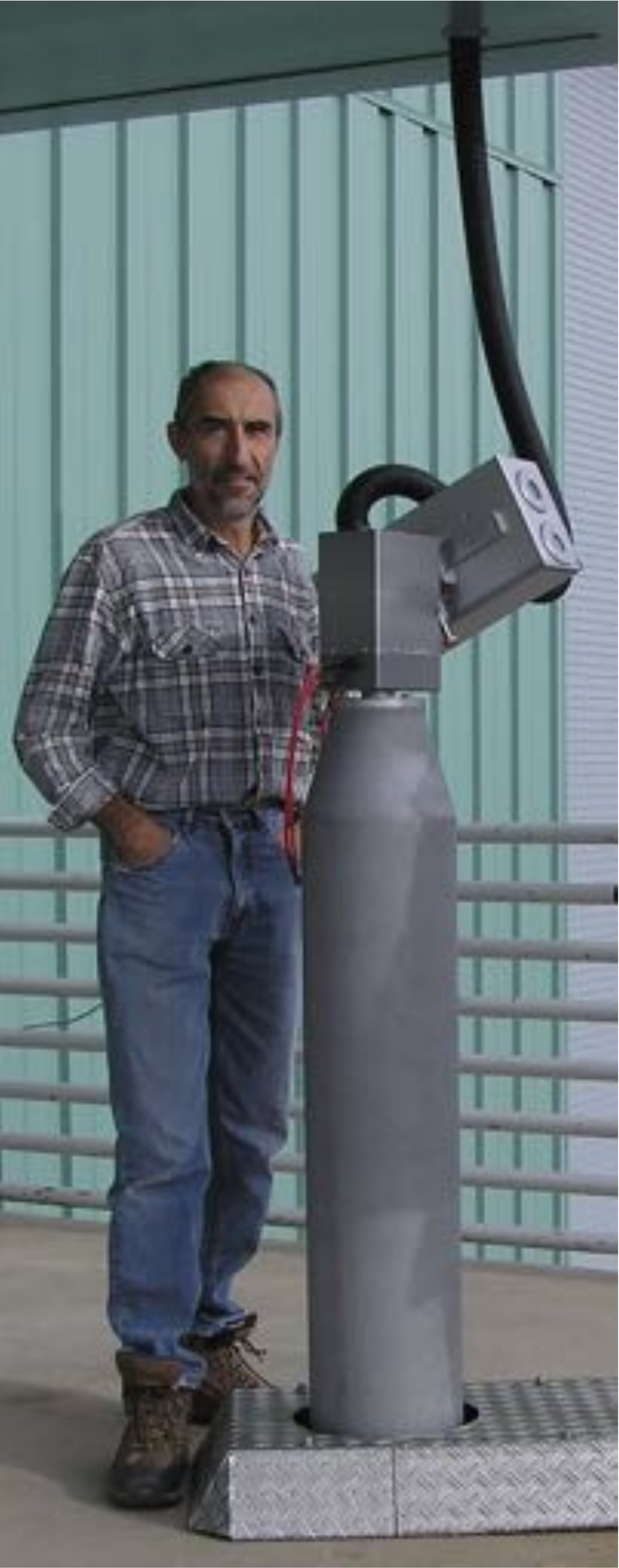}
%\end{center}
\caption{\emph{a}) Optical design of one of the two SDI telescopes.  Left is the 10\,mm diameter f/3.1 triplet lens, right is the 300\,$\mu$m fibre entrance. The total length of the telescope is just 31\,mm. \emph{b}) Mechanical design of the telescope. \emph{c}) Assembled unit at LBTO. The lower entrance aperture is the guiding telescope, the upper aperture holds both SDI telescopes. }\label{F-SDI-optdesign}
%\vspace{-2mm}
\end{figure}

\subsection{Guiding system}

The mount of the SDI-telescope is a modified Alt-Az SPTRV5 Sun tracker system from Dr.\,Schulz \& Partner GmbH. The system is mounted on a 1.5-m pier on the LBTO kitchen balcony. We modified the system by replacing its commercial visor with a guiding telescope feeding its own CCD, a Baseler camera. Two operation modes are possible. The first mode allows full manual operation for engineering purposes. In the second mode the Sun is centered to a pre-calculated position and aided by the signal from the guiding telescope. The RPC interface makes it possible to operate the Sun-guider system remotely in almost real time. The tracker is connected to a server computer via the RS232 port. On this server an RPC server program is running which can receive the remote calls from another computer via ethernet. In this way, we can verify at any time whether clouds had dimmed or blocked the sunlight.

\subsection{Telescopes and fibres}

The two telescopes are 10\,mm f/3.1 air-split triplets and are designed to feed a pair of 300\,$\mu$m-core circular fibres. Figure~\ref{F-SDI-optdesign} shows the optical design. The entrance aperture is stopped down to an 8.5\,mm beam. The system collects 95\,\% of the energy within a centroid of 14\,$\mu$m in diameter. A sapphire window protects the optics. The pair of 40-m fibres redirects the light into the calibration unit on the PEPSI support bridge where it is available as one of the light sources for further distribution. Another pair of 300\,$\mu$m fibres (39\,m length each) bring the light up to the two PFUs where it is converted to f/15 and injected into the respective 100\,$\mu$m octagonal science fibres (44\,m) that feed the spectrograph just like any other night target.

The geometric light loss at the 300\,$\mu$m$\longrightarrow$100\,$\mu$m junction is close to a factor 10. The total loss due to the fibre length of 125\,m plus the two fibre-coupling losses add up to a grand total of over a factor 10. With a PEPSI photon flux of one photon per second for a $V$ = 0-mag star at 540\,nm, we expect a flux of 1$\times$10$^{10}$ photons/s from SDI. Therefore, we still need a neutral filter for additional attenuation. This filter is currently a metallic film with an optical density of 3.8\,dex for the wavelength range 250--2000\,nm from Baader Planetarium, Inc., but will be replaced with a custom-made wedged optical filter. With the current filter the photon flux transforms to integration times of between 10\,ms--2\,s per exposure (2\,s for the bluest wavelength range).

%------------------------------ F full frame for one CD
\begin{SCfigure*}
\includegraphics[angle=0,width=120mm, clip]{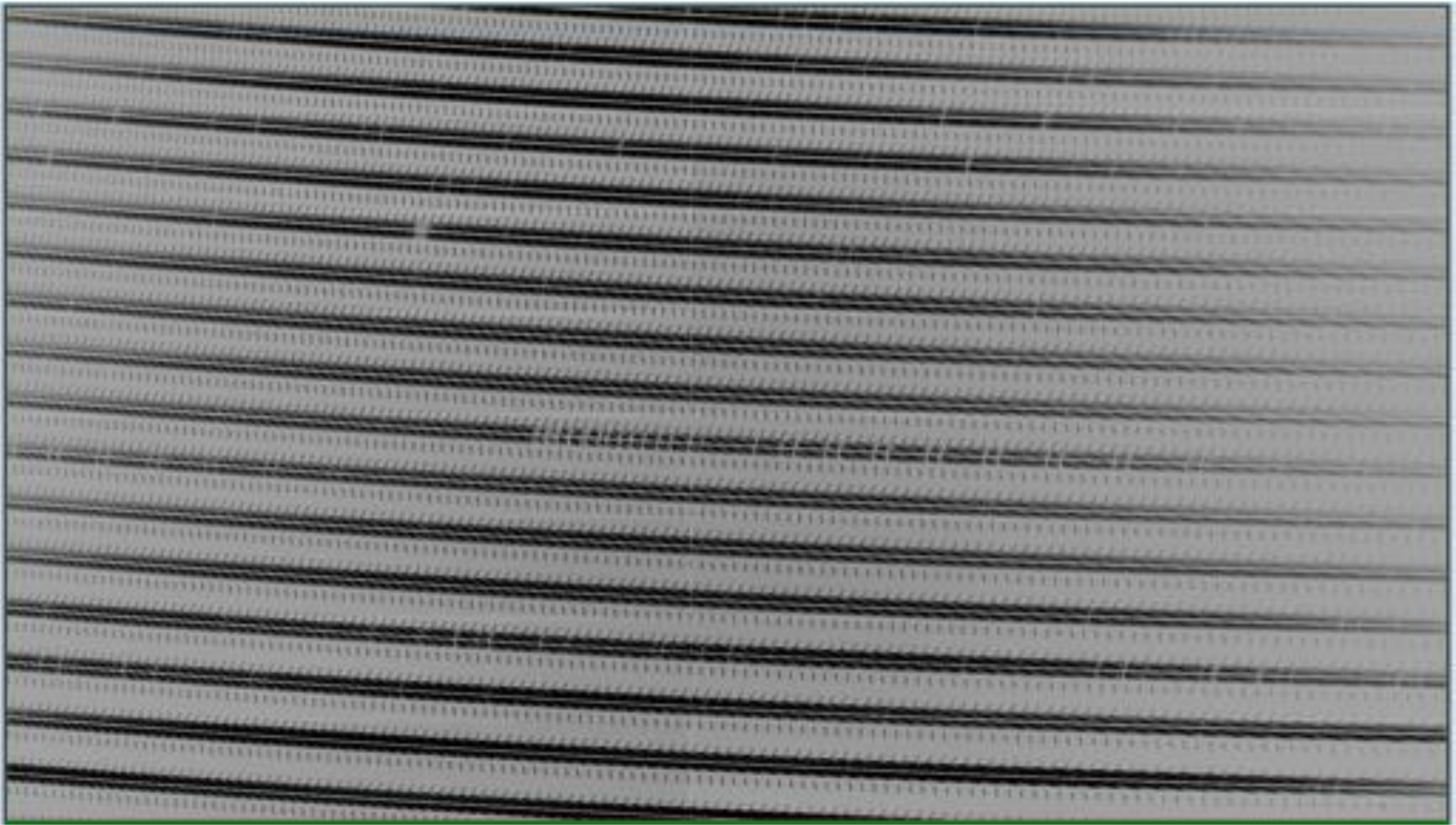}
\caption{Full frame readout (inverted intensities). Shown is a full 10k$\times$10k solar CCD image from the red arm with cross disperser~V for the wavelength range 627.8\,nm (top left corner) to 714.9\,nm (bottom right corner). The image is expanded in dispersion direction for better visibility. The two solar spectra are the two dark lanes flanked above and below by a Fabry-P\'erot spectrum. The broad line in the upper left quarter is \Halpha . The edge of the telluric absorption series of the O$_2$ 1--0 and 2--1 bands are seen approximately in the middle. }\label{F-fullframe}
\end{SCfigure*}

%------------------------------ F trace of the solar spectrum  >> LAM coverage 5230 Angs
\begin{figure*}
\includegraphics[angle=0,width=\textwidth, clip]{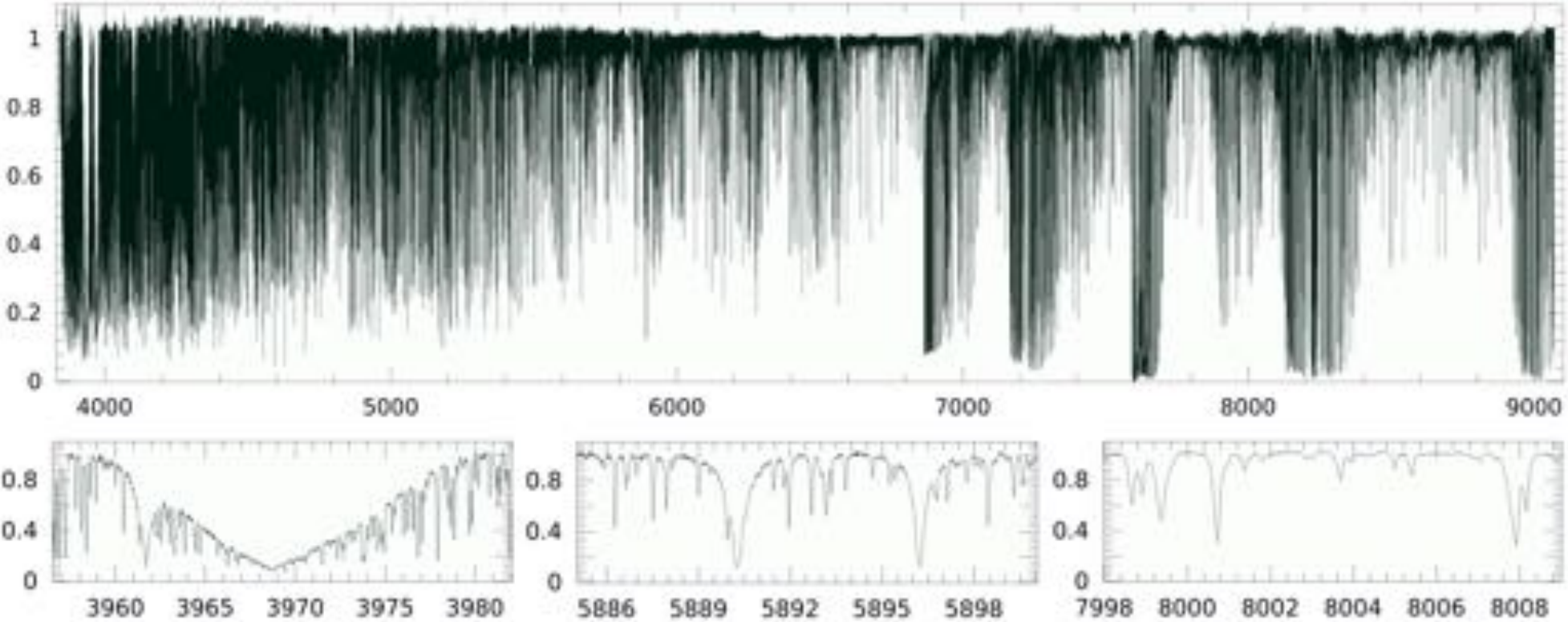}
\caption{Full solar spectrum from 383--907\,nm (top panel). The spectrum is made up of all six cross-disperser settings and was combined with the help of the PEPSI data-reduction package S4S. Its average wavelength dispersion is 7\,m\AA\ per pixel and comprises a total of $\approx$\,737\,000 pixels. Note that only the middle slice of the seven slices of one of the two SDI spectra is shown. The \emph{lower panels} show zoomed subsets for three selected wavelength regions. }\label{F-solarspec}
\end{figure*}

\subsection{Daily operation}

Operation of SDI is fully automated. The position of the Sun is pre-calculated based on the geographical coordinates and the date and time provided by the internal clock. The telescope will move to this position once a height above the horizon of 15\degr\ is reached. If the Sun is recognized, a fine positional adjustment is carried out with the parallel mounted acquisition and guiding telescope. If successfully acquired, the guiding loop is closed. If not, the telescopes keeps tracking and the procedure is repeated once per second. The sky limit in within the Sun is observed is again set by the minimum Sun height, currently 3\degr\ in the west and 15\degr\ in the east (values between 1--15\degr\ are possible). If the minimum Sun height is reached the guiding loop is interrupted, the azimuth of the Sun rise for the next day calculated, and the telescope moved there to wait until the Sun rises above that limit on the next day. If for one day the Sun is not reaching the limit height (due to e.g. clouds) the system moves to the southern horizon and waits until the Sun reaches the limit value again. The system continues working in this operation mode even after switching the device off and on, e.g. after an electrical power outage. In case of an emergency the system can be brought into its safe mode by pressing the Start/Stop button.

%------------------------------ F zoom into frames for all three R modes
\begin{figure*}
{\it a) \hspace{85mm} b)}

\includegraphics[angle=0,width=169mm, clip]{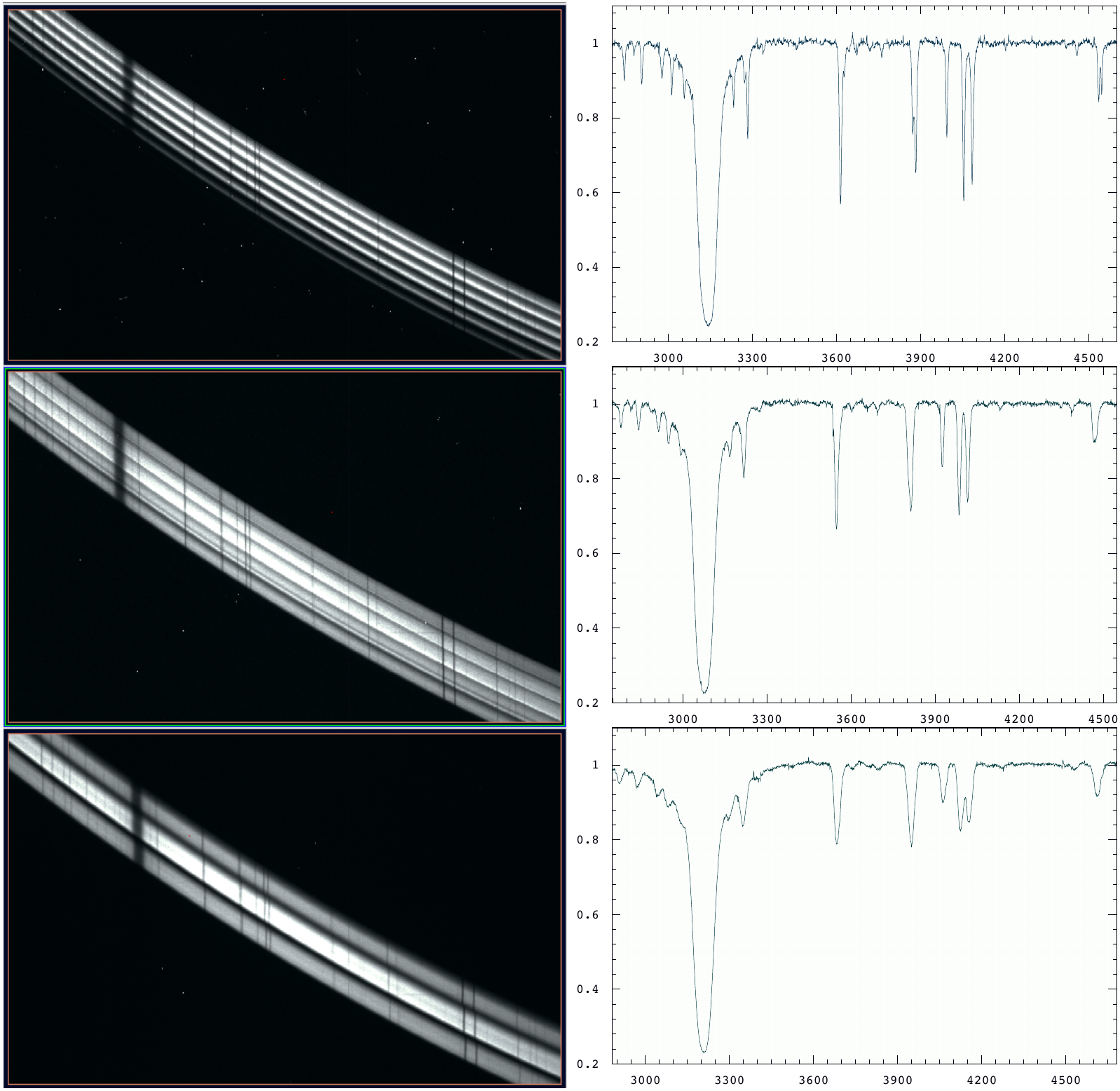}
\caption{Zoom of a fraction of one selected \'echelle order (\#93 in CD\,V) for all three resolution modes when in VATT feed. The $x$-axis is in pixels at a dispersion of 12\,m\AA/pix. Note that a DX target spectrum is not recorded when the VATT is used (single eye). Target is the G8.5V star $\tau$~Ceti. \emph{a}) Two dimensional spectra redwards of \Halpha, \emph{b}) one-dimensional spectra of the same range. From top to bottom; 100\,$\mu$-fibre and seven slicer mode, 200\,$\mu$-fibre and five slicer mode, 300\,$\mu$-fibre and three slicer mode. S/N ratio for the region shown is, from top to bottom,  S/N = 260\,:1, 290\,:1, and 410\,:1, respectively. }\label{F-framezoom}
\vspace{1mm}
\end{figure*}

The RPC interface makes it possible to also operate the Sun-tracker system remotely. The tracker is connected to a server computer via its RS232 port. On this server, a RPC server program is running which can receive the remote calls from another computer via ethernet. In this way, we may execute solar observing proposals beyond the daily monitoring program.

\newpage

\subsection{Data acquisition and storage}

Daily monitoring of the Sun is SDI's default science operation. Forty consecutive exposures with 10\,s integration time (40\,s in the blue), 55\,s CCD readout time, 9\,s transfer and writing time, 4\,s CCD cleaning and 1\,s overhead take $\approx$\,60\,min. Three such sequences for all cross-dispersers are required to cover the entire wavelength range of the spectrograph. This takes 3$\times$60\,+\,5 = 185\,min, including 5\,min overheads for the cross-disperser selection. Then the whole cycle is repeated depending on the weather conditions. In routine operation and clear sky, we thus make 2$\times$300 solar exposures per day with each exposure consisting of one 223\,MB CCD image resulting in 600 images occupying 134\,GB of disk storage per day. A S/N ratio of $\approx$\,1000\,:1 per spectrum will sum up to $\approx$\,5000\,:1 in the  co-added spectrum.

The SAN Data Storage System amounts to 32\,TB of disk space. The slow 27\,TB storage is used for SDI data for about 200 days. The fast 5\,TB storage (100 MB/s) is used for immediate image processing. The raw and processed data are stored to a Magnetic Tape Library. A PowerVault Magnetic Library LT4000 with up to 48 cartridges loaded in four drives has the capacity of 120\,TB with \hbox{LTO-6} media (2.5\,TB per cartridge). The tape library capacity is enough to collect 600~days of uncompressed data with a rate of 200\,GB per day.

%------------------------------------------------------------------------------------------------------------------------------
\section{Pre-commissioning spectra}

\subsection{Solar spectra with SDI}

Figure~\ref{F-fullframe} is an example of a raw single-frame readout with image slicer \#\,1 in $R$ = 270\,000 mode. It shows the recorded \'echelle spectrum from both SDI telescopes in the middle. The red arm with CD\,V and the 100-$\mu$m fibre was used. Note again that the solar spectrum is recorded twice (one spectrum from each telescope) flanked above and below by one Fabry-P\'erot spectrum. The set-up of this first-light exposure did not allow the same separation in cross dispersion for both Fabry-P\'erot spectra with respect to the solar spectra but this is corrected in the second image slicer stack, called slicer \#2. In the exposure shown, the strongest solar absorption line is \Halpha . The edge of the absorption series of the O$_2$ 1--0 and 2--1 bands (part of the B-band) at \,687\,nm is seen approximately in the center of the image. The strongest of these lines reach zero intensity.

A one-dimensional trace of the full solar spectrum as seen by PEPSI is presented in Fig.~\ref{F-solarspec}. It comprises of spectra from all six cross dispersers with a total of 92 \'echelle orders which were merged with the S4S data-reduction package into a single spectrum of some 737\,000 pixels in length. The average dispersion is 12\,m\AA\ per pixel and the wavelength coverage is from 383.7 to 906.7\,nm. The spectrum was taken on 2014 November~12. Integration time per exposure was 1\,s for CD\,VI, 10\,s for CDs\,III-V, 20\,s for CD\,II, and 40\,s for the very blue cross disperser CD\,I. Count rate is nearly equal in all CDs. Average S/N ratio for a single spectrum is 1500\,:1 per pixel (S/N $\approx$ 500\,:1 per pixel per slice in the continuum). The seven slices per fibre are not equal in intensity and the S/N multiplication factor is therefore not ${\sqrt{2{\times} 7}=3.7}$ but $\approx$\,3.0. Note that the continuum setting in the very blue and very red parts of the spectrum is not final yet. Also note that only the mid slice of the seven slices of one of the two telescopes is used for the plot in Fig.~\ref{F-solarspec}. The wavelength solution is based on contemporaneous \hbox{Th-Ar} spectra.

The lower panels in Fig.~\ref{F-solarspec} show three selected wavelength regions from the full solar  spectrum. From left to right, the Ca\,{\sc ii} H line at 396.8\,nm, the sodium D lines at 589\,nm, and the region near 800.3\,nm with several CN lines of interest. It shall just demonstrate the spectrum morphology across our wavelength coverage. Note that telluric lines were not removed in the PEPSI spectrum.

%------------------------------ F ISM spectra of alpha Cyg
\begin{figure}[!b]
\includegraphics[angle=0,width=82mm, clip]{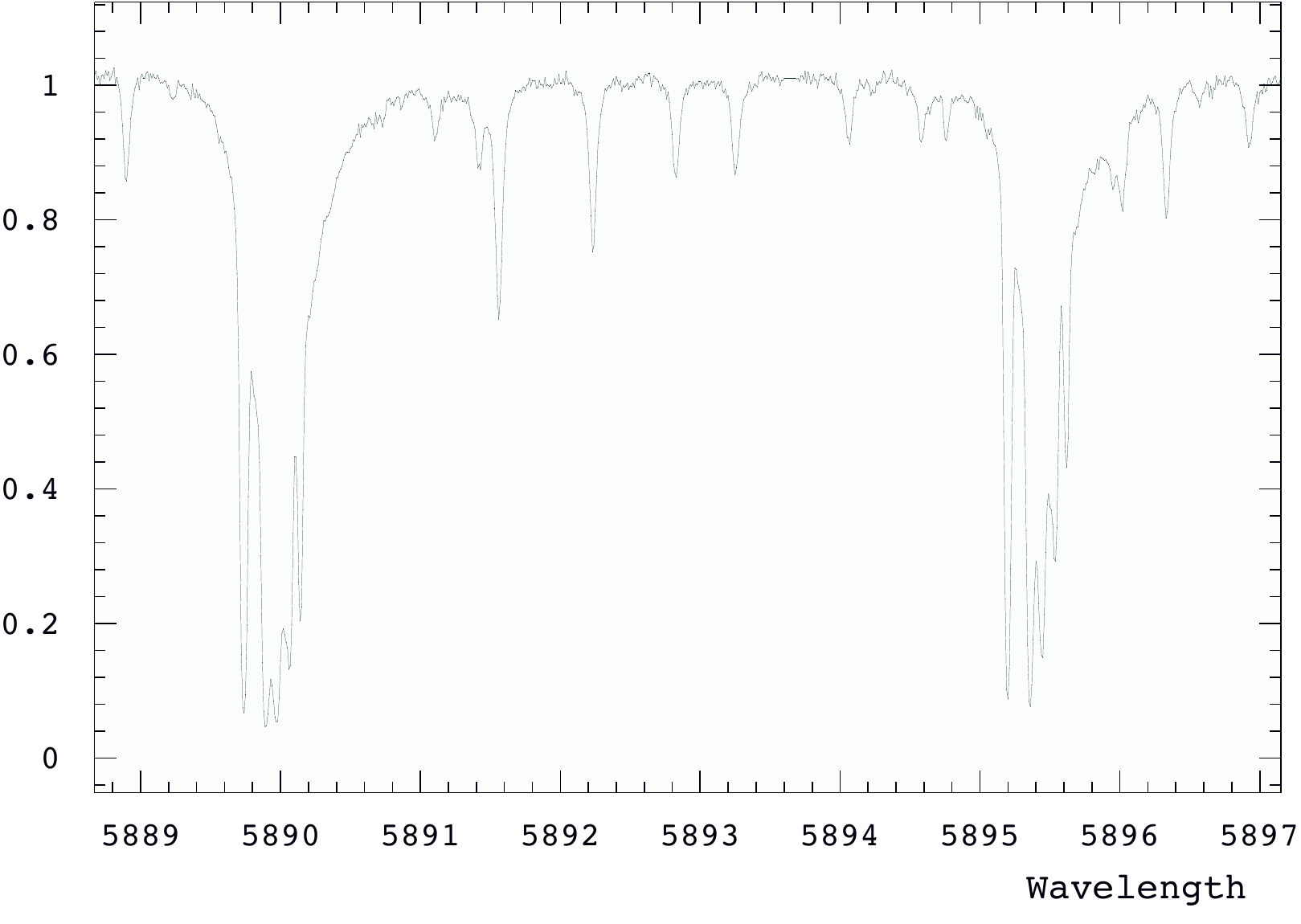}
\caption{The Na\,{\sc i} $D_1$ and $D_2$ interstellar line system towards $\alpha$\,Cyg. The spectrum was taken with the VATT but with a 200\,$\mu$m circular fibre step-down into the 100\,$\mu$m octagonal fibre. Spectral resolution in this case is $\approx$\,200\,000. Note that the lines around and between $D_1$ and $D_2$ are all of telluric origin.}\label{F-alphaCyg}
\vspace{-2mm}
\end{figure}

%\newpage

\subsection{VATT observations}

A first pre-LBT commissioning run was carried out 2014 September~15--22 with the 500-m fibre link to the VATT.  This run was prior to the arrival of the final field lens and the final image-slicer block and thus prior to final spectrograph alignment. Nevertheless, its test spectra could be used for a first system definition and initial trouble shooting. Note that when in VATT mode, PEPSI records only a single target spectrum due to the VATT's single-eye nature and thus also only one calibration-light spectrum, if enabled at all. Unfortunately, out of the seven nights just two partial nights were available due to clouds and/or high humidity. However, the one better night offered occasionally a peak seeing of around 0.4\arcsec . Our targets were \object{$\tau$\,Cet}, \object{$\alpha$\,Cyg}, \object{$\delta$\,Cyg}, and \object{HD\,194937}.

Figure~\ref{F-framezoom} compares a zoomed section of a single \'echelle order from the three resolution modes taken with the VATT. The target is the G8.5V-star $\tau$\,Ceti and the \'echelle order is \#93 which contains the \Halpha\ line. The different number of slices per order is evident in Fig.~\ref{F-framezoom}a. Note that for these spectra no simultaneous calibration light was recorded and therefore the \'echelle order appears a bit ``lonely'' on the 10k CCD (660\,pixels are available per \'echelle order in cross dispersion direction). Again, the spectral resolutions achieved with the VATT set-up are not the nominal ones because of the (engineering) fibre step-downs in between. The spectra shown have resolutions of approximately 200\,000, 100\,000, and 30\,000, from top to bottom. $\tau$\,Ceti was observed with all cross dispersers and with all fibre modes. Signal-to-noise is between 280--560\,:1, being worst for the bluest wavelengths where the 500-m fibre link transmits only about 5\,\%\ of the stellar light and is not offered in routine operation.

Figure~\ref{F-alphaCyg} is a spectrum of the Na\,{\sc i} D lines of $\alpha$\,Cygni (A2Ia). All of the visible spectral lines within the D$_1$ and D$_2$ profiles are of interstellar origin. Our engineering set-up with the VATT allowed for a maximum spectral resolution of $\approx$\,200\,000 (1.5\,\kms ). Note that the VATT fibre coupling is laid out for the 200-$\mu$m fibre (maximum ${R\approx 120\,000}$) and does not support the other fibre-core diameters. Feeding its light into the 100-$\mu$m octagonal fibre needed for the ultra-high resolution is only recommended for engineering purposes because of the large light losses and other issues like f/ratio changes. Even in this preliminary stage and even without direct Gaussian fitting, both D-line spectra in Fig.~\ref{F-alphaCyg} detect 7 of the 11 cloud components that, e.g., Welty et al. (\cite{welty94}) had deduced from a $R$ = 600\,000 spectrum.  Similar is the case for the sodium D features in $\delta$\,Cyg (B9.5IV), where we also see the main cloud component. Besides, our spectrum in Fig.~\ref{F-alphaCyg} is not corrected for telluric lines.

%------------------------------ F hi-res spectrum of HD194937
\begin{figure}
{\it a)}

\includegraphics[angle=0,width=83mm, clip]{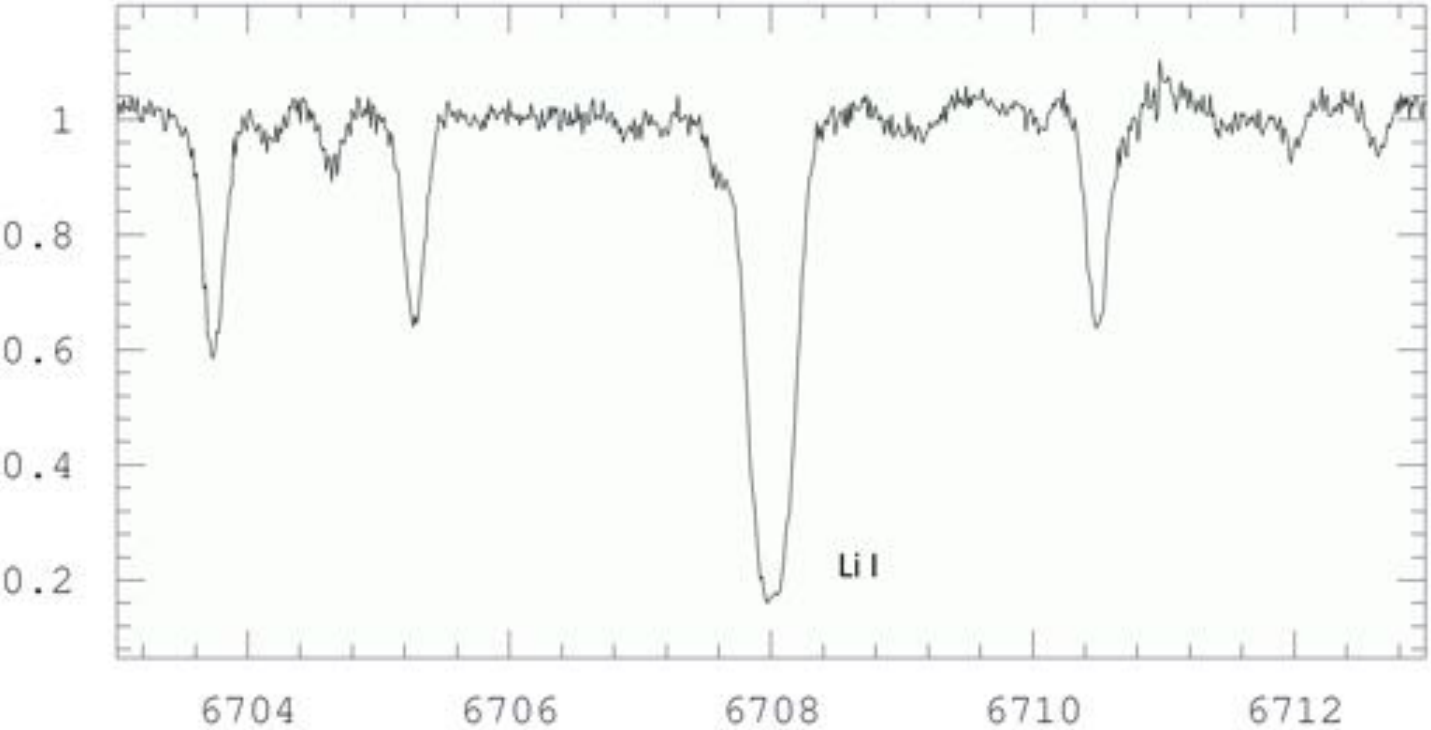}\\
{\it b)}

\includegraphics[angle=0,width=83mm, clip]{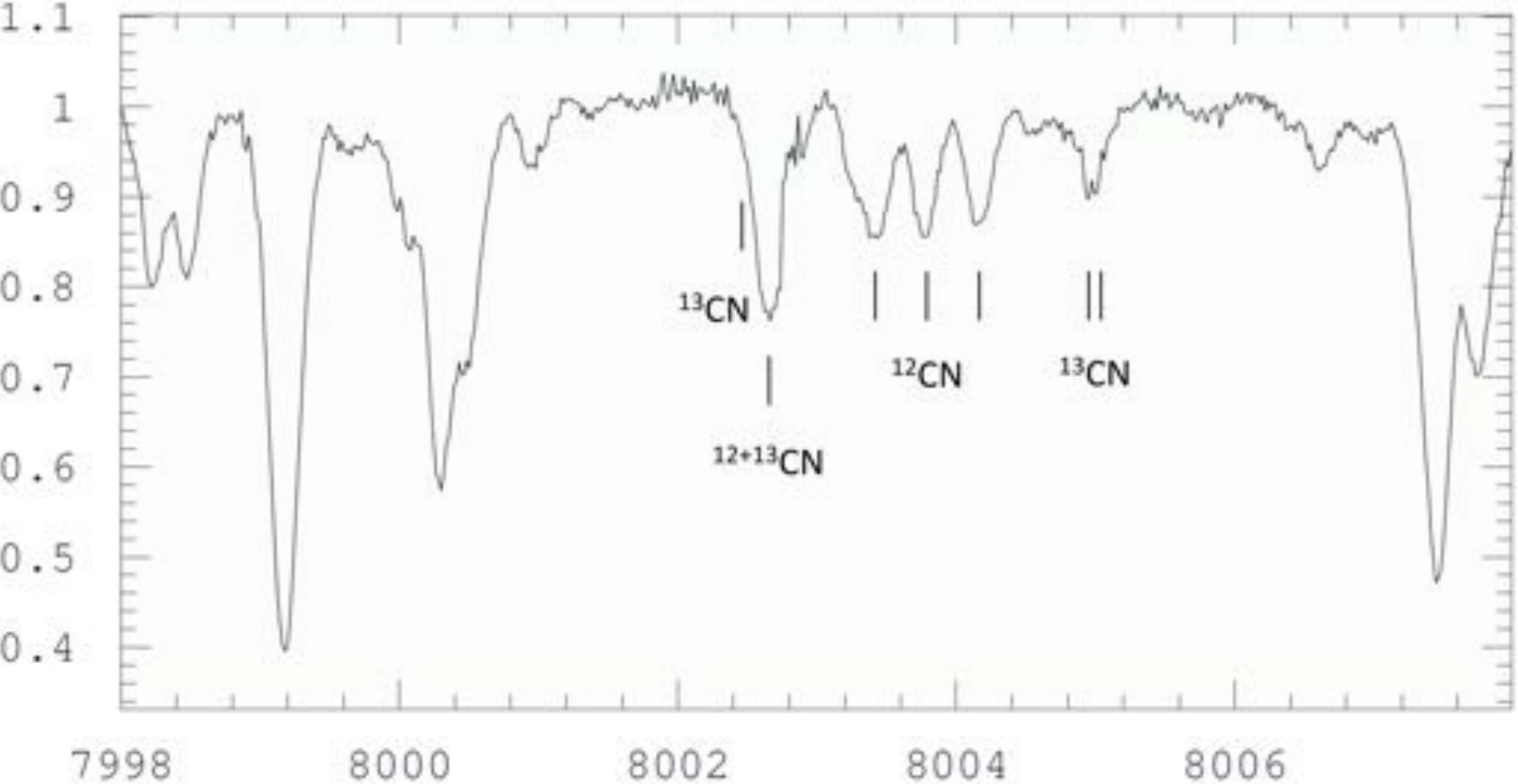}\\
\caption{A $R\approx100\,000$ spectrum of the super Li-rich red giant HD\,194937 obtained with the VATT telescope and a 500-m fibre. \emph{a}) The Li\,{\sc i} 670.8-nm region (note the enormous Li line strength with a residual core intensity of less than $\approx$0.2) and, \emph{b}), the 800.4-nm region with several $^{12}$CN and $^{13}$CN bands used for the determination of the carbon isotope ratio. }\label{F-hd194}
\vspace{-2mm}
\end{figure}

Several spectra of HD\,194937 (G9.5III) were taken to inspect the longest wavelength ranges of PEPSI. This 6\fm2-star is a super lithium-rich giant with a logarithmic Li abundance of $\log n$ = 3.2, as re-determined recently from a high-resolution ($R$ = 60\,000) spectrum by Kumar et al. (\cite{kum11}). According to Luck \& Heiter (\cite{luc:hei}), the star is a first-ascent red giant within the bump region but had no $^{12}$C/$^{13}$C  determination so far which could validate its evolutionary position in the H-R diagram. 

In this paper, we present and quickly analyze a spectrum with ${R\approx 100\,000}$ taken with the VATT on 2014 September~20--21 (Fig.~\ref{F-hd194}). Following Fekel (\cite{fekel}), we determine the rotational broadening from the FWHM of several weak iron and calcium lines in the red spectral regions to 4.5$\pm$0.5\,\kms , assuming a macroturbulence broadening of 3\,\kms . The value of our $^{12}$C/$^{13}$C measurement is based on a few molecular lines from $^{13}$C$^{14}$N and $^{12}$C$^{14}$N in the 800.35--800.46\,nm region using a similar procedure to that described in Strassmeier et al. (\cite{hde}) (instead of a full LTE analysis). The given line broadening does not allow to cleanly separate the two $^{13}$C$^{14}$N lines at 800.6\,nm, despite the high spectral resolution, which makes the isotope ratio a bit uncertain but a most probable value of $^{12}$C/$^{13}$C = 4$\pm$1 is obtained. Note that the error is the internal measuring error. If confirmed, this value would be among the lowest carbon isotope ratios ever measured for an RGB star.

%As a rule of thumb, a photon noise error of 1 m/s (or S/N = 100) can be achieved for a 6th
%magnitude G-dwarf in 60 seconds in the HARPS mode.
%Rule of thumb; rms(RV) = 100 / S/N

%------------------------------------------------------------------------------------------------------------------------------
\section{Summary}

This paper presents a detailed technical description of all major components of the new high-resolution spectrograph and polarimeter for the Large Binocular Telescope. At the time of writing the instrument is still under commissioning and the two polarimeters not delivered yet. We do not present nominal-quality spectra at this point in time but show the functionality with the aid of sunlight and of several bright stars via a 500-m fibre link to the 1.8-m VATT.

PEPSI's unique capabilities are its very-high resolution mode achieved with a sky aperture of the median seeing of the LBT site and its full Stokes-vector capability at $R$ = 120\,000 spectral resolution. Unlike instruments like HARPS-S and HARPS-N, where the entire spectrograph is within a vacuum vessel, PEPSI achieves its stability through an actively temperature and pressure-controlled chamber and the simultaneous recording of the target spectrum and the fringes from a sealed Fabry-P\'erot etalon.

\acknowledgements
PEPSI was co-funded by the German BMBF-Verbundforschung with grants 05AL2BA1/3 and 05A08BAC which are kindly acknowledged here. Numerous technical experts at various companies and research institutes contributed to the final success of PEPSI, in particular we want to thank Michael Andersen, now in Copenhagen, Ramona Eberhardt and her crew from IOF-Jena, and our late head of administration at AIP, Peter A. Stolz, who passed away much too early. The state ministry of science, research, and culture (MWFK) of Brandenburg is thanked for their decade-long support without which such a project would not be doable. We also thank the staff of the Semiconductor Technology Associates and the University of Arizona Imaging Technology Laboratory for their camera/controller development activities. It is also a real pleasure to thank the entire LBTO staff who helped in the long-haul of the realization of this project, in particular John Little, Joar Brynnel, R. Mark Wagner, John Hill, Richard Green, and Christian Veillet. We also thank the other LBT instrument teams, most notably the LUCI team and Walter Seifert, who always shared their experience with us. Last but not least, we thank Paul Gabor and his staff from the Vatican Observatory in Tucson for enabling the fibre link between the VATT and PEPSI. We also mention and thank PepsiCo for allowing us to use and alter their brand logo. KGS sincerely thanks his wife Elfi and kids for not ice-bucketing him earlier in the project (nor later, or similar).

%  -------------------------------------  R e f e r e n c e s


\begin{thebibliography}{}

\bibitem[2010]{avila2012}
Avila, G. 2012, SPIE, 8446-9

\bibitem[2010]{avila2010}
Avila, G., Singh, P., \& Chazelas, B. 2010, SPIE, 7735-268

\bibitem[1996]{elodie}
Baranne, A., Queloz, D., Mayor, M., et al. 1996, A\&AS, 119, 373

\bibitem[2012]{echmod}
Barnes, S. I. 2012, SPIE, 8550-1

\bibitem[2001]{bau:wal}
Baudrand, J., \& Walker, G. A. H. 2001, PASP, 113, 851

\bibitem[2008]{beckert}
Beckert, E., Strassmeier, K. G., Woche, M., et al.
%%Eberhardt, R., T\"unnermann, A., \& Andersen, M. I. 
2008, SPIE, 7018-82

\bibitem[2003]{mike}
Bernstein, R. A., Shectman, S. A., Gunnels, S. M., Mochnacki, S., \& Athey, A. E. 2003, SPIE, 4841-1694

\bibitem[2012]{sta}
Bredthauer, R., Boggs, K., Bredthauer, G., \& Lesser, M. 2012, SPIE, 8453-1M

%\bibitem[2004]{steles}
%Castilho, B. V., Delabre, B. \& Gneiding, C. D. 2004, SPIE, 5492, 433

\bibitem[2011]{igor}
Di Varano, I., Strassmeier, K. G., Ilyin, I., Woche, M., \& Kaercher, H. J. 2011, SPIE, 8336, 83360W

\bibitem[1997]{fekel}
Fekel, F. C. 1997, PASP, 109, 514

\bibitem[2014]{golota}
Golota, T., De La Pena, M. D., Biddick, C., et al. 2014, SPIE, 9152, 9152-89

\bibitem[1994]{sarg}
Gratton, R. G., Bhatia, R. K., \& Cavazza, A. 1994, SPIE, 2198, 309

\bibitem[2012]{hill}
Hill, J. M., Green, R. F., Ashby, D. S., et al. 2012, SPIE, 8444-1

\bibitem[1962]{hop}
Hopkins, H. H.  1962, Proc. of the Physical Society, 79, 889

\bibitem[2000]{ii2000}
Ilyin, I. 2000, PhD Thesis, Univ. of Oulu

\bibitem[2012]{ii}
Ilyin, I. 2012, \an, 333, 213

\bibitem[2011]{ilya:etal}
Ilyin, I., Strassmeier, K. G., Woche, M., Diones, F., \& Di Varano, I. 2011, \an, 332, 753


%\newpage

\bibitem[2011]{kum11}
Kumar, Y. B., Reddy, B. E., \& Lambert, D. L. 2011, ApJ, 730, L12

\newpage

\bibitem[2012]{itl}
Lesser, M. 2012, SPIE, 8453-1L

\bibitem[2014]{itl_qe}
Lesser, M. 2014, SPIE, 9154, 9154-18

\bibitem[2006]{lov:pep}
Lovis, C., Pepe, F., Bouchy, F., et al. 2006, SPIE, 6269, 23

\bibitem[2007]{luc:hei}
Luck, R. E., Heiter, U. 2007, AJ, 133, 2464

\bibitem[1973]{nor}
Norlen, G. 1973, Phys. Scripta, 8, 249

\bibitem[1983]{pal:eng}
Palmer, B. A., \& Engleman, R. Jr. 1983, {Atlas of the Thorium Spectrum},
ed. H. Sinoradzky, Los Alamos National Laboratory

\bibitem[2000]{harps}
Pepe, F., Mayor, M., Delabre, B., et al. 2000, SPIE, 4008, 582

%\bibitem[2013]{espresso}
%Pepe, F., Cristiani, S., Rebolo, R., et al. 2013, Messenger, 153, 6

\bibitem[2010]{mods}
Pogge, R. W., Attwood, B., Belville, S. R., et al. 2006, SPIE, 6269-16

\bibitem[2007]{reif}
Reif, K., M\"uller, P., Klink, G., Polder, M., \&  Poschmann, H. 2007, \an, 328, 711

%\bibitem[2014]{rey:kos}
%Reynolds, R. O., \& Kost, A. 2014, SPIE Montreal, in press

\bibitem[2015]{sab}
Sablowski, D., Pl\"uschke, D., Weber, M., \& Strassmeier, K. G. 2015, \an, submitted

\bibitem[2004]{sam}
Samoylov, A.V., Samoylov, V.S, Vidmachenko, A.P., \& Perekhod,
A.V. 2004, \jqsrt, 88, 319

%\bibitem[2010]{chiron}
%Schwab, C., Spronck, J. F. P., Tokovinin, A., \& Fischer, D. A., 2010, SPIE 7735-4G

\bibitem[2010]{seif:luci}
Seifert, W., Ageorges, N., Lehmitz, M,, et al. 2010, SPIE, 7735-256

\bibitem[2010]{agw12}
Storm, J., Hill, J., Miller, D., et al. 2010, SPIE, 7733-167

\bibitem[2015]{hde}
Strassmeier, K. G., Carroll, T. A., Weber, M., \& Granzer, T. 2015, A\&A, 574, A31

\bibitem[2004]{sci:case}
Strassmeier, K. G., Pallavicini, R., Rice, J. B., Andersen, M., \& Zerbi, F. 2004, \an, 325, 278

\bibitem[2008]{str:spie}
Strassmeier, K. G., Woche, M., Ilyin, I., et al. 2008, SPIE, 7014-22

\bibitem[2014]{apf}
Vogt, S. S., Radovan, M., Kibrick, R. et al. 2014, PASP, 126, 359

%\bibitem[1998]{nso-fts}
%Wallace, L., Hinkle, K., \& Livingston, W. 1998, NSO Tech. Report 98-001

\bibitem[1994]{welty94}
Welty, D. E., Hobbs, L. M., \& Kulkarni, V. P. 1994, ApJ, 436, 152


\end{thebibliography}
\end{document}